\documentclass[twocolumn,twocolappendix]{aastex7}
\usepackage{amsmath}
\usepackage{float}
\usepackage{booktabs}

\newcommand{\swift}{\emph{Swift}}
\newcommand{\chandra}{\emph{Chandra}}
\newcommand{\wise}{\emph{WISE}}
\newcommand{\galex}{\emph{GALEX}}

\begin{document}

\title{Multiwavelength Analysis of Six Luminous Fast Blue Optical Transients}


\author[0009-0003-2780-704X]{Cassie Sevilla}
\affiliation{Department of Astronomy, Cornell University, Ithaca, NY 14853, USA}
\email{jms949@cornell.edu}

\author[0000-0002-9017-3567]{Anna~Y. Q. Ho}
\affiliation{Department of Astronomy, Cornell University, Ithaca, NY 14853, USA}
\email{ayh24@cornell.edu}

\author[0000-0002-8070-5400]{Nayana A.J.}
\affiliation{Department of Astronomy, University of California, Berkeley, CA 94720-3411, USA}
\affiliation{Berkeley Center for Multi-messenger Research on Astrophysical Transients and Outreach (Multi-RAPTOR), University of California,
Berkeley, CA 94720-3411, USA}
\email{nayan89deva@gmail.com}

\author[0000-0001-6797-1889]{Steve Schulze}
\affiliation{Center for Interdisciplinary Exploration and Research in
             Astrophysics (CIERA),
             Northwestern University,
             1800 Sherman Ave, Evanston, IL 60201, USA}
\email{steve.schulze@fysik.su.se, steve.t.schulze@gmail.com}

\author[0000-0001-8472-1996]{Daniel A.~Perley}
\affiliation{Astrophysics Research Institute, Liverpool John Moores University, IC2, Liverpool Science Park, 146 Brownlow Hill, Liverpool L3 5RF, UK}
\email{D.A.Perley@ljmu.ac.uk}

\author{Michael Bremer}
\affiliation{Institut de Radio Astronomie Millimétrique (IRAM), 300 rue de la Piscine, 38406 Saint Martin d’Hères, France}
\email{bremer@iram.fr}


\author[0000-0002-8977-1498]{Igor Andreoni}
\affiliation{Department of Physics and Astronomy, University of North Carolina at Chapel Hill, Chapel Hill, NC 27599-3255, USA}
\email{Igor.Andreoni@unc.edu}

\author{Ivan Altunin}
\affiliation{Department of Physics, University of Nevada, Reno NV 89557, USA}
\email{ialtunin@unr.edu}

\author[0000-0001-5955-2502]{Thomas G.~Brink}
\affiliation{Department of Astronomy, University of California, Berkeley, CA 94720-3411, USA}
\email{tgbrink@berkeley.edu}

\author[0000-0001-5593-0721]{Michael Camilo}
\affiliation{Department of Astronomy, Cornell University, Ithaca, NY 14853, USA}
\email{mc2923@cornell.edu}

\author[0000-0002-0844-6563]{Poonam Chandra}
\affiliation{National Radio Astronomy Observatory, 520 Edgemont Road, Charlottesville, VA 22903-2475, USA}
\email{pchandra@nrao.edu}

\author[0000-0003-0853-6427]{Ping Chen}\affil{Institute for Advanced Study in Physics, Zhejiang University, Hangzhou 310027, China}\affil{Institute for Astronomy, School of Physics, Zhejiang University, Hangzhou 310027, China}\affil{Department of Particle Physics and Astrophysics, Weizmann Institute of Science, 234 Herzl St, 7610001 Rehovot, Israel}
\email{chenp1220@gmail.com}

\author[0000-0001-9842-6808]{Ashley A. Chrimes}
\affiliation{European Space Agency (ESA), European Space Research and Technology Centre (ESTEC), Keplerlaan 1, 2201 AZ Noordwijk, the Netherlands}
\affiliation{Department of Astrophysics/IMAPP, Radboud University, PO Box 9010, 6500 GL Nijmegen, The Netherlands}
\email{ashley.chrimes@esa.int}

\author[0000-0002-8262-2924]{Michael W. Coughlin}
\affiliation{School of Physics and Astronomy, University of Minnesota, Minneapolis, MN 55455, USA}
\email{cough052@umn.edu}

\author[0000-0001-8372-997X]{Kaustav K.~Das}
\affiliation{Cahill Center for Astrophysics, California Institute of Technology, MC 249-17, 
1200 E California Boulevard, Pasadena, CA, 91125, USA}
\email{kdas@astro.caltech.edu}

\author{Andrew Drake}
\affiliation{Cahill Center for Astrophysics, California Institute of Technology, MC 249-17, 1200 E California Boulevard, Pasadena, CA, 91125, USA}
\email{ajd@astro.caltech.edu}

\author[0000-0003-3460-0103]{Alexei V. Filippenko}
\affiliation{Department of Astronomy, University of California, Berkeley, CA 94720-3411, USA}
\email{afilippenko@berkeley.edu}

\author[0000-0002-4223-103X]{Christoffer Fremling}
\affiliation{Caltech Optical Observatories, California Institute of Technology, Pasadena, CA 91125, USA}
\affiliation{Division of Physics, Mathematics and Astronomy, California Institute of Technology, Pasadena, CA 91125, USA}
\email{fremling@caltech.edu}

\author[0009-0006-7990-0547]{James~Freeburn}
\affiliation{Centre for Astrophysics and Supercomputing, Swinburne University of Technology, Victoria 3122, Australia}
\affiliation{ARC Centre of Excellence for Gravitational Wave Discovery (OzGrav), Victoria 3122, Australia}
\affiliation{Department of Physics and Astronomy, University of North Carolina at Chapel Hill, Chapel Hill, NC 27599-3255, USA}
\email{jfreeburn@swin.edu.au}

\author[0000-0002-3653-5598]{Avishay Gal Yam}
\affiliation{Department of Particle Physics and Astrophysics, Weizmann Institute of Science, 234 Herzl St, 7610001 Rehovot, Israel}
\email{avishay.gal-yam@weizmann.ac.il}

\author[0009-0009-5187-4123]{Mary Gerhart}
\affiliation{Department of Astronomy, University of California, Berkeley, CA 94720-3411, USA}
\email{mary.gerhart@berkeley.edu}

\author[0000-0002-3168-0139]{Matthew J.~Graham}
\affiliation{Division of Physics, Mathematics and Astronomy, California Institute of Technology, Pasadena, CA 91125, USA}
\email{mjg@caltech.edu}

\author[0000-0003-3367-3415]{George Helou}
\affiliation{Division of Physics, Mathematics and Astronomy, California Institute of Technology, Pasadena, CA 91125, USA}
\email{gxh@ipac.caltech.edu}

\author[0000-0002-0129-806X]{K-Ryan Hinds}
\affiliation{Astrophysics Research Institute, Liverpool John Moores University, IC2, Liverpool Science Park, 146 Brownlow Hill, Liverpool L3 5RF, UK}
\email{khinds@caltech.edu}

\author[0009-0008-8062-445X]{Natalya Johnson}
\affiliation{Department of Physics, Drexel University, Philadelphia, PA 19104, USA}
\email{np699@drexel.edu}

\author[0000-0002-5619-4938]{Mansi M.~Kasliwal}
\affiliation{Division of Physics, Mathematics and Astronomy, California Institute of Technology, Pasadena, CA 91125, USA}
\email{mansi@astro.caltech.edu}

\author[0000-0003-0871-4641]{Harsh Kumar}
\affiliation{Center for Astrophysics \textbar{} Harvard \& Smithsonian, 60 Garden Street, Cambridge, MA 02138-1516, USA}
\affiliation{The NSF AI Institute for Artificial Intelligence and Fundamental Interactions, USA}
\email{harshkumar@fas.harvard.edu}

\author[0000-0003-2451-5482]{Russ R. Laher}
\affiliation{IPAC, California Institute of Technology, 1200 E. California Blvd, Pasadena, CA 91125, USA}
\email{laher@ipac.caltech.edu}

\author[0000-0002-2249-0595]{Natalie LeBaron}
\affiliation{Department of Astronomy, University of California, Berkeley, CA 94720-3411, USA}
\affiliation{Berkeley Center for Multi-messenger Research on Astrophysical Transients and Outreach (Multi-RAPTOR), University of California, Berkeley, CA 94720-3411, USA}
\email{nlebaron@berkeley.edu}

\author[0009-0001-6911-9144]{Maggie L.~Li}
\email{maggieli@caltech.edu}
\affiliation{Cahill Center for Astrophysics, California Institute of Technology, MC 249-17, 1200 E California Boulevard, Pasadena, CA, 91125, USA}
\email{maggieli@caltech.edu}

\author[0000-0002-7866-4531]{Chang~Liu}
\affil{Department of Physics and Astronomy, Northwestern University, 2145 Sheridan Rd, Evanston, IL 60208, USA}
\affil{Center for Interdisciplinary Exploration and Research in Astrophysics (CIERA), Northwestern University, 1800 Sherman Ave, Evanston, IL 60201, USA}
\affil{NSF-Simons AI Institute for the Sky (SkAI), 172 E. Chestnut St., Chicago, IL 60611, USA}
\email{ptg.cliu@u.northwestern.edu}

\author[0000-0001-8405-2649]{Ben~Margalit}
\affiliation{School of Physics and Astronomy, University of Minnesota, Minneapolis, MN 55455, USA}
\email{margalit@umn.edu}

\author[0000-0003-3658-6026]{Yu-Jing Qin}
\affiliation{Cahill Center for Astrophysics, California Institute of Technology, MC 249-17, 1200 E California Boulevard, Pasadena, CA, 91125, USA}
\email{yujingq@caltech.edu}

\author[0000-0002-5683-2389]{Nabeel~Rehemtulla}
\affiliation{Department of Physics and Astronomy, Northwestern University, 2145 Sheridan Rd, Evanston, IL 60208, USA}
\affiliation{Center for Interdisciplinary Exploration and Research in Astrophysics (CIERA), Northwestern University, 1800 Sherman Ave, Evanston, IL 60201, USA}
\affiliation{NSF-Simons AI Institute for the Sky (SkAI), 172 E. Chestnut St., Chicago, IL 60611, USA}
\email{nabeelrehemtulla2027@u.northwestern.edu}

\author[0000-0002-1637-7668]{Sophia Risin}
\affiliation{Department of Astronomy, University of California, Berkeley, CA 94720-3411, USA}
\email{sbrisin@berkeley.edu}

\author[0000-0003-4725-4481]{Sam Rose}
\affiliation{Division of Physics, Mathematics and Astronomy, California Institute of Technology, Pasadena, CA 91125, USA}
\email{samantharose131@gmail.com}

\author{Rupak Roy}
\affiliation{Institute of Astronomy Space and Earth Science (IASES), P 177, CIT Road, Scheme 7m, Kolkata-700054, West Bengal, India}
\email{rupakroy1980@gmail.com}
             
\author[0000-0001-7648-4142]{Ben Rusholme}
\affiliation{IPAC, California Institute of Technology, 1200 E. California Blvd, Pasadena, CA 91125, USA}
\email{rusholme@ipac.caltech.edu}

\author[0000-0001-9915-8147]{Genevieve~Schroeder}
\affiliation{Department of Astronomy, Cornell University, Ithaca, NY 14853, USA}
\email{gms279@cornell.edu}

\author[0000-0003-1546-6615]{Jesper~Sollerman}
\affiliation{The Oskar Klein Centre, Department of Astronomy, AlbaNova,  Stockholm University, SE-106 91 Stockholm , Sweden}
\email{jesper@astro.su.se}

\author[0000-0002-6428-2700]{Gokul P. Srinivasaragavan}
\affiliation{Department of Astronomy, University of Maryland, College Park, MD 20742, USA}
\affiliation{Joint Space-Science Institute, University of Maryland, College Park, MD 20742, USA}
 \affiliation{Astrophysics Science Division, NASA Goddard Space Flight Center, 8800 Greenbelt Rd, Greenbelt, MD 20771, USA}
 \affiliation{Division of Physics, Mathematics and Astronomy, California Institute of Technology, Pasadena, CA 91125, USA}
\email{gsriniv2@umd.edu}

\author[0009-0000-4756-4223]{Kailai Wang}
\affiliation{Department of Astronomy, Cornell University, Ithaca, NY 14853, USA}
\email{kw395@g.cornell.edu}

\author[0000-0003-0733-2916]{Jacob L. Wise}
\affiliation{Astrophysics Research Institute, Liverpool John Moores University, IC2, Liverpool Science Park, 146 Brownlow Hill, Liverpool L3 5RF, UK}
\email{J.L.Wise@2022.ljmu.ac.uk}

\author[0000-0002-6535-8500]{Yi Yang}
\affiliation{Department of Astronomy, University of California, Berkeley, CA 94720-3411, USA}
\affiliation{Department of Physics, Tsinghua University, Qinghua Yuan, Beijing 100084, China}
\email{yiyangtamu@gmail.com}

\author[0000-0001-6747-8509]{Yuhan Yao}
\affiliation{Miller Institute for Basic Research in Science, 206B Stanley Hall, Berkeley, CA 94720, USA}
\affiliation{Department of Astronomy, University of California, Berkeley, CA 94720-3411, USA}
\email{yuhanyao@berkeley.edu}

\author[0000-0002-2636-6508]{WeiKang Zheng}
\affiliation{Department of Astronomy, University of California, Berkeley, CA 94720-3411, USA}
\email{zwk@astro.berkeley.edu}

\begin{abstract}
    We present \added{early ($\lesssim1\,$yr)} multiwavelength observations and analysis of six luminous fast blue optical transients (LFBOTs). We identified these LFBOTs \added{in Zwicky Transient Facility (ZTF) survey data} from their fast light-curve evolution ($t_{1/2}\leq 12\,$d), blue colors at peak brightness ($g-r\leq-0.\added{2}\,$mag), a visible host galaxy, high optical luminosity ($M_g<-20\,$mag), and an X-ray \added{and/or} radio detection.  
    With the exception of AT2024aehp (ZTF24abygbss), these transients exhibit peaks in their $10\,$GHz radio light curves at $t_{\text{rest}} \approx 50$--100\,d, with peak radio luminosities ranging from $10^{38}$--$10^{40}$\,erg\,s$^{-1}$. Modeling the radio emission as synchrotron radiation indicates a fast ($v=0.1$--$0.3c$) shock in a dense ($n_e\approx10^{3}$--$10^{4}$\,cm$^{-3}$)  medium. The X-ray emission varies by $\approx2$ orders of magnitude in luminosity ($10^{42}$--$10^{44}$\,erg\,s$^{-1}$) at $t_{\text{rest}}\sim20\,$d. 
    Analysis of the host-galaxy photometry and spectroscopy for each transient shows that they are predominantly nonnuclear (a few kpc offset) with star-forming host galaxies of stellar masses $10^{9}$--$10^{11}\,M_\odot$. 
    Unlike all other LFBOTs to date, AT2024aehp exhibited a luminous ($M<-19\,$mag) plateau in the optical light curve\added{ and its }6--15\,GHz radio emission brightened by over an order of magnitude from $t_{\text{rest}} \approx70\,$d to $t_{\mathrm{rest}} \approx130\,$d.
    The mostly consistent radio behavior between \added{these} LFBOTs implies a similar circumburst medium, leading us to prefer a progenitor scenario in which mass is lost in a consistent way shortly prior to the terminal event.
\end{abstract}

\section{Introduction}
The advent of high-cadence optical surveys such as the Asteroid Terrestrial-impact Last Alert System (ATLAS; \citealt{atlas}) and the Zwicky Transient Facility (ZTF; \citealt{ztf1}, \citealt{ztf2}, \citealt{ztf3}) has led to the discovery of new classes of extragalactic transients.  Some \added{of these} transients are referred to as fast blue optical transients (FBOTs), most famously AT2018cow \citep[e.g.,][]{prentice2018}.  FBOTs are typically defined as transients with blue colors ($g-r\leq -0.2$ mag), rapid time evolution ($<12$ days above half-maximum brightness), and extragalactic nature \citep{Drout2014,pursiainen2018,inserra2019}. 

In this paper, we focus on a particular \added{subset} of FBOT\added{s} dubbed ``AT2018cow-like'' transients \citep{ho2023} or ``luminous fast blue optical transients'' (LFBOTs; \citealt{metzger2022, Klencki2025}). The term LFBOT is empirical and has been used to describe optically discovered transients with very similar optical, radio, and X-ray light curves to AT2018cow (e.g., \citealt{coppejans2020, Gutierrez2024, perley2021, ho2022xnd, bright2022, lebaron2025, nayana2025, Perley2026}) as well as blue optical counterparts to gamma-ray bursts (e.g., \citealt{Pieterse2026}). In this paper, we use the term LFBOT to refer to an FBOT with high optical brightness ($M_g < -20\,$mag) and luminous X-ray and/or radio emission.

There are currently eight published transients \added{discovered by optical surveys} that fall under this \added{empirical} classification (Table~\ref{tab:summary}). 
It is theorized that the peak of their optical light curves results from shock breakout from a dense medium \citep{ofek2010, pursiainen2018} or from a jet cocoon \citep{gottlieb2022}. The radio emission is found to be consistent with synchrotron radiation emanating from \added{a fast to} mildly relativistic shock \citep{margutti,coppejans2020,ho2020koala}.  The X-ray radiation is suggested to come from a central X-ray source, separate from the synchrotron radiation that powers the radio \citep{margutti,ho2019}. For objects with high-cadence X-ray observations (AT2018cow, AT2020mrf, AT2022tsd, and AT2024wpp) it was found that the X-rays had fast variability on a timescale of minutes to days \citep{sandoval2018,margutti,yao2022, ho2023tsd, nayana2025,Perley2026}.

Based on the first few LFBOTs discovered, \citet{metzger2022} argued that progenitor models must reconcile two key characteristics: a dense surrounding medium and a long-lived central engine. Some propositions include tidal disruption events (TDEs) involving an intermediate-mass black hole \citep{perley2019,kuin2019,Gutierrez2024}, the direct collapse (failed explosion) of a supergiant star \citep{margutti,perley2019,Chrimes2025}, and the tidal disruption of a star by a stellar-mass black hole resulting from a merger \citep{metzger2022, Klencki2025} or random interactions in a cluster \citep{Kremer2021,Kremer2023}. 


\added{Our understanding of LFBOTs has been limited in part by the small sample size, particularly the number that have multiwavelength data. Some behavior has only been observed in a single object, such as luminous UV emission years after AT2018cow \citep{inkenhaag2025}, late-time luminous X-ray emission in AT2020mrf \citep{yao2022}, and minutes-duration optical flares in AT2022tsd \citep{ho2023tsd}---and it is not clear how prevalent this behavior is among LFBOTs.} The rarity of these transients and the recency of the development of optical surveys like ZTF has limited the discovery rate.

\added{Here we present early ($\lesssim1\,$yr) multiwavelength observations of five new LFBOTs, as well as further radio observations on a sixth (AT2023fhn), increasing the sample size by over 50\%.} 
Section \ref{sec:methods} describes the observations and data reduction. 
\added{In Section \ref{sec:radio-analysis} we model the radio emission as self-absorbed synchrotron radiation.  In Section~\ref{sec:host-analysis} we analyze the offsets from each LFBOT to its host galaxy, and each host galaxy's properties. Throughout, we use methods that are similar to those employed in previous LFBOT papers in order to make effective comparisons.} 
\added{We present our discussion and conclusions in Section~\ref{sec:discussion}, including suggestions for future LFBOT studies.} 

Throughout this paper, we assume a flat $\Lambda$CDM cosmology with H$_0=67.4$ km\,s$^{-1}$\,Mpc$^{-1}$ and $\Omega_m=0.315$ \citep{planck2018}.  All usages of $t_{\text{obs}}$ refer to the time after the optical light curve peak in the observer frame, while $t_{\mathrm{rest}}$ refers to this time in the rest frame.  All dates are in UTC and all magnitudes are in AB.

\begin{deluxetable*}{ccccccccc}[!htb]\label{tab:summary}
\setlength{\tabcolsep}{0.25em}
\tablewidth{0.9\textwidth}
\tabletypesize{\footnotesize}
\tablewidth{0pt}
\tablecaption{LFBOT Summary}
\tablecolumns{7}
\tablehead{
\colhead{\vspace{-2mm}TNS name}&\colhead{Survey name}
& \colhead{Redshift} & \colhead{RA}& \colhead{Dec} & \colhead{Offset} & \colhead{$t_0$} & \colhead{Ref.}\\
\colhead{}&\colhead{}
& \colhead{} & \colhead{(J2000)}& \colhead{(J2000)} & \colhead{(kpc)} & \colhead{(UTC)} & \colhead{}}
\startdata
AT2022abfc &ZTF22abvrxkk& 0.212&4:51:19.200 &$-$26:58:41.59& 2.0 &2022-11-21 08:15:48 & This work \\
AT2023fhn&ZTF23aaeozpp/ATLAS23hnk
&0.24&10:08:03.805 & 21:04:26.70& 17 &2023-04-12 05:22:39& [1,2]; This work \\
AT2023hkw&ZTF23aaimsja &0.339&10:42:17.748 & 52:29:19.03&15&2023-04-30 04:08:52 & This work \\
AT2023vth &ZTF23ableqsp&0.0747&17:56:34.403 & 8:02:37.30& 3 &2023-10-20 03:08:39 & This work \\
AT2024qfm & ZTF24aaxhxhf/PS24hhm & 0.2270 & 23:21:23.450 & +11:56:31.99 & 3.7 & 2024-07-27 08:18:46 & This work; [14] \\
AT2024aehp & ZTF24abygbss & 0.1715& 8:21:07.474 & 28:44:22.17 & 0.47&2024-12-19 10:02:49 & This work \\
\hline\hline
---&CSS161010&0.033&04:58:34.396&$-08$:18:03.95&0.28&2016-10-10 11:31:12& [3,4] \\
AT2018lug&ZTF18abvkwla&0.2714&02:00:15.194&+16:47:57.32&1.9&2018-09-12 09:46:48 & [5] \\
AT2018cow&ATLAS18qqn/ZTF18abcfcoo/Gaia18bqa&0.014145&16:16:00.220&+22:16:04.91&1.7&2018-06-16 10:35:02 & [6] \\
AT2020mrf&ZTF20abfhyil/ATLAS20paa&0.1353&15:47:54.163&+44:39:07.41&1.19&2020-06-15 08:55:40 & [7] \\
AT2020xnd&ZTF20acigmel&0.2433&22:20:02.030&$-02$:50:25.30&0.8&2020-10-12 04:06:03 & [8] \\
AT2022tsd&ZTF22abftjko/PS22jvi&0.2564&03:20:10.863&+08:44:55.63&6&2022-09-07 11:21:22 & [9] \\
AT2024wpp& \added{\begin{tabular}{@{}c@{}}GOTO24gjk/ZTF24abjjpbo/ \\ PS24jhj/ATLAS24ond\end{tabular}}& 0.0868 & 02:42:05.499 & $-16$:57:22.90 &5&2024-09-26 01:24:35& [10--13] \\
\enddata
\vspace{2.5mm}   
\tablecomments{Basic information on each LFBOT, such as position and redshift. 
The six LFBOTs discussed in this paper are above the line, while the rest of the known LFBOT population is shown below.  $t_0$ is chosen to be the peak of the optical light curve.  The redshift errors are to the last reported digit.  Positions of the six LFBOTs presented in the paper and AT2020xnd were determined using the methodology in Section~\ref{sec:offset}.}
\tablerefs{[1] \citet{chrimes2024}, [2] \citet{chrimesfhn}, [3] \citet{coppejans2020}, [4] \citet{Gutierrez2024}, [5] \citet{ho2020koala}, [6] \citet{prentice2018}, [7] \citet{yao2022}, [8] \citet{perley2021}, [9] \citet{ho2023tsd}, [10] \citet{Pursiainen2025}, [11] \citet{nayana2025}, [12] \citet{lebaron2025}, [13] \citet{Perley2026}, [14] Nayana A.J. et al. (in prep)}
\end{deluxetable*}

\section{Observations} \label{sec:methods}

The basic characteristics for all LFBOTs are listed in Table~\ref{tab:summary}.  The discovery details for AT2022abfc, AT2023fhn, AT2023hkw, AT2023vth, AT2024qfm, and AT2024aehp are given in Appendix~\ref{sec:discovery}. Here we provide details on our multiwavelength follow-up observations. 

\subsection{Optical Photometry}
\label{sec:optical-photometry}

After each LFBOT was identified in ZTF photometry (outlined in Appendix~\ref{sec:discovery}), we obtained follow-up optical photometry 
using the Liverpool Telescope (LT; \citealt{lt}) and the SED Machine (SEDM; \citealt{sedm1, sedm2, sedm3}) on the Palomar 60-inch (1.5\,m) telescope (P60; \citealt{p60}). For AT2024aehp, we also have ultraviolet (UV) data from the {\it Neil Gehrels Swift Observatory} (\swift; \citealt{Gehrels2004})$\,$ Ultraviolet/Optical Telescope \citep{swift_uvot}.  
For AT2024qfm, we also present photometry from the Alhambra Faint Object Spectrograph and Camera (ALFOSC) on the 2.56\,m Nordic Optical Telescope (NOT; \citealt{not}), the Inamori-Magellan Areal Camera \& Spectrograph (IMACS; \citealt{imacs}) mounted on the 6.5\,m Magellan-Baade telescope, the Large Monolithic Imager (LMI) on the 4.3\,m Lowell Discovery Telescope (LDT; \citealt{ldt}), and the Goodman High Throughput Spectrograph (GHTS; \citealt{ghts}) camera on the 4.1\,m Southern Astrophysical Research (SOAR) Telescope\footnote{SOAR2024B-021; PI I. Andreoni}. 
\added{We correct for Milky Way extinction using the \texttt{extinction} Python package\footnote{\href{https://github.com/sncosmo/extinction}{https://github.com/sncosmo/extinction}}, using the prescription of \citet{schlafly} and $R_V=3.1$.}

\begin{figure*}[!htb]
 \centering
  \includegraphics[width=\linewidth]{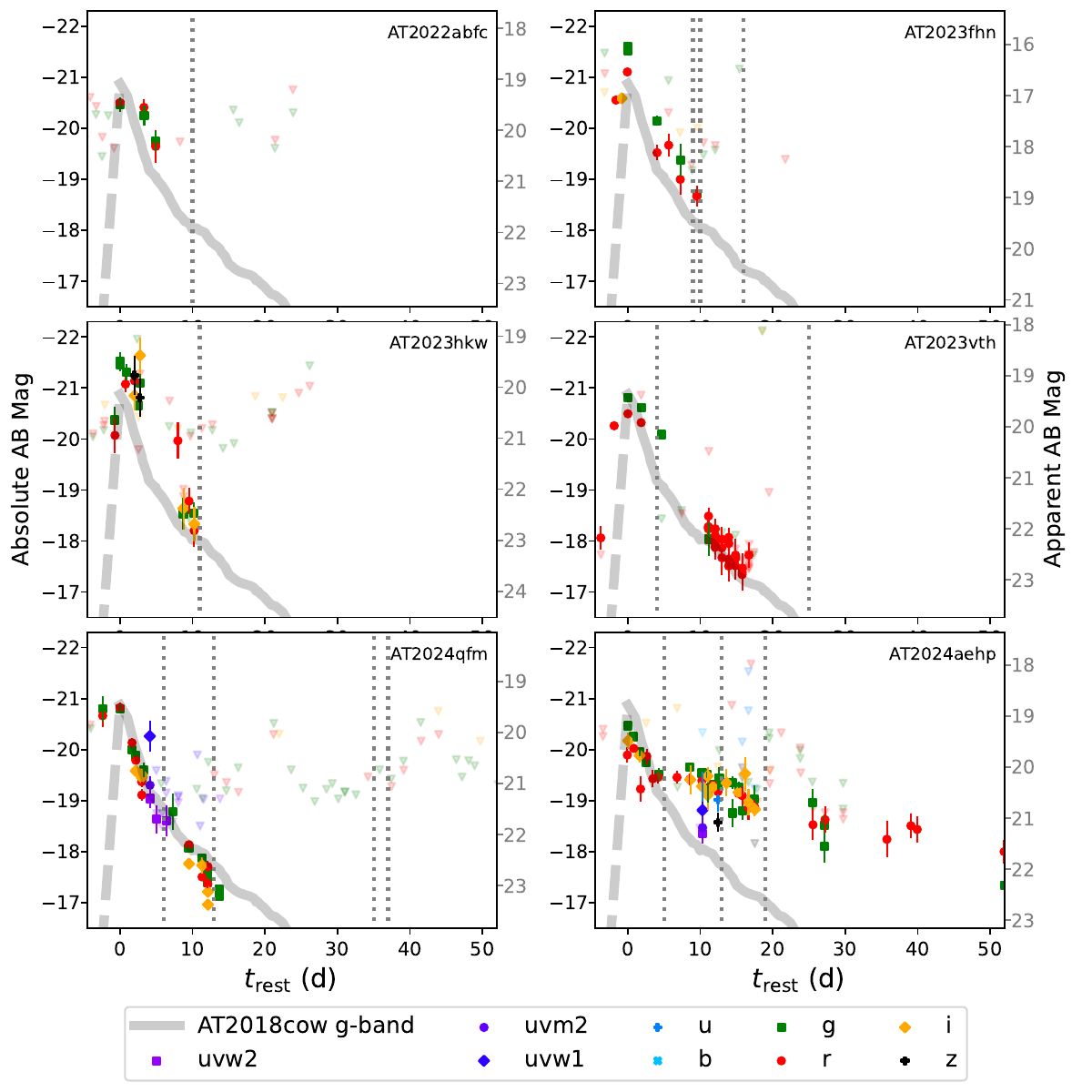}
\caption{Optical light curves of each LFBOT (points), as well as the light curve of AT2018cow (thick gray line, where dashed lines mark the connection to the last upper limit).  The left-hand ordinate values display the $E(B-V)$ corrected absolute AB magnitude \citep{abmag} and are shared for all plots.  The right-hand ordinate displays the apparent AB magnitude.  Each filter has a unique marker color and style. Upside-down triangles represent 3-$\sigma$ upper limits. The vertical dotted lines show when spectra of the transient and/or its host galaxy were taken.  These spectra are displayed in Figure~\ref{fig:spec}.}
\label{fig:optical}
\end{figure*}

The optical light curves of each object are shown in Figure~\ref{fig:optical}.  For each object, $t_{\text{rest}}=t_{\text{obs}}=0$ was chosen to match 
the peak of the transient's optical light curve. To convert to absolute magnitude, we apply the conversion

\begin{equation}
    M = m_{\rm obs} - 5 \log_{10}\left(
    \frac{D_L}{10 \,\mathrm{pc}}
    \right)+2.5 \log_{10}(1+z)\,.
\end{equation}


\subsection{Optical Spectroscopy}
\label{sec:optical-spectroscopy}

As part of our follow-up campaigns, we  collected optical spectra at $t_{\text{obs}}=4$--37 days. We utilized the following facilities: the Gemini Multi-Object Spectrograph (GMOS; \citealt{gmos-n}, \citealt{gmos-s}), the Keck Low-Resolution Imaging Spectrometer (LRIS; \citealt{lris}) on the 10\,m Keck-I telescope, the Keck Cosmic Web Imager (KCWI; \citealt{kcwi}) and the Deep Imaging Multi-Object Spectrograph (DEIMOS; \citealt{deimos}) on the 10\,m Keck-II telescope, the Double Spectrograph on the Palomar 5\,m telescope (DBSP; \citealt{p200_dbsp}), and the Binospec instrument \citep{Fabricant2019} on the 6.5\,m Multiple Mirror Telescope (MMT). \added{The basic details for each spectroscopic observation are provided in Table~\ref{tab:spectra} within Appendix~\ref{sec:appendix-tables}.} 

\begin{figure*}[t]
 \centering
  \includegraphics[width=0.85\textwidth]{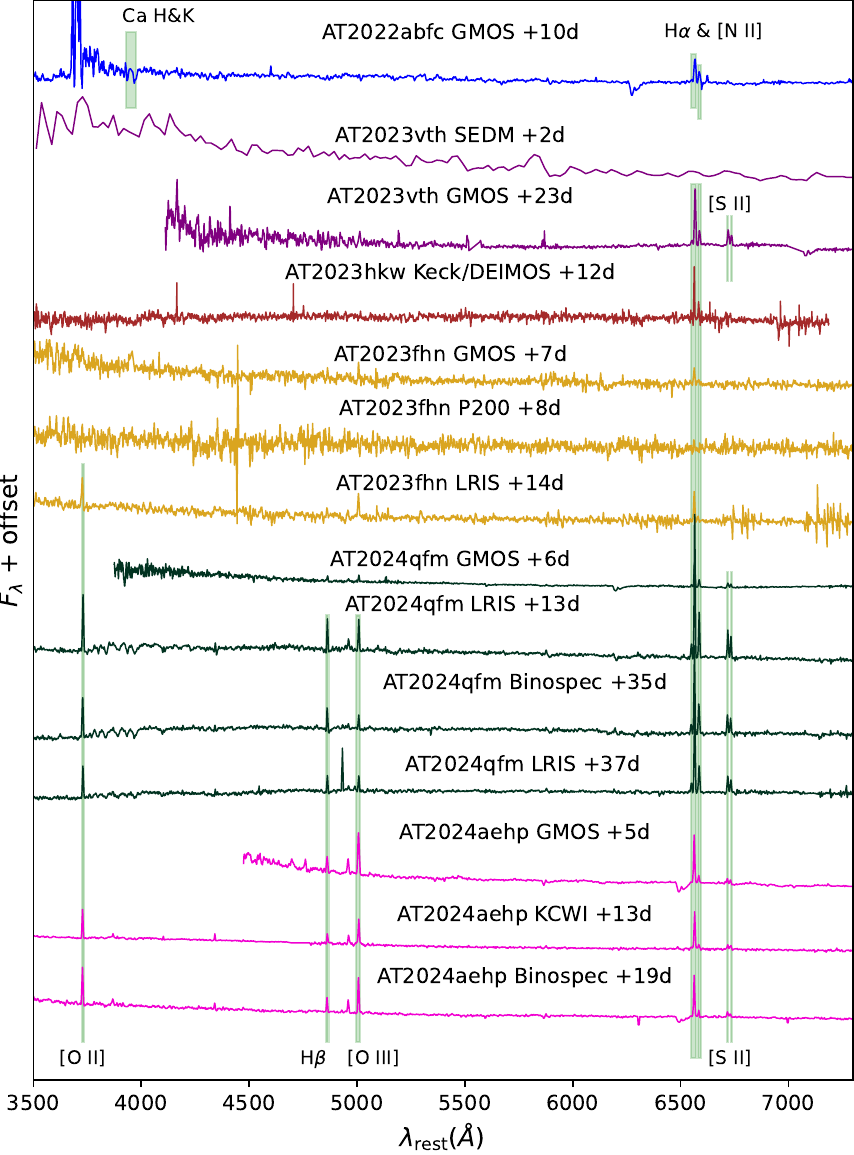}
\caption{Spectra of each LFBOT and/or its host galaxy. For each spectrum we indicate the observer-frame epoch and instrument used.  We bin all spectra with a bin size of  3 \AA.  The green shaded rectangles show  galactic narrow emission lines that were used to determine a redshift. The spectra are displayed with an arbitrary vertical offset.
}
\label{fig:spec}
\end{figure*}

\begin{figure*}[]
 \centering
  \includegraphics[width=\linewidth]{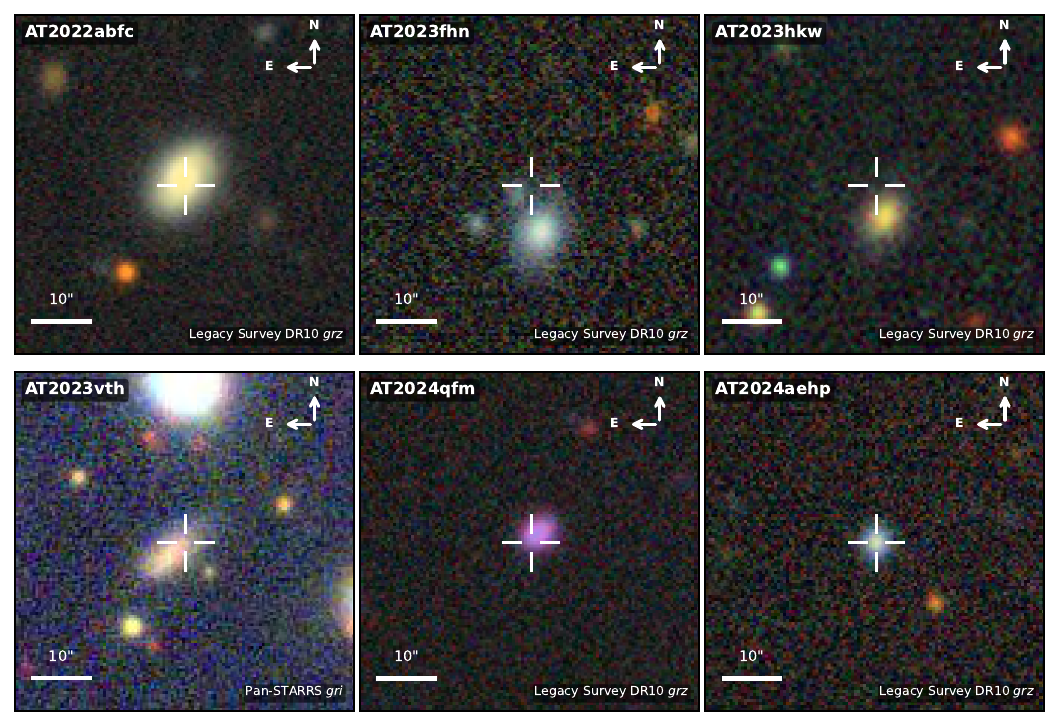}
\caption{ Optical cutouts of the six LFBOT host galaxies we discuss in this paper. In each panel we mark the position of the LFBOT with cross hairs.}
\label{fig:cutouts}
\end{figure*}

\added{Spectra are shown in Figure~\ref{fig:spec}. In all cases narrow emission lines are observed from the LFBOT host galaxies, (e.g., \ion{Ca}{2} H\&K, [\ion{N}{2}], [\ion{O}{2}], [\ion{O}{3}], and [\ion{S}{2}]). The Balmer H$\alpha$ line is prominent for all six objects. Spectra were obtained at the position of the transient (Figure~\ref{fig:cutouts}) at a time when the transient and the host galaxy's brightness are comparable (Figure~\ref{fig:optical}) so they typically contain contributions from both the transient and its host galaxy. All narrow emission lines we identified are well-known galactic emission lines; we do not identify any clear features from the transient itself, which is unsurprising given that at these phases LFBOT spectra are dominated by a blue continuum that is well approximated by a blackbody (e.g., \citealt{margutti,perley2021}).}

\subsection{Swift X-ray Observations}

\begin{figure*}[t]
 \centering
  \includegraphics[width=0.7\linewidth]{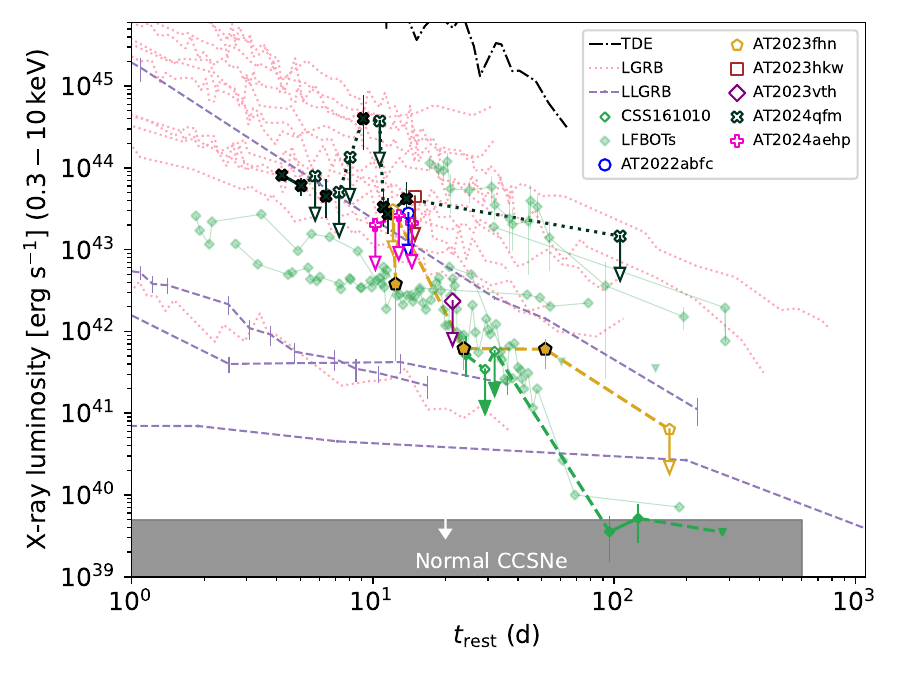}
\caption{0.3--10\,keV X-ray light curves of the six LFBOTs in this paper, in comparison to those of other LFBOTs, long gamma-ray bursts (LGRBs), and low-luminosity gamma-ray bursts (LLGRBs). Open symbols with downward-facing triangles mark 3-$\sigma$ upper limits.  Data are from \citet{pian2000}, \citet{tiengo2004}, \citet{campana2006b}, \citet{soderberg2006b}, \citet{blanchard2017}, \citet{ho2020blt}, \citet{coppejans2020}, \citet{hinkle2021}, \citet{perley2021},\citet{yao2022}, \citet{ho2022}, \citet{perley2023abvizsw}, \citet{ho2023tsd}, \citet{chrimes2024}, \added{\citet{yao2024}}, and \citet{Perley2026}, as well as public \swift\ data.}
\label{fig:xray}
\end{figure*}

X-ray observations were obtained with the \emph{Swift} X-ray Telescope (XRT; \citealt{swift_xrt}) \added{in the 0.3--10\,keV range} by 
ToO triggers shortly after discovery of each LFBOT.  Count rates for the XRT instrument were obtained from the UKSSDC \swift\ light-curve generator\footnote{\href{https://www.swift.ac.uk/user_objects/lc_docs.php}{UKSSDC \swift\ Light-Curve Generator}}.  Utilizing tools from NASA HEASARC, we converted these count rates to an X-ray flux accounting for the \ion{H}{1} column density in the Milky Way. For upper limits (all objects except AT2024qfm), we assume a power-law photon index of $\Gamma=2$, a typical value for previous LFBOTs \citep{sandoval2018, margutti, bright2022,matthews2023}. 

For AT2024qfm, \swift\ obtained 12 observations, from $t_{\text{obs}}\sim8$--133\,d after the first optical detection. We visually inspected the images and discarded one observation (ID 00016748006) because the source landed on the XRT detector's hot column.  The X-ray light curve shows a general decline, with the exception of one observation 
(ID 00016748007) which showed a significant rebrightening. This observation was extremely short, with 233\,s of raw exposure but only 40\,s of clean data. Two source counts were detected at the position of AT2024qfm; using the \citet{Kraft1991} approach of the \swift\ tools, the implied detection significance is 99.83\% (just over 3-$\sigma$). Closer inspection of the data revealed no clear reason why this would be an erroneous measurement.

For purpose of comparison and for completeness, we also analyze three early \emph{Swift} XRT observations of the LFBOT CSS161010. Late-time (100\,d) observations were presented in \citet{coppejans2020}, while upper limits for these early observations were presented in \citet{Gutierrez2024} (see their Appendix~A.1). We find a very marginal (3.3-$\sigma$) detection in the first epoch, and non detections in the subsequent two epochs. Our upper limits are within a factor \added{of three of those in \citet{Gutierrez2024}, who presented binned measurements}.  We also show the \added{0.2--10\,keV} \chandra\ measurements of AT2023fhn published by \citet{chrimesfhn}, using the revised values in \citet{nayana2025}.

The \swift\ XRT observations are summarized in Table~\ref{tab:xray} in Appendix~\ref{sec:appendix-tables}.  Five of the \added{new} LFBOTs had a \swift\ non-detection, resulting in upper limits for the photon count rate and the X-ray flux.  These upper limits are compared to other objects with \swift\ detections in Figure~\ref{fig:xray}. 

\subsection{Radio and Millimeter Observations}

 We obtained radio observations with the National Science Foundation's Karl G. Jansky Very Large Array (VLA; \citealt{vla, evla}), the Giant Metrewave Radio Telescope (GMRT; \citealt{gmrt}), and the Northern Extended Millimeter Array (NOEMA).  All of these observations are listed in Table~\ref{tab:radio} in Appendix~\ref{sec:appendix-tables}.
 
\subsubsection{VLA}

\begin{figure}[!h]
 \centering
  \includegraphics[width=\linewidth]{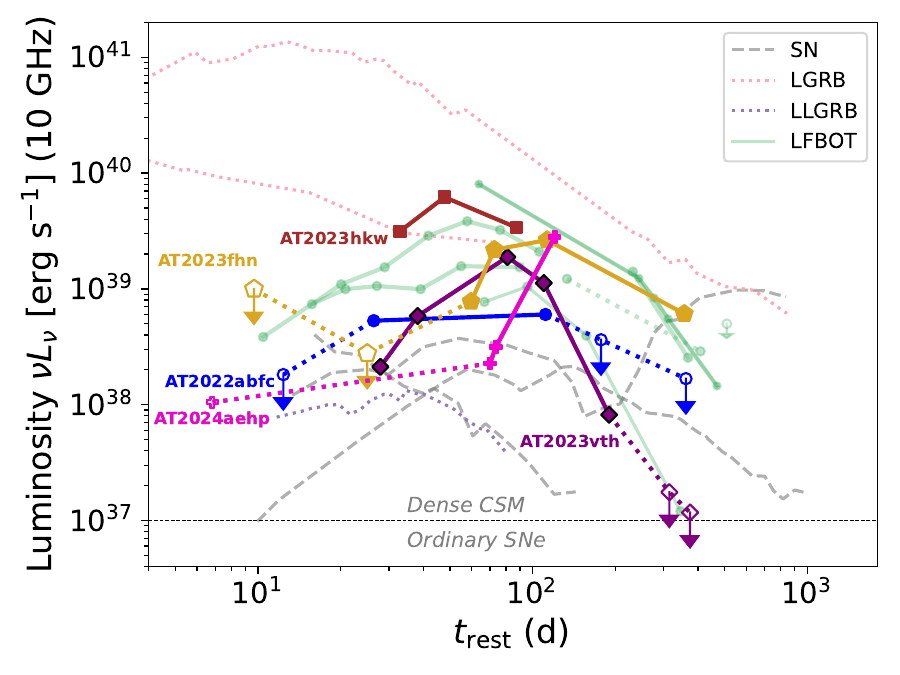}
\caption{Radio light curves at 10 GHz of the LFBOTs with data presented in this paper, compared with other known LFBOTs, long GRBs (LGRBs), low-luminosity GRBs (LLGRBs), and core-collapse supernovae with non-relativistic ejecta (SNe). We measure time after peak light in the transient's rest frame.  Data from \citet{1998bwradio}, \citet{berger2003}, \citet{soderberg2006}, \citet{vanderhorst2008}, \citet{soderberg2010}, \citet{salas2013}, \citet{margutti}, \citet{coppejans2020}, \citet{ho2022xnd}, \citet{yao2022}, \citet{perley2021}, \citet{ho2023tsd}, and \citet{nayana2025}.}
\label{fig:radio-lc}
\end{figure}

\begin{figure*}[]
 \centering
  \includegraphics[width=\linewidth]{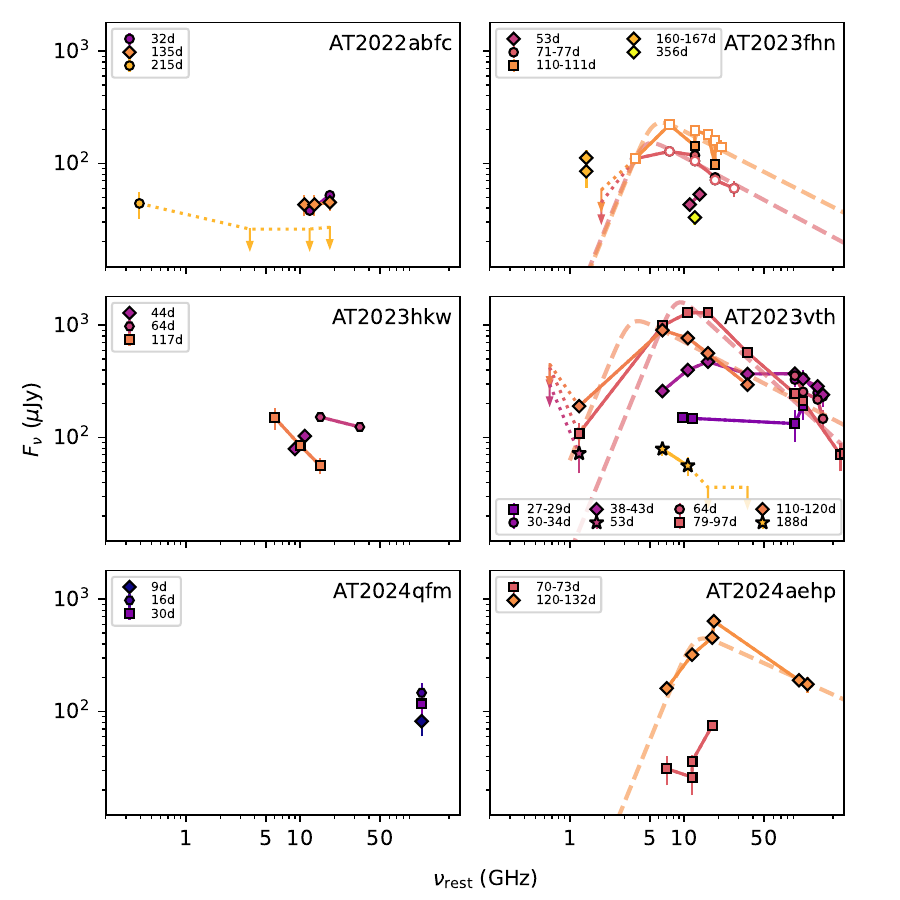}
\caption{The radio and millimeter SED of each LFBOT at different observer-frame epochs.  Observations taken within a few days are grouped into one epoch.  Non-detections are signified by a downward pointing arrow.  For clarity, we omit some non-detections.  For epochs with sufficient data points, we display the broken power-law fit to the SED as a dashed line.  Data from \citet{chrimes2024} for AT2023fhn have markers filled in with white.}
\label{fig:radio-sed}
\end{figure*}

We present VLA observations for five LFBOTs\footnote{AT2022abfc was observed under programs 22A-405 (PI D. Perley), 23A-393, 23A-422, and 23B-138;  AT2023fhn under programs 23A-403 and 23B-138;  AT2023vth under program 23B-138;  AT2023hkw under program 23A-415; AT2024aehp under programs 24B-338 and 25A-367 (PI A. Ho)}, all of which were detected. Radio data for AT2024qfm will be published by Nayana A.J. et al. in prep \added{as per a prior arrangement}. Observations were obtained at mid-frequencies of 3 GHz (S-band), 6 GHz (C-band), 10 GHz (X-band), 16 GHz (Ku-band), and 33 GHz (Ka-band).
Based on the behavior of past sources \added{\citep{yao2022, ho2022xnd, chrimes2024}}, 10 GHz (X-band) gives the most reliable chances of detection, so every epoch features an observation in this band. We calibrate VLA data using the automated pipeline in Common Astronomy Software Applications package version 6.5.4-9 (CASA; \citealt{casa}).   We use Briggs weighting with \texttt{robust = 0.5} \citep{briggs1995}.  When the image quality is degraded by bright artifacts from nearby luminous sources, we perform phase-only self-calibration on these other sources. Flux measurements were determined by using the Cube Analysis and Rendering Tool (CARTA; \citealt{carta}). For detections, the flux density is the maximum intensity at the transient's position, and the uncertainty is the measured root-mean square (RMS) in the background. For non-detections, we take the upper limit to be 3 $\times$ RMS. The radio light curves are shown in Figure~\ref{fig:radio-lc} and the radio SED evolution is shown in Figure~\ref{fig:radio-sed}.

\subsubsection{GMRT}
We also present GMRT observations of  AT2022abfc, AT2023fhn, AT2023hkw, and AT2023vth. Both AT2023fhn and AT2023vth were detected. These observations were spread across three frequency bands: Band-3 (250-500 MHz), Band-4 (550-850 MHz), and Band-5 (1000-1460 MHz)\footnote{AT2022abfc was observed under program ID 44\_099.  AT2023fhn was observed under programs 45\_117 and ddtC299.  AT2023hkw was observed under 45\_117 and ddtC294.  AT2023vth was observed under program 45\_113 (PI Nayana A.J.).}.
The GMRT data were reduced through CASA and AIPS.  We use Briggs weighting with \texttt{robust = 0.5}.  The primary beam correction was done following the method described in the CASA documentation\footnote{\href{https://casadocs.readthedocs.io/en/latest/examples/community/Example_Wideband_PrimaryBeamCorrection.html\#Imaging-With-Custom-Primary-Beams}{CASA: Imaging with Custom Primary Beams}}.  We measure the flux density using the \texttt{imfit} method in the CASA documentation, which provides the maximum intensity for a fitted Gaussian.  If the measured flux is not 3-$\sigma$ above the background flux RMS, we record an upper limit of 3$\times\,$the RMS. 

\subsubsection{NOEMA}

\begin{figure}[!h]
 \centering
  \includegraphics[width=\linewidth]{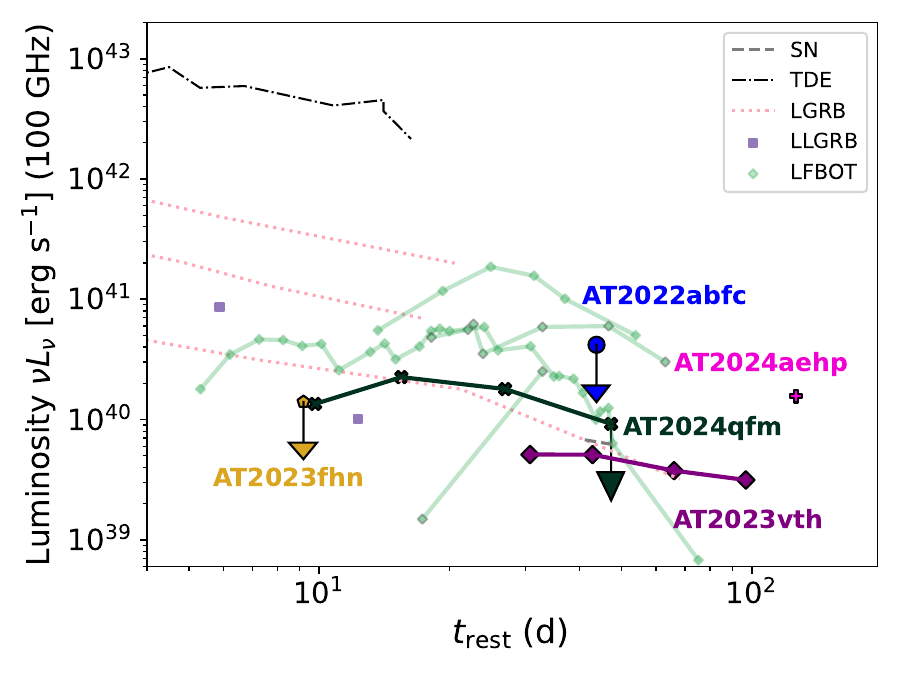}
\caption{Millimeter-wave (100\,GHz) light curves in comparison to other LFBOTs, long-duration gamma-ray bursts (LGRBs), low-luminosity GRBs (LLGRBs), and tidal disruption events (TDEs).  AT2022abfc and AT2023fhn only have an upper limit. AT2023fhn has been shifted slightly to the left for clarity. Data are from \citet{1998bwradio, zauderer2011, perley2013_130427a, corsi2014, laskar2016, perley2017_171205a, laskar2018, margutti, bright2022, ho2023tsd, nayana2025}.}
\label{fig:mm}
\end{figure}

We observed five LFBOTs using NOEMA\footnote{Program IDs W22BT (AT2023fhn and AT2022abfc), W24EW (AT2024aehp), S23BF (AT2023vth), D24AA (Director's Discretionary Time; AT2024qfm), all PI A. Ho.} (all but AT2023hkw), at times ranging from $t=10$--100\,d, resulting in detections of three objects (AT2024aehp, AT2024qfm, AT2023vth) and non-detections of two objects (AT2022abfc, AT2023fhn). The data reduction was done with Continuum and Line Interferometer Calibration (\texttt{CLIC}) and \texttt{MAPPING}, which are part of the \texttt{GILDAS} software package \citep{gildas}\footnote{\href{https://www.iram.fr/IRAMFR/GILDAS}{https://www.iram.fr/IRAMFR/GILDAS}}. Unless otherwise specified, all uncertainties are 1$\sigma$, and \added{the systematic error of the} flux scale is 10\% in the 3\,mm band and 15\% at 1.3\,mm. The NOEMA data are provided in Table~\ref{tab:radio} and displayed in Figure~\ref{fig:mm}. 

AT2024aehp was observed on May 23, 2025 ($>100\,$d after optical discovery; we triggered mm observations following the cm-wave detection) in the 10D configuration under mediocre weather conditions, with MWC349 as the flux calibrator. Reduction was performed by hand and we estimate a 15\% systematic calibration uncertainty for the absolute flux scale. 

AT2023fhn was observed on April 22, 2023 in the 12C configuration under good conditions, with LKHA101 as the flux calibrator. The image included a nearby bright source, and subtracting the contribution from the bright source did not result in a detection of AT2023fhn at the phase center. 

An observation of AT2022abfc was triggered on December 30, 2022; an attempt was made on January 7, 2023, but could not be carried out owing to bad weather. A successful observation was obtained on January 10 in the 12A configuration, with LKAH101 as the flux calibrator. The source was not detected, but limits were weak because the source was at high airmass and the array was in an extended configuration.

AT2023vth was first observed on November 20, 2023 in the 9C configuration, under cloudy conditions, with the flux calibrator MWC349. The source was detected, so we pursued multifrequency follow-up observations. The timing of these observations was affected by weather conditions, particularly at the highest frequencies. 

NOEMA observed AT2024qfm on four
dates, in compact configuration (D), with primary flux calibrator MWC349. The first observation was on August 5, 2024. The source was detected in the first three epochs. 

\subsection{Search for Gamma-Ray Bursts}
We performed an archival search for GRB detections in the  NASA General Coordinates Network Circulars\footnote{\href{https://gcn.nasa.gov/circulars/}{https://gcn.nasa.gov/circulars/}}.  We looked for GRBs detected within a week of the transient's optical light-curve peak \added{that had a localization that was within 3-$\sigma$ of the transient's coordinates}. No such GRBs were reported.
\newline
\newline
\subsection{Search for X-ray Precursors}

\added{One way to distinguish between LFBOT progenitor models could be to detect precursor emission: this might be expected in a model with runaway mass transfer leading up to the dynamical merger of a stripped star and compact-object companion \citep{metzger2022,Tsuna2024} but not in e.g., a core collapse scenario.} We searched for X-ray precursor emission in the eROSITA-DE Data Release 1 (DR1). eROSITA \citep{eROSITA} is the soft X-ray instrument onboard the orbital observatory Spectrum-Roentgen-Gamma (SRG; \citealt{SRG2021}). DR1 comprises data from the first six months, which ended in June 2020 \citep{Merloni2024,TubinArenas2024}.  

Of the thirteen LFBOTs in Table~\ref{tab:summary}, three were in the footprint of eROSITA-DE DR1: AT2022abfc, CSS161010, and AT2024wpp. For each object, we used the products from \citet{TubinArenas2024} and \citet{Merloni2024} to determine if there was a detection at the source position, and if not, obtained a 0.2--2.3\,keV upper limit (the most sensitive energy range for eROSITA).  \added{As each field was only imaged once in this data release, we only check for consistency in position and allow the observation to be at any time.}  The closest catalogued sources were \added{offset by} $5.5^{\prime}$ (AT2022abfc), $3^{\prime}$ (CSS161010), and $2^{\prime}$ (AT2024wpp), too distant to be associated \added{(the spatial resolution of eROSITA is $<10^{\prime\prime}$, \citealt{eROSITA})}.

For AT2022abfc and AT2024wpp (the two objects for which the eROSITA observations were prior to the LFBOT), the limits are $9.1\times10^{42}$\,erg\,s$^{-1}$ corresponding to 2.7\,yrs pre-LFBOT, and $9.0\times10^{41}$\,erg\,s$^{-1}$ corresponding to 4.7\,yrs pre-LFBOT, respectively. For CSS161010, the eROSITA limit is $1.5\times10^{41}$\,erg\,s$^{-1}$, 3.4\,yrs after the LFBOT. The next eROSITA data release (scheduled for mid-2026) will provide more stringent upper limits (i.e., closer to the LFBOT time).  

\subsection{Host-Galaxy Photometry}
IR and UV photometry were obtained for all LFBOT host galaxies from the Near-Earth Object Wide-field Infrared Survey Explorer mission (NEOWISE; \citealt{neowise}) on the {\em Wide-field Infrared Survey Explorer} space telescope (\wise), and the {\em Galaxy Evolution Explorer} (\galex; \citealt{galex}), respectively.  Optical photometry for the host galaxies was obtained from the Legacy Survey Data Release 10 \citep{legacy}. The host galaxy of AT2023vth does not have Legacy Survey optical photometry, so we used optical photometry from Pan-STARRS DR2 (PS1; \citealt{panstarrs}).  We also used {\it Hubble Space Telescope (HST)} photometry from \citet{chrimesfhn} for AT2023fhn.  All the {\galex} data used in this paper can be found in MAST: \dataset[10.17909/zktz-mn64]{http://dx.doi.org/10.17909/zktz-mn64}.  For the photometry in all bands, we required a signal-to-noise ratio (SNR) greater than 5 for a detection.  \added{We choose this SNR to include several marginal GALEX and WISE detections, while excluding a few measurements with errors of over 0.5 mag.  Choosing a stricter cutoff would make the fit far more uncertain as we lose a significant number of data points, but a looser cutoff would have little effect as the new points would have large errors, which the fitting process accounts for.  We display cutouts of the host galaxies in Figure~\ref{fig:cutouts}. }

We used the Python package \texttt{gPhoton} \citep{gphoton} to retrieve \galex\ near-UV and far-UV photometry for the LFBOT host galaxies.  In \texttt{gPhoton}, we chose the observation with the longest exposure time and recorded the background-corrected magnitudes.  We chose an aperture radius of 6.12 arcseconds for the source, and measured the background from an annulus with an inner radius of 10.8 arcseconds and an outer radius of 21.6 arcseconds.  \added{We estimate that this size would exclude all of the host galaxy's emission and include as much background as possible for an accurate background flux estimate without including any flux from other sources.  We apply this annulus size to most of the LFBOT hosts to maintain consistency.}  In the cases of AT2023hkw and AT2023vth, there were \added{nearby UV-bright sources that required us to use a smaller annulus}.  The inner and outer annulus radii are 10.8 and 16.2 arcseconds for AT2023hkw, and 7.2 and 10.8 arcseconds for AT2023vth.  Legacy Survey and {\it NEOWISE} photometry was obtained with the Python package \texttt{noaodatalab}\footnote{\href{https://github.com/astro-datalab/datalab}{https://github.com/astro-datalab/datalab}} version 2.20.0.  Pan-STARRS DR2 {\em grizy} photometry was \added{retrieved through} the online catalog search tool\footnote{\href{https://catalogs.mast.stsci.edu/panstarrs/}{https://catalogs.mast.stsci.edu/panstarrs/}}.

The host-galaxy photometry is provided in Table~\ref{tab:galaxy} in Appendix~\ref{sec:appendix-tables}.



\section{Radio Analysis}
\label{sec:radio-analysis}

\subsection{Radio Synchrotron Modeling}
\label{sec:synchro}

\begin{figure*}[!htb]
 \centering
  \includegraphics[width=\linewidth]{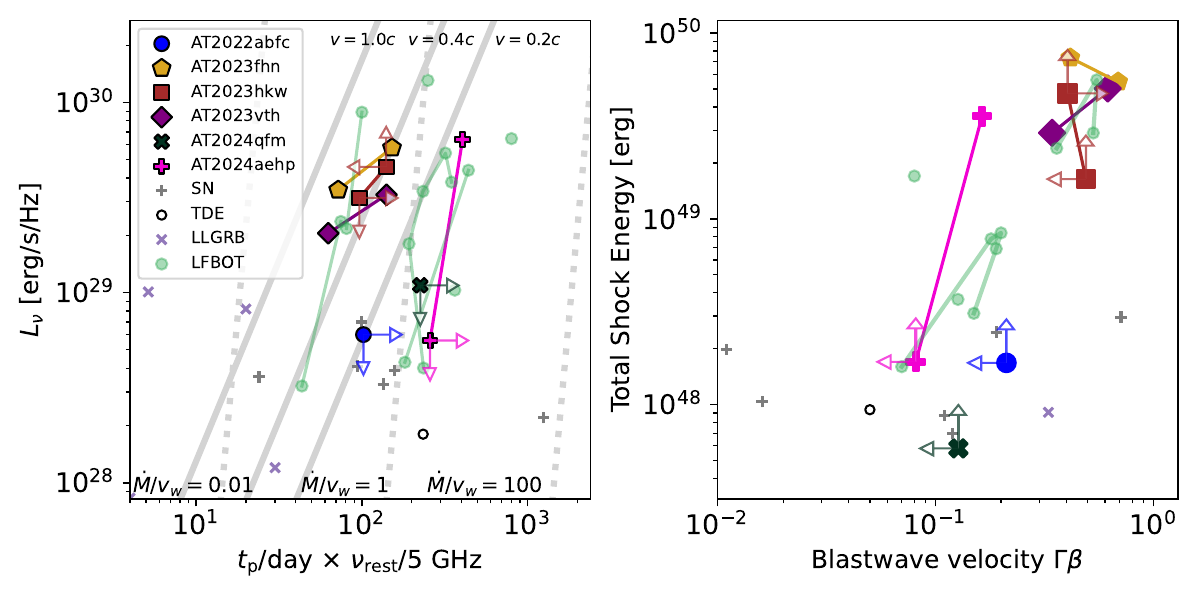}
\caption{Inferred parameters from our synchrotron self-absorption modeling.  Some epochs do not include the self-absorption peak and merely constrain its location; the corresponding upper limits on parameters are marked with arrows. Values for other objects are obtained from \citet{ho2019} and \citet{ho2023tsd}; and \citet{Perley2026} for AT2024wpp.  We label each object and the observer-frame time after peak light. \emph{Left:} Luminosity density at $\nu_p$ versus peak frequency and time after explosion.  We also display lines of constant shock velocity and $\dot{M}/v_w$, where $v_w=1000$\,km s$^{-2}$ and $\dot{M}$ is in units of $10^{-4}\,M_{\odot}\,\mathrm{yr}^{-1}$. \emph{Right}: Total shock energy derived from Equation~\ref{eq:energy} against shock velocity in terms of $\Gamma \beta$.  }
\label{fig:synchrotron}
\end{figure*}

We model the SED
at radio and millimeter wavelengths as the result of synchrotron emission.  Synchrotron radiation with a self-absorbed component forms a broken power-law radio SED. We follow the prescription of \citet{chevalier} to derive properties of the shock and the medium it traverses, such as the shock radius, ambient magnetic field, average shock velocity, pre-shock density, and total shock energy.  This is the same procedure used in modeling the radio SEDs for other AT2018cow-like FBOTs (e.g., \citealt{margutti,ho2019,chrimesfhn,nayana2025,Perley2026}). We note that in some of these works, some equations are given in terms of angular diameter distance $D_\theta$, but should be given in terms of luminosity distance ($D_L$; private communication) as in the expressions below. 
\added{We also note that the following prescription assumes a nonrelativistic shock and does not account for relativistic effects (see, e.g., \citealt{margalit2024, ferguson2025}), which become relevant for $\Gamma \beta \sim1$. 
For self-consistency, we estimate the $\Gamma \beta$ this model outputs and we find that some objects do approach the trans-relativistic regime (Fig~\ref{fig:synchrotron}), so we caution against taking our presented results at face value for the higher-velocity events. Still, we use this method for comparison with previous LFBOTs, which is the main goal of this paper's analysis.} 

As in \citet{soderberg2010b}, we assume equipartition of energy between protons, electrons, and the magnetic field;  thus, the energy densities in electrons ($\epsilon_e$) and magnetic fields ($\epsilon_B$) are given by $\epsilon_e=\epsilon_B=1/3$.  If we model the source of the emission as a disk with radius $R$ and thickness $S$, then its total volume is $\pi R^2 S$.  We then parameterize this as equivalent to a spherical volume $4\pi f R^3/3$, where $f$ is the filling factor.  We take $f=0.5$ as in previous LFBOT studies \citep{yao2022, bright2022}, but the dependences of the equations on this parameter is weak. We assume an electron power-law index of 3, following \citet{chevalier2006}; this is a typical value for previous LFBOTs \added{as determined from fits to the slope of the self-absorbed end of the synchrotron SED \citep{coppejans2020, ho2019}.  We do not have more than two points in the self-absorbed regime for most of our SEDs, so we use this fiducial value with the recognition that the results do not strongly depend on the optically thin power-law index \citep{chevalier}.}

Finally, we verify that the self-absorption frequency is less than the cooling frequency, which is defined as the frequency where the electrons have lost an amount of energy equal to their total energy through synchrotron cooling. We use Equation~(C13) of \citet{ho2022xnd}, which gives the ratio between the self-absorption and cooling frequencies, and confirm that the self-absorption frequency is lower for all cases.

Under these assumptions, Equations (13) and (14) from \citet{chevalier} reduce to these expressions for the shock radius and magnetic field:
\begin{equation}
\begin{split}
    R_p =& \,8.8 \times 10^{15}
    \left( \frac{\epsilon_e}{\epsilon_B}\right)^{-1/19}
    \left( \frac{f}{0.5}\right)^{-1/19}
    \left( \frac{F_{\rm p}}{(1+z)\, \text{Jy}}\right)^{9/19} \\
    &\times
    \left( \frac{D_L}{\text{Mpc}}\right)^{18/19}
    \left( \frac{\nu_{\rm p}}{5 \,\text{GHz}}\right)^{-1} \mathrm{cm}\, ,
\end{split}
\end{equation}

\begin{equation}
\begin{split}
    B =& \,0.58
    \left( \frac{\epsilon_e}{\epsilon_B}\right)^{-4/19}
    \left( \frac{f}{0.5}\right)^{-4/19}
    \left( \frac{F_{\rm p}}{(1+z)\,\text{Jy}}\right)^{-2/19} \\
    &\times
    \left( \frac{D_L}{\text{Mpc}}\right)^{-4/19}
    \left( \frac{\nu_{\rm p}}{5 \,\text{GHz}}\right) \mathrm{G}\, ,
\end{split}
\end{equation}
where $D_L$ is the luminosity distance, $\nu_{\rm p}$ \added{and $F_{\rm p}$  are the peak frequency and flux density at peak frequency of the radio SED in the transient's rest frame, respectively.}

Following \citet{ho2019}, this can then be combined into an equation for total energy $U=U_B/\epsilon_B$:

\begin{equation}
\begin{split}
    U =& \, \frac{B^2}{8\pi}\frac{4 \pi f R^3}{3}\frac{1}{\epsilon_B}\\
    =& \,1.9\times10^{46} \left( \frac{1}{\epsilon_B} \right)
    \left( \frac{\epsilon_e}{\epsilon_B}\right)^{-11/19}
    \left( \frac{f}{0.5}\right)^{8/19}\\
    &\times
    \left( \frac{F_{\rm p}}{(1+z)\,\text{Jy}}\right)^{23/19} 
    \left( \frac{D_L}{\text{Mpc}}\right)^{46/19}
    \left( \frac{\nu_{\rm p}}{5 \,\text{GHz}}\right)^{-1} \mathrm{erg}\, .
\end{split}
\label{eq:energy}
\end{equation}

We estimate the average velocity as $R/( t_{\mathrm{obs}}/(1+z))=R/ t_{\mathrm{rest}}=\Gamma \beta c$, where $\beta=v/c$ and $\Gamma$ is the Lorentz factor.

Following \citet{chrimesfhn}, we derive formulas for the CSM electron number density $n_e$ and the wind density $\dot M/v_w$.  This assumes the CSM resulted from a spherical stellar wind with mass-loss rate $\dot M$ and wind velocity $v_w$, where $\dot M/v_w=4 \pi m_p R^2 n_e$ and $m_p$ is the proton mass. Thus,

\begin{equation}
\begin{split}
    n_e =& \,1.02
    \frac{1}{\epsilon_B}
    \left( \frac{\epsilon_e}{\epsilon_B}\right)^{-6/19}
    \left( \frac{f}{0.5}\right)^{-6/19}
    \left( \frac{F_{\rm p}}{(1+z)\,\text{Jy}}\right)^{-22/19} \\
    &\times
    \left( \frac{D_L}{\text{Mpc}}\right)^{-44/19}
    \left( \frac{\nu_{\rm p}}{5 \,\rm GHz}\right)^{4} 
    \left( \frac{t_{\text{rest}}}{\text{days}}\right)^{2} \mathrm{cm}^{-3}\, ,
\end{split}
\end{equation}

\begin{equation}
\begin{split}
    \frac{\dot M}{v_w}\left( \frac{1000\,\mathrm{km}\,\mathrm{s}^{-1}}{10^{-4}\,M_{\odot}\,\mathrm{yr}^{-1}}\right) =& \,2.5\times10^{-5} \left(\frac{1}{\epsilon_B}\right)
    \left( \frac{\epsilon_e}{\epsilon_B}\right)^{-8/19} \\
    &\times \left( \frac{f}{0.5}\right)^{-8/19}
    \left( \frac{F_{\rm p}}{(1+z)\,\text{Jy}}\right)^{-4/19} \\
    &\times
    \left( \frac{D_L}{\text{Mpc}}\right)^{-8/19}
    \left( \frac{\nu_{\rm p}}{5 \,\text{GHz}}\right)^{2} \\
    &\times \left( \frac{t_{\text{rest}}}{\text{days}}\right)^{2}\, .
\end{split}
\end{equation}

The peak frequency and flux density of the SED at a given epoch are measured by fitting a smoothed broken-power law of the form
\begin{equation}
F(\nu)=F_{\rm p}\left((\nu/\nu_{\rm p})^{-sa_1}+(\nu/\nu_{\rm p})^{-sa_2}\right)^{-1/s}.
\label{eq:broken}
\end{equation}

We choose $s=1$ for the smoothness parameter, as was done in previous LFBOT studies \citep{bright2022, chrimes2024, nayana2025}.  We also set $a_1=5/2$ to fix the power-law slope of the self-absorbed half of the SED, as required by Chevalier's formulation. When an epoch involves observations from multiple \added{observatories}, \added{we apply an extra error in quadrature equal to 10\% of the flux density to account for systematic effects between arrays}.  A fit is performed only when we have detections on either side of the SED's peak.  The best-fit parameters from the broken power-law fit are provided in Table~\ref{tab:synchro_fit} in Appendix~\ref{sec:appendix-tables}.  If we are unable to perform a broken power-law fit, we use the values at the detection closest to the peak's implied location in place of the frequency and flux density at peak, and calculations based on these values will instead set limits on the synchrotron parameters.  
The inferred synchrotron parameters are listed in Table~\ref{tab:synchro} in Appendix~\ref{sec:appendix-tables}, and the results are displayed in Figure~\ref{fig:synchrotron}.  This includes epochs where we do fit a broken-power law and have point estimates for parameters, and epochs where we can only constrain the peak's location and place limits on parameters instead.  We find that the implied CSM densities for these six LFBOTs are comparable to those derived in the literature for previous LFBOTs, with values of 10--100 $\mathrm{cm}^{-3}$ at radii of $10^{17} \mathrm{cm}$.  

\section{Host-Galaxy Analysis}
\label{sec:host-analysis}

\subsection{Host-Galaxy Offsets}\label{sec:offset}
\added{With an expanded sample of 13 LFBOTs (double the previous size in the literature), we also examine the distribution of LFBOT offsets from the host-galaxy nuclei.  We choose the host galaxy as the galaxy closest in offset to the transient, and then confirm that there are no lines at conflicting redshifts in spectroscopy of the LFBOT and its host.  To measure offsets, we require accurate localizations of each LFBOT and its host galaxy.}  For AT2023fhn \added{and its host}, we used the localization from {\it HST} imaging presented by \citet{chrimesfhn}.  \added{For the other LFBOTs,} we measure the host galaxy's position from Legacy Survey Data Release 10, \added{and the transient's position from a Gaussian fit in the image of the LFBOT's brightest radio detection. We also do this for AT2020xnd, which did not have a published offset.  The exceptions are} AT2023vth's host (which was measured from \added{the} Pan-STARRS Data Release 2), \added{and AT2024qfm's localization which we measure from the ZTF data (as we do not present AT2024qfm's radio data in this paper).}  These offsets are recorded in Table~\ref{tab:summary}. 

\begin{figure}[!h]
 \centering
  \includegraphics[width=\linewidth]{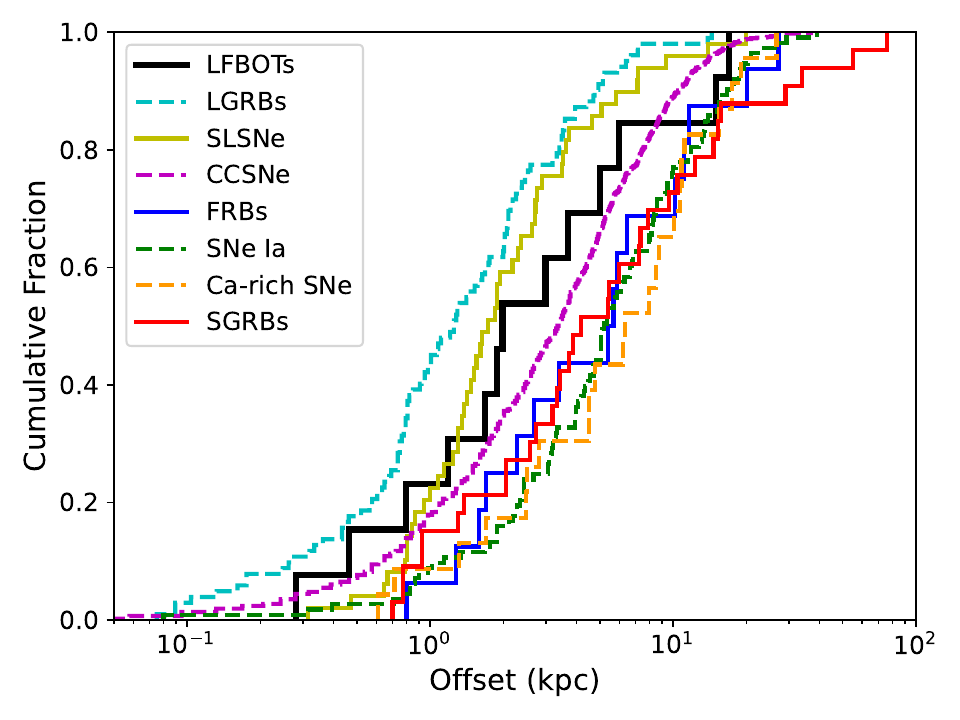}
\caption{The cumulative distribution of host-galaxy offsets for LFBOTs, as well as fast radio bursts (FRBs; \citealt{bhandari2022}), LGRBs  \citep{blanchard2016, lyman2017}, short GRBs (SGRBs; \citealt{fong2022}), CCSNe \citep{kelly2012, schulze2021}, Type Ia SNe \citep{wang2013}, SLSNe \citep{lunnan2015, schulze2021}, and Ca-rich transients \citep{de2020}.  The offsets for previous LFBOTs are from \citet{prentice2018, coppejans2020, ho2020koala, yao2022, chrimesfhn}.}
\label{fig:offset}
\end{figure}

In Figure~\ref{fig:offset}, we display the cumulative distribution of all LFBOT offsets in kpc compared with other classes of extragalactic explosive transients. The offset distribution appears to be intermediate to that of CCSNe and that of LGRBs and SLSNe.  



\begin{figure*}[!tb]
 \centering
  \includegraphics[width=\linewidth]{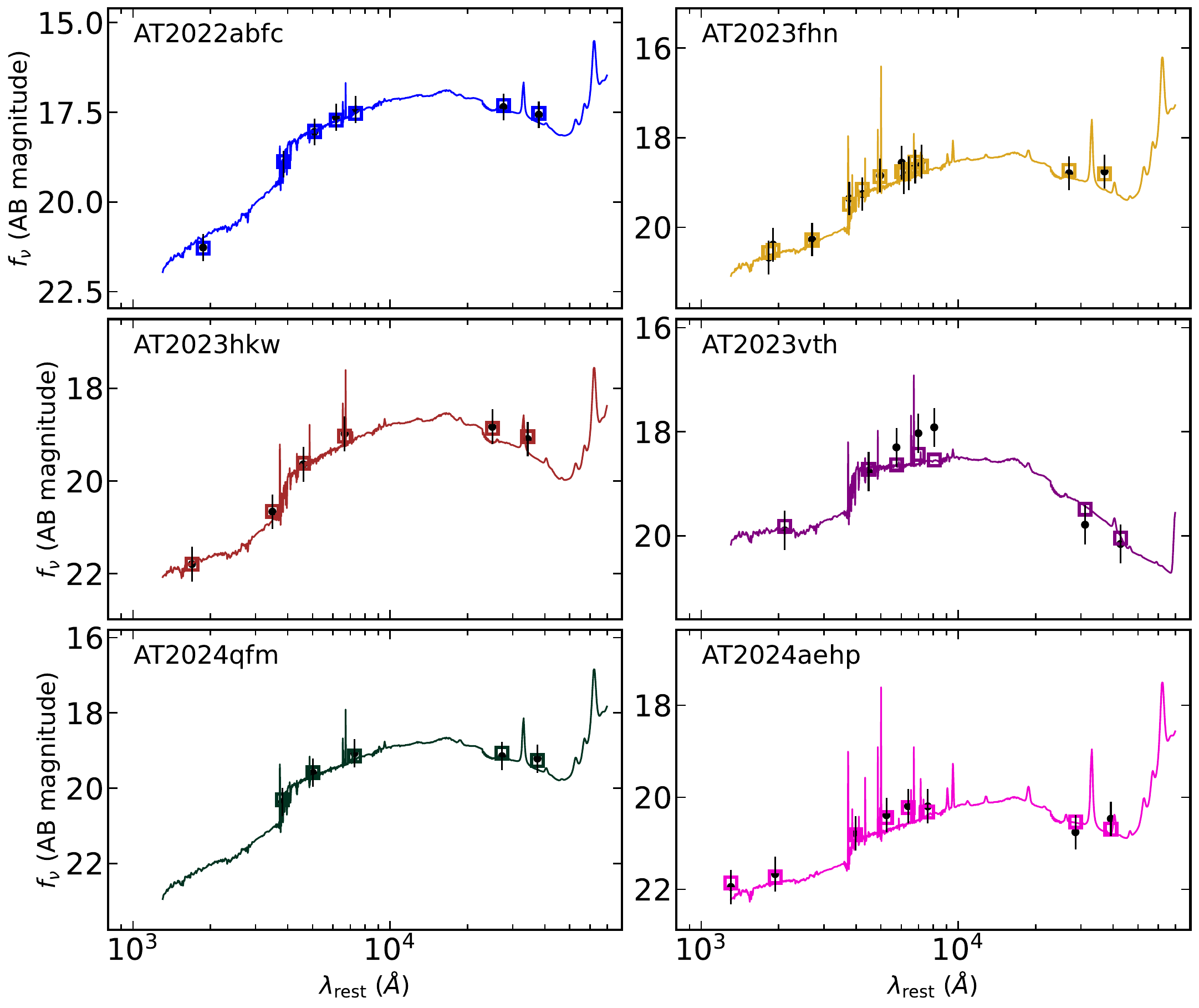}
\caption{Best-fit host-galaxy spectra and photometry from \texttt{prospector} (color) as well as the input photometry (black circles with error bars).  Note that the parameters reported in Table~\ref{tab:galaxy_prop} are the median, 16th percentile, and 84th percentile.  The best-fit host galaxy spectra displayed here corresponds to the maximum a posteriori model, which may not necessarily have the median parameters.  The abscissa 
displays rest-frame wavelength.}
\label{fig:spec_fit}
\end{figure*}

\begin{figure}[!h]
 \centering
  \includegraphics[width=\linewidth]{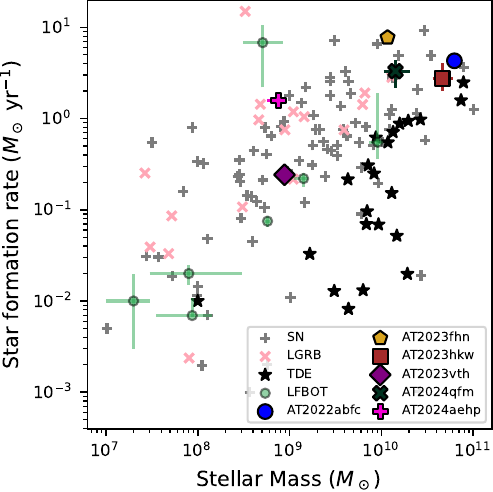}
\caption{Median mass and 
SFR for each LFBOT's host galaxy in comparison to the host galaxies of other energetic transients.  Colored lines show error bars, though they may be smaller than the marker for some points.  \added{Data for the other transients from \citep{lawsmith2017, french2020, taggart2021, Somalwar2025}.}}
\label{fig:host_galaxy}
\end{figure}

\subsection{Host-Galaxy Properties}\label{sec:host}

We used the Python package \texttt{prospector} \citep{prospector1, prospector2}\footnote{\url{https://github.com/bd-j/prospector}} to derive properties such as stellar mass and star-formation rate (SFR) from the host-galaxy photometry.  The \texttt{prospector} package utilizes the Flexible Stellar Population Synthesis\footnote{\href{https://dfm.io/python-fsps/current/}{https://dfm.io/python-fsps/current/}} (FSPS) package to model the emission from each galaxy given a set of parameters for the galaxy's composition and stellar population. \added{It also utilizes dynamic nested sampling through the package \texttt{dynesty}\footnote{\href{https://github.com/joshspeagle/dynesty}{https://github.com/joshspeagle/dynesty}} to fit galaxy parameters to the observed galaxy photometry under Bayesian inference.}  The basic parametric star-formation history template is adopted along with the parameters for nebular emission.  We chose the Chabrier initial mass function \citep{chabrier2003}, the Calzetti dust attenuation model \citep{calzetti2000}, and a star-formation history of the form $t e^{-t/\tau}$.  \added{These procedures were chosen to match LFBOT host galaxy analysis in previous papers as closely as possible \citep{yao2022, ho2023tsd, chrimes2024}.} A model was first run with a fixed metallicity of $\log (Z/Z_{\odot})=-0.2$.  Then, we used the median mass and the mass-metallicity relation of \citet{gallazzi2005} to fix the metallicity to a more accurate value, and ran a final model.
The fitted parameters are listed in Table~\ref{tab:galaxy}, the best-fit photometry and spectroscopy are displayed in Figure~\ref{fig:spec_fit}, and the fitted host-galaxy properties are listed in Table~\ref{tab:galaxy_prop}.  We show the median host-galaxy masses and SFRs in Figure~\ref{fig:host_galaxy}.  Our host-galaxy properties derived for the host of AT2023fhn agree with \citet{chrimes2024} to within 1-$\sigma$ for \added{the} host-galaxy mass and 2-$\sigma$ for \added{the} SFR. 

The host galaxies of our LFBOTs are all star-forming galaxies. Whereas several previous events were in low-mass dwarf galaxies \citep{ho2020koala, coppejans2020,chrimesfhn,margutti}, the host galaxies of four of our new LFBOTs have higher masses ($10^{10}$--$10^{11}\,M_{\odot}$, Figure~\ref{fig:host_galaxy}). However, our selection criteria for identifying LFBOT candidates do require there to be a visible host galaxy (to avoid contamination by dwarf novae), so a bias to higher mass galaxies \added{(and therefore higher star-formation rate)} is to be expected. 

\begin{figure*}[!htb]
 \centering
  \includegraphics[width=1.0\linewidth]{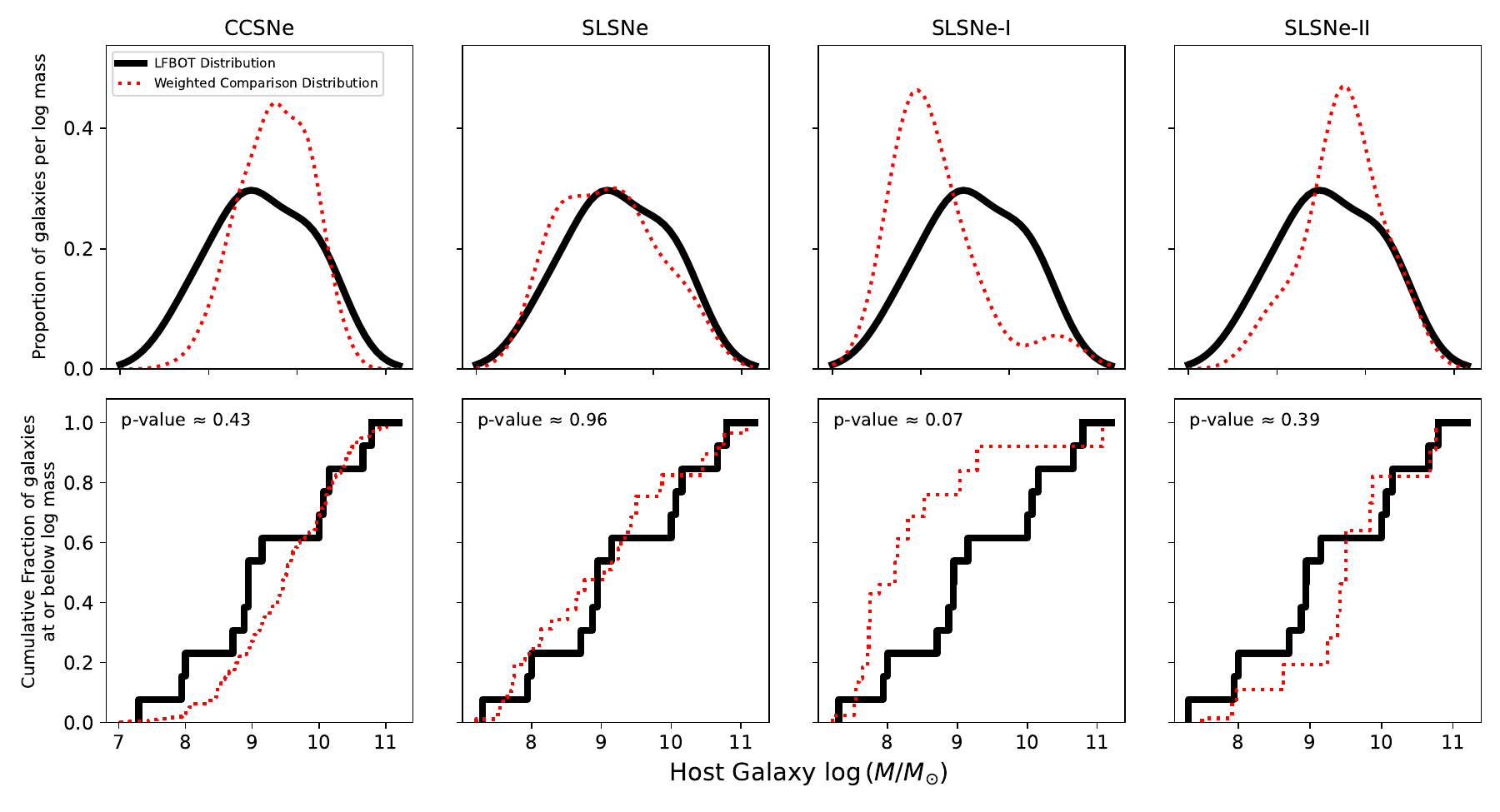}
\caption{Gaussian KDEs for the LFBOT host-galaxy masses (black) in comparison to the Gaussian KDEs for each of the weighted mass distributions from the \citet{taggart2021} CCSN/SLSN host-galaxy population (dashed red).  \bf{We also list the p-values from our Kolgomorov-Smirnov test.}}
\label{fig:host_galaxy_sim}
\end{figure*}

To compare the \emph{intrinsic} distribution of host-galaxy mass to that of CCSNe, we used the untargeted sample of 150 CCSN host galaxies from \citet{taggart2021} as representative of the true CCSN host galaxy population.  \added{We prioritized having an unbiased sample over a large comparison sample, as our uncertainties are limited by our small LFBOT sample.}  For each of the thirteen LFBOT host galaxies, we calculate the minimum absolute magnitude for it to have been detected by the Legacy Survey at that redshift, which corresponds to the apparent magnitude cutoff of $\mathrm{m}_r\leq23$.  Then, for each CCSN host galaxy, we apply a weight equal to the proportion of the thirteen cutoffs that the CCSN host galaxy would pass. We do not account for any systematic differences in host galaxy properties between those at redshift $z\approx 0.01$ as in the Taggart sample, and those at $z\approx 0.1$--0.2 as with most of the LFBOTs.  We plot the Gaussian kernel density estimation (KDE) for the weighted distribution of CCSN host-galaxy masses Figure~\ref{fig:host_galaxy_sim} as a dotted red line.  The Gaussian KDE for the LFBOT host-galaxy masses is shown in black. We repeated the exercise for SLSNe, which are known to have a preference for lower galaxy masses than regular CCSNe \citep{taggart2021}.

We use the \texttt{Ecume} package \citep{ecume} in the programming language \texttt{R} \citep{r} to perform a Kolmogorov-Smirnov (KS) test comparing the distribution of LFBOT host-galaxy masses with the weighted CCSN and SLSN populations.  These simulations imply that the LFBOT host-galaxy population is consistent with both the populations of CCSN and SLSN host galaxies at a significance of $\alpha=0.05$ (\added{which means that the p-value must be less than $0.05$ for us to conclude that the underlying distributions are different with 95\% confidence}).  \added{We also display the results of the KS test in Figure~\ref{fig:host_galaxy_sim}}. The SLSN host-galaxy distributions appears especially similar to that of LFBOTs. However, the SLSNe are subdivided into SLSNe-I and SLSNe-II.  These two classes are markedly different, and this is reflected in their host-galaxy masses---SLSN-I host galaxies are clustered around $\log \,(\mathrm{M}/\mathrm{M}_{\odot})\approx 8$ while SLSN-II hosts are $\log \,(\mathrm{M}/\mathrm{M}_{\odot})\approx 9$--10.  Despite the visible differences, the sample sizes of SLSN-I and SLSN-II hosts are too small to find a statistically significant difference at $\alpha=0.05$.  





\section{Discussion}
\label{sec:discussion}

\subsection{Observational Comparison}
\label{sec:comparison}

The objects presented in Section~\ref{sec:methods} show relative homogeneity in the optical (Figure~\ref{fig:optical}) and radio (Figure~\ref{fig:radio-lc}) light curves. To some extent the similarity of the optical light curves to those of AT2018cow is by selection, as these objects were identified on the basis of a rapid rise and decay \added{within the 2--3 days of the optical peak,} along with high luminosities. 
The rapid decline of 3--4 mag over $t_\mathrm{rest}\approx10\bf{-15}$\,d observed in AT2023fhn, AT2023vth, AT2023hkw, and AT2024qfm --- very similar to AT2018cow --- was not selected for.  
\added{We note that AT2024aehp differs in many of these properties.  This section will focus on the other LFBOTs in our sample; AT2024aehp will be discussed in more detail in Section~\ref{sec:24aehp}.}

\added{While the other optical light curves could appear similar due to selection effects, selection effects do not account for the similarity between the LFBOT radio light curves as we only require one detection for classification.} Figure~\ref{fig:radio-lc} shows that the peak $\nu L_{\nu}$ values mostly lie within the range of $10^{39}$--$10^{40}$\,erg\,s$^{-1}$. The time evolution for the radio emission is also similar across the sample.  The 10\,GHz luminosity gradually rises, plateaus between $t_{\mathrm{rest}}\sim50$--100\,d after the initial discovery, and then begins to fade. During the rise, the radio emission is typically optically thick (has a positive spectral index; Figure~\ref{fig:radio-sed}). The \added{decay} rates of the radio emission vary, and in some cases are poorly constrained (e.g., for AT2022abfc). The fast fade of AT2023vth is similar to that of CSS161010 \citep{coppejans2020}. \added{Meanwhile}, the mm-wave light curves appear to peak close \added{at times close} to $t_{\text{rest}}=20-30\,$d (Figure~\ref{fig:mm}).  \added{All SEDs either have or are consistent with a broken-power law shape.  The frequency is near 50\,GHz during the first detections at $t_{\mathrm{rest}}\sim30$\,d, then gradually shift to lower frequencies, passing 10\, GHz at roughly $t_{\mathrm{rest}}\sim100$\,d.}

The uniformity of the radio luminosity and evolution of all AT2018cow-like FBOTs is striking. Of course, the radio emission had to be sufficiently luminous to be detected in the first place, but we did not require the radio luminosity to brighten over time, nor was there a requirement about the timing of the peak. The radio behavior is in sharp contrast to CCSN 
radio light curves, which span over four orders of magnitude in luminosity, with peak times ranging from a few days to thousands of days \citep{Bietenholz2021}.  \added{This diversity suggests that core collapse supernovae occur with a wide range of CSM densities and spatial distributions.  As this is not the case for LFBOTs, we conclude that the CSM properties must be tied to the LFBOT progenitor channel.  We discuss the implications of this further in Section~\ref{sec:discussion-progenitors}.}

We have fewer constraints on the X-ray behavior, as several of the objects only have upper limits on their luminosity (the calculated X-ray luminosity upper limits are driven mostly by their redshift, as \swift\ determined similar photon count upper limits for each non-detection).  \added{But as seen in Figure~\ref{fig:xray},} these upper limits appear to rule out X-ray emission similar to those of jetted TDEs or classical LGRBs while allowing for the possibility of X-ray emission similar to that of AT2018cow.  \added{In addition}, the order-of-magnitude jump in X-ray luminosity seen in AT2024qfm over just two days is much more pronounced than the X-ray variability observed in past LFBOTs. 

\subsection{Implications for Progenitors}\label{sec:discussion-progenitors}

The consistent radio light curves imply consistent properties of the surrounding medium on scales of $\sim 10^{16}$--$10^{17}\,$cm. \added{Assuming this medium was produced by material ejected from the progenitor at 1000\,km\,s$^{-1}$, it probes timescales of a few to tens of years prior to the transient's explosion. Therefore, the radio behavior favors progenitor theories that can explain the presence of a dense ambient medium that is fairly similar from event to event and is produced well in advance of the explosion. This is hard to reconcile with a TDE scenario.  \citet{margutti} finds that the IMBH invoked to explain LFBOTs are unable to sustain outflows sufficient to produce this medium.}  
Difficulties also arise in the massive-star core-collapse model: the stars that we usually suspect to be the progenitors of luminous explosions and energetic SNe are not expected to undergo significant mass loss in such a consistent way just prior to the explosion (indeed, SN radio light curves are highly diverse).  
On the other hand, progenitor models involving mergers \citep[e.g.,][]{metzger2022,Tsuna2025,Klencki2025} naturally explain the dense medium as material ejected from the merger.   \added{Each proposed merger scenario involves mass transfer from a massive donor star to a compact object.  These papers estimate the circumburst medium density from calculations of the stellar wind from the donor star and mass-loss during the merger, and have found them to be consistent with density estimates from past LFBOT papers.  These estimates are also expected to hold for each instance of such a merger, explaining the consistency across the LFBOT sample.}  

\subsection{AT2024aehp}\label{sec:24aehp}
AT2024aehp has several characteristics that distinguish it from the other LFBOTs in this paper. Its optical light curve exhibited a plateau \added{at $M=-19$\,mag from 5--50\,d, behavior we can rule out in AT2023vth and AT2024qfm but not in AT2022abfc, AT2023fhn, and AT2023hkw. AT2024aehp is also an outlier in terms of radio evolution. As shown in Figure~\ref{fig:radio-lc} and Figure~\ref{fig:radio-sed}, AT2024aehp dramatically brightens in radio luminosity at a later time than the other transients we have discussed, by over an order of magnitude in all frequencies between the $t_{\mathrm{rest}}\sim70\,$d and $t_{\mathrm{rest}}\sim120\,$d epochs. In contrast, the radio emission of other LFBOTs has already peaked by $t_{\mathrm{rest}}\sim120\,$d and begun to shift to lower frequencies.}  \added{AT2024aehp is also one of only two LFBOTs to have a localization that is consistent with it being a nuclear transient (the other being CSS161010)}. 

\added{The properties of AT2024aehp} are reminiscent of TDEs. 
\added{Most observed TDEs occur in the nucleus as they involve the galaxy's central supermassive black hole \citep{gezari2021}},  exhibit late-time radio brightening \citep{horesh2021, cendes2024}, and have a slower optical light-curve decay following $t^{-5/3}$ \citep{evans1989}.  While known TDEs have optical rises lasting weeks \citep{blagorodnova2017}, a TDE from an intermediate-mass black hole could have a much shorter rise \citep{kuin2019,perley2019,Gutierrez2024}.
Indeed, the transient AT2024puz---which produced luminous optical/UV evolving on a $\approx20\,$d timescale as well as luminous ($10^{44}\,$erg\,s$^{-1}$) X-ray emission---was posited to be an object intermediate between LFBOTs and TDEs \citep{Somalwar2025}. Observations of AT2024aehp are ongoing, and a detailed analysis of this transient will be presented in future work.   


\subsection{Considerations for Future Work}\label{discussion-improvements}

From our work discovering and performing multiwavelength observations of LFBOTs, we have identified areas to improve on for future analysis, summarized below. 

{\em Observations} -- 
\added{It is unclear how prevalent optical plateaus and late-time radio rebrightenings (as was seen in AT2024aehp) are in LFBOTs.} The optical light curve plateau of AT2024aehp motivates long-term ($>10\,$d) optical monitoring of LFBOTs. As shown in Figure~\ref{fig:optical}, this extended follow-up was not present in our sample's early LFBOTs \added{AT2022abfc, AT2023fhn, and AT2023hkw}, which limits our ability to constrain the presence of a similar plateau.  \added{In addition, late-time radio follow-up observations should be obtained to search for  rebrightenings on timescales of hundreds of days.}

Another gap in our follow-up observations is the lack of X-ray detections for most of these LFBOTs. \swift\ observations were not sensitive enough to rule out levels of X-ray emission that have already been detected in previous LFBOTs such as CSS161010 \citep{coppejans2020} and AT2018cow \citep{perley2019}.  Going forward, deeper {\em Chandra} observations should be obtained for each object, as was done in, e.g., AT2023fhn \citep{chrimes2024}.  \added{X-ray detections will let us verify that the X-ray emission is in excess of the extrapolated synchtrotron SED, confirming that it originates from a different source, like in previous LFBOTs \citep[e.g.][]{margutti, bright2022}.  This also allows us to track the ratio of X-ray to UV/optical flux over time.  Under the proposed scenario that the UV/optical flux is reprocessed high-energy radiation \citep{margutti, lebaron2025}, this would enable us to track the optical depth of the ejecta's reprocessing layer.  Finally, as the X-ray flux is thought to originate from a central object \citep{ho2019, metzger2022, nayana2025}, it is one of the few probes of the regions closest to the transient's center.}

{\em Modeling} -- The models we use in this paper's analysis could be improved upon or have assumptions that could be relaxed to explore a wider range of parameter space.  

We use the model of \citet{chevalier} to derive properties of the shock and ambient medium from our radio and millimeter observations.  This formulation requires assuming that certain conditions hold, such as the equipartition of energy, a non-thermal electron energy distribution, and that the shock is spherical and nonrelativistic.  The nonrelativistic assumption will introduce some inaccuracies, as we calculate some values of $\Gamma \beta$ that are greater than 0.5.  We also assume values for the filling factor $f$, $\epsilon_e$, $\epsilon_B$, and the electron power-law index, and our results do depend on these parameters. 
A future analysis could relax some of these assumptions or fit for parameters that we set to be fixed.  Some of these flaws can be inferred from our fitting results.  According to \citet{chevalier}, the broken power-law fit should have a $\nu^{5/2}$ dependence in the optically thick regime and a $\nu^{-(p-1)/2}$ dependence in the optically thin regime, where $p$ is the electron power-law index.  We assume $p=3$, so the optically thin dependence should be as $\nu^{-1}$.  However, for most epochs we do not have enough detections to confirm a $\nu^{5/2}$ dependence in the optically thick regime (Figure~\ref{fig:radio-sed}).  In addition, from the results shown in Table~\ref{tab:synchro_fit}, the observed slopes in the optically thin regime are not all consistent with $\nu^{-1}$.  Work has already been done to extend this model to account for thermal electrons \citep{margalit2021} and relativistic effects \citep{margalit2024, ferguson2025}.  Generalizing and refining the synchrotron model would help obtain more rigorous estimates of shock and ambient medium parameters.  \added{For example, relativistic effects become significant at $\Gamma \beta \gtrsim1$---which some of our LFBOTs approach---where results can differ by an order of magnitude \citep{ferguson2025}.  Our conclusions would still be applicable as they rely on the similarity of the radio emission between LFBOTs, but our numerical results can be significantly affected.  As our paper's scope is to present new observational data for six LFBOTs and use techniques analogous to those in previous LFBOTs papers to compare properties across this sample, we defer the application of improved synchrotron radiation models to future work.} 

In our \texttt{prospector} fits for the LFBOT host-galaxy properties, we used a single prescription for the SFH, dust model, and initial mass function, but a future analysis should ensure that results are robust to different choices for these parameters.  In particular, we use a parametric SFH to \added{match the work of previous LFBOT papers, but internal testing with a non-parametric SFH produced results that differ by about 1-$\sigma$}.  We also omit the spectra when performing our fits, as it is difficult to standardize spectra from different observing conditions, processed through different software, and possessing different levels of contamination from the transient itself.  This required us to use the mass-metallicity relation in \citet{gallazzi2005} to fix a metallicity in our models.  \added{As the mass is largely set by the galaxy's overall luminosity, we expect that this would lead to little change in our mass estimates, but will bias our metallicity estimates to more moderate values.  We have also observed that the star-formation rate estimate is insensitive to changes in metallicity.  So, our method does introduce a caveat to our metallicity estimates, but we still have confidence in our host-galaxy analysis as it relies on the mass and star-formation rate.  Still, }incorporating a host-galaxy spectrum obtained after the transient faded would produce a more accurate result.  \added{After the submission of this paper, \citet{Nugent2026} performed a \texttt{prospector} LFBOT host-galaxy analysis with a non-parametric star-formation history and utilized host-galaxy spectrscopy when available.  They find similar results for the hosts of AT2023fhn and AT2024qfm, but have larger host-galaxy masses by a factor of a few for AT2023hkw and AT2023vth.}

{\em Selection Bias} -- \added{
Our requirements for LFBOT classification introduce biases that are difficult to account for.}  For example, the requirement of a visible host galaxy biases our sample towards more luminous hosts.  
While we did account for some bias in our LFBOT sample in our study on host-galaxy masses, we \added{caution that we }did not perform any such correction for our analysis of host-galaxy offsets. 
\added{Another source of bias is our requirement of an X-ray and radio detection.  While this would exclude LFBOTs that are X-ray and radio faint, our procedure's success rate of detecting X-rays or radio emission from promising LFBOT candidates implies that the loss is not significant.  Finally, all of these LFBOTs were selected on the basis of their optical light curves.  As we discussed in Section \ref{sec:comparison}, this would bias our LFBOTs to have similar optical light curve properties, such as their absolute luminosities and rise/\added{decay} rates.  This could lead us to miss any diversity in LFBOT optical characteristics (e.g., \citealt{Somalwar2025}).}  

\emph{Future Surveys} -- The development of the Vera C. Rubin Observatory \citep{lsst} and its deep, wide-field optical imaging survey is expected to increase the transient discovery rate by orders of magnitude. This would cause the number of known LFBOTs to increase significantly, paving the way for more robust analysis. The results presented in this paper inform how to best handle each new LFBOT discovery and serve as a blueprint for statistical analysis of an expanded LFBOT sample. 

Rubin data, along with the advent of high-cadence wide-field UV surveys such as the Ultraviolet Transient Astronomy Satellite (ULTRASAT; \citealt{Shvartzvald2024}) and the Ultraviolet Explorer (UVEX; \citealt{Kulkarni2021}), will enable LFBOTs to be identified and followed up much earlier in their evolution.  \added{Rubin's sensitivity can allow for detections before the optical peak, allowing us to identify rapid rising light curves within a couple of days.  Photometric classifiers can potentially screen out LFBOT false positives, mitigating the dependence on the presence of a host galaxy for classification.    Meanwhile, ULTRASAT's rapid cadence (5 minutes) means that an LFBOT's fast brightening can be identified in a matter of hours.   Faster discoveries enable earlier radio observations at $t_\mathrm{obs}\leq5$ days, probing smaller shock radii and more recent mass-loss history.  
Early UV observations, in conjunction with NIR observations, can show evidence of a dust echo, providing an independent measure of the surrounding CSM \citep{metzger2023}.  The UV can also provide better estimates of the early photospheric radius and effective temperatures, as fits to only optical detections can be uncertain \citep{Pursiainen2025}.  Additionally, time-domain surveys outside the optical band can provide a new avenue to find LFBOTs.  
These surveys may show that LFBOTs are part of a more diverse sample that we have only begun to study.  In all, the ability to detect and study LFBOTs over a broader suite of observatories and surveys can remedy some of the challenges of our current studies.}  


\section{Summary}

We presented multiwavelength data---including optical photometry, optical spectroscopy, \emph{Swift} X-ray observations, and VLA/NOEMA/uGMRT radio observations---of six LFBOTs (AT2022abfc, AT2023fhn, AT2023hkw, AT2023vth, AT2024qfm, and AT2024aehp) at redshifts spanning $z=0.0747$--0.339. We identified all of these objects on the basis of fast light-curve evolution in ZTF survey data, and confirmed their similarity to the prototype AT2018cow via radio detections. Five of the six LFBOTs (all except AT2023fhn) are presented here for the first time. Our work increases the known number of LFBOTs by 50\%. 

The optical light curves of four \added{of these} events fade by 3--4\,mag over $t_\mathrm{rest}\approx10\,$d, similarly to AT2018cow. A fifth event, \added{AT2022abfc, did not have deep photmetric follow-up to confirm fading over this timescale, but had a similar decay rate over the first five days.  The last,} AT2024aehp, exhibited a plateau at $M\approx-19\,$mag over tens of days, after an initial fast decline. The optical spectra do not show any clear features from the transients themselves, including a spectrum of AT2024aehp during its plateau phase. 

The X-ray luminosity of AT2024qfm is among the highest of all LFBOTs to date, at $10^{44}\,$erg\,s$^{-1}$ at $t_\mathrm{rest}\approx1\,$week. The X-rays also appeared to show a brightening by an order of magnitude over just a few days. \emph{Swift} X-ray observations of the other objects resulted in non-detections that were not deep enough to rule out emission identical to AT2018cow itself. AT2023fhn was detected by Chandra at similar levels to AT2018cow \citep{chrimesfhn,nayana2025}. 

The 10\,GHz radio light curves peak at $t_\mathrm{rest}\approx50$--100\,d, with peak luminosities primarily in the range of $10^{39}$--$10^{40}$\,erg\,s$^{-1}$. During the rise, the emission appears to be self-absorbed, confirmed by multi-band VLA observations, as well as by NOEMA 100\,GHz detections for AT2024qfm and AT2023vth. We measured basic shock parameters using a simple equipartition analysis, and found---as for previous LFBOTs---fast but sub-relativistic shock speeds. The exception to the radio behavior is AT2024aehp, which, starting at $t_\mathrm{rest}\approx70\,$d, displayed rapid brightening. This late-time brightening is reminiscent of TDEs. Observations of AT2024aehp are ongoing, and will be presented in future work. 

For all objects, we measured the offset of the transient position from the host-galaxy nucleus, as well as the basic host-galaxy properties (e.g., stellar mass, star formation rate or SFR). Combining our results with all literature objects, for a total sample of 13 objects, we find that the offset distribution appears intermediate to CCSNe and SLSNe. We find that the host galaxy masses and SFRs trace the full CCSN sequence, and that---even correcting for the fact that we are biased towards detecting LFBOTs in higher-mass galaxies, due to our requirement of a host-galaxy detection---the intrinsic mass distribution appears to be consistent with both SLSNe and CCSNe. 

The similar radio light curves imply a similar circumburst medium for LFBOTs, and therefore a progenitor scenario that can produce a consistent medium timed with the terminal event, such as the merger of a massive star with a compact object. This argument was made when the sample size of LFBOTs was much smaller \citep{metzger2022}, but here we confirm it with a much larger number of events. 
\added{For future LFBOT studies, we would suggest} longer-term optical monitoring (to confirm or rule out the presence of a plateau as observed in AT2024aehp), deeper X-ray observations (i.e., with \emph{Chandra}), and selection in other bands of the electromagnetic spectrum to mitigate some of the biases present here. \added{LFBOT studies are also poised to benefit from new and upcoming surveys.  The deep sensitivity of the Legacy Survey of Space and Time on the Rubin Observatory and the development of very-high-cadence surveys like ULTRASAT can simultaneously expand our number of LFBOTs and enable us to study them in greater depth. These observatories, among others such as UVEX and the Argus Array, can provide breakthroughs in our understanding of LFBOTs.}

\section{Data Availability}
\added{All data and code underlying this paper are available at \href{https://github.com/CassieSev/Multiwavelength-Analysis-of-Six-LFBOTs}{https://github.com/CassieSev/Multiwavelength-Analysis-of-Six-LFBOTs} and at Zenodo: \dataset[10.5281/zenodo.20819447]{10.5281/zenodo.20819447}.}

\section{Acknowledgements}
We thank the reviewer and the journal editors for their insightful and constructive feedback.

We thank Dusán Tubín for assistance using the eROSITA upper limit service; Ryan Chornock for providing a AT2024qfm LRIS spectrum; and Raffaella Margutti and Joe Bright for useful discussions.

A.Y.Q.H., C.S., \added{M.C., and G.S. acknowledge support from a Sloan Research Fellowship (Award Number FG-2024-21320) from the Alfred P. Sloan Foundation, a Packard Fellowship from the David and Lucile Packard Foundation}, National Aeronautics and Space Administration (NASA) grant 80NSSC24K0377, an LSST Scialog Early Science grant from the Research Corporation for Science Advancement, {\it Hubble Space Telescope} ({\it HST}) grant HST-GO-17477.006-A. 

M.W.C. acknowledges support from the U.S. National Science Foundation (NSF) with grant numbers PHY-2409481, PHY-2308862, and PHY-2117997.
A.V.F.'s group at UC Berkeley received financial assistance from the Christopher R. Redlich Fund, as well as donations from Gary and    
Cynthia Bengier, Clark and Sharon Winslow, Alan Eustace and Kathy Kwan, William Draper, Timothy and Melissa Draper, Briggs and Kathleen Wood, Sanford Robertson (W.Z. is a Bengier-Winslow-Eustace Specialist in Astronomy, T.G.B. is a Draper-Wood-Robertson Specialist in Astronomy, Y.Y. was a Bengier-Winslow-Robertson Fellow in Astronomy), and numerous other donors.  
A.G.Y.'s research is supported by ISF, IMOS and BSF grants, as well as the André Deloro Institute for Space and Optics Research, the Center for Experimental Physics, a WIS-MIT Sagol grant, the Norman E Alexander Family M Foundation ULTRASAT Data Center Fund, and Yeda-Sela;  A.G.Y. is the incumbent of the The Arlyn Imberman Professorial Chair.
C.L. is supported by DoE award \#\,DE-SC0025599.
B.M. is supported in part by the National Science Foundation under grant number AST-2508620.
N.R. is supported by a Northwestern University Presidential Fellowship Award. Zwicky Transient Facility and MMT Observatory access for N.R. was supported by Northwestern University and the Center for Interdisciplinary Exploration and Research in Astrophysics (CIERA).
This work is supported by the U.S. NSF 
under Cooperative Agreement PHY-2019786 (The NSF AI Institute for Artificial Intelligence and Fundamental Interactions; http://iaifi.org/).

We thank the staff of the various observatories at which data were obtained.

Based in part on observations obtained with the Samuel Oschin Telescope 48-inch and the 60-inch
Telescope at the Palomar Observatory as part of the Zwicky Transient Facility project. ZTF is
supported by the National Science Foundation under Grants No. AST-1440341, AST-2034437, and
currently Award \#2407588. ZTF receives additional funding from the ZTF partnership. Current
members include Caltech, USA; Caltech/IPAC, USA; University of Maryland, USA; University of
California, Berkeley, USA; University of Wisconsin at Milwaukee, USA; Cornell University, USA;
Drexel University, USA; University of North Carolina at Chapel Hill, USA; Institute of Science and
Technology, Austria; National Central University, Taiwan, and OKC, University of Stockholm,
Sweden. Operations are conducted by Caltech's Optical Observatory (COO), Caltech/IPAC, and the
University of Washington at Seattle, USA.

SED Machine is based upon work supported by the National Science Foundation under Grant No.
1106171.

Some of the data presented herein were obtained at W. M. Keck Observatory, which is a private 501(c)3 nonprofit organization operated as a scientific partnership among the California Institute of Technology, the University of California, and NASA. 
The Observatory was made possible by the generous financial support of the W. M. Keck Foundation.
This research has made use of the Keck Observatory Archive (KOA), which is operated by the W. M. Keck Observatory and the NASA Exoplanet Science Institute (NExScI), under contract with NASA.
This work was enabled by observations made from the Gemini North telescope, located within the Maunakea Science Reserve and adjacent to the summit of Maunakea. We are grateful for the privilege of observing the Universe from a place that is unique in both its astronomical quality and its cultural significance.  The authors wish to recognize and acknowledge the very significant cultural role and reverence that the summit of Maunakea has always had within the indigenous Hawaiian community. We are most fortunate to have the opportunity to conduct observations from this mountain.

This research has made use of the NASA/IPAC Extragalactic Database (NED), which is funded by NASA and operated by the California Institute of Technology, as well as data and software provided by the High Energy Astrophysics Science Archive Research Center (HEASARC), which is a service of the Astrophysics Science Division at NASA/GSFC.

The Liverpool Telescope is operated on the island of La Palma by Liverpool John Moores University in the Spanish Observatorio del Roque de los Muchachos of the Instituto de Astrofisica de Canarias with financial support from the UK Science and Technology Facilities Council.

Observations reported here were obtained in part at the MMT Observatory, a joint facility of the Smithsonian Institution and the University of Arizona.
Based in part on observations made with the Nordic Optical Telescope, owned in collaboration by the University of Turku and Aarhus University, and operated jointly by Aarhus University, the University of Turku and the University of Oslo, representing Denmark, Finland and Norway, the University of Iceland and Stockholm University at the Observatorio del Roque de los Muchachos, La Palma, Spain, of the Instituto de Astrofisica de Canarias. The NOT data were obtained under program ID P68-501.

These results made use of the Lowell Discovery Telescope (LDT) at Lowell Observatory.  Lowell is a private, nonprofit institution dedicated to astrophysical research and public appreciation of astronomy and operates the LDT in partnership with Boston University, the University of Maryland, the University of Toledo, Northern Arizona University and Yale University.  The Large Monolithic Imager was built by Lowell Observatory using funds provided by the U.S. NSF (AST-1005313).
This paper includes data gathered with the 6.5\,m Magellan Telescopes located at Las Campanas Observatory, Chile. 
Based in part on observations obtained at the Southern Astrophysical Research (SOAR) telescope, which is a joint project of the Minist\'{e}rio da Ci\^{e}ncia, Tecnologia e Inova\c{c}\~{o}es (MCTI/LNA) do Brasil, the U.S. NSF's NOIRLab, the University of North Carolina at Chapel Hill (UNC), and Michigan State University (MSU).

Based in part on observations processed using DRAGONS (Data Reduction for Astronomy from Gemini Observatory North and South), obtained at the international Gemini Observatory, program of NSF NOIRLab,  which is managed by the Association of Universities for Research in Astronomy (AURA) under a cooperative agreement with the U.S. NSF on behalf of the Gemini Observatory partnership: the U.S. NSF (United States), National Research Council (Canada), Agencia Nacional de Investigaci\'{o}n y Desarrollo (Chile), Ministerio de Ciencia, Tecnolog\'{i}a e Innovaci\'{o}n (Argentina), Minist\'{e}rio da Ci\^{e}ncia, Tecnologia, Inova\c{c}\~{o}es e Comunica\c{c}\~{o}es (Brazil), and Korea Astronomy and Space Science Institute (Republic of Korea).

The National Radio Astronomy Observatory is a facility of the U.S. NSF operated under cooperative agreement by Associated Universities, Inc.
This work is based on observations carried out under project numbers D24AA, W22BT and S23BF with the IRAM NOEMA Interferometer. IRAM is supported by INSU/CNRS (France), MPG (Germany) and IGN (Spain).
%
GMRT is run by the National Centre for Radio Astrophysics of the Tata Institute of Fundamental Research.

We acknowledge the use of public data from the \swift\  data archive.
This research is based on observations made with the {\em Galaxy Evolution Explorer} and the {\em Neil Gehrels Swift Observatory}, obtained from the MAST data archive at the Space Telescope Science Institute, which is operated by the Association of Universities for Research in Astronomy, Inc., under NASA contract NAS 5–26555.

The Pan-STARRS1 Surveys (PS1) and the PS1 public science archive have been made possible through contributions by the Institute for Astronomy, the University of Hawaii, the Pan-STARRS Project Office, the Max-Planck Society and its participating institutes, the Max Planck Institute for Astronomy, Heidelberg and the Max Planck Institute for Extraterrestrial Physics, Garching, The Johns Hopkins University, Durham University, the University of Edinburgh, the Queen's University Belfast, the Harvard-Smithsonian Center for Astrophysics, the Las Cumbres Observatory Global Telescope Network Incorporated, the National Central University of Taiwan, the Space Telescope Science Institute, NASA under grant  \#NNX08AR22G issued through the Planetary Science Division of the NASA Science Mission Directorate, NSF grant AST–1238877, the University of Maryland, Eotvos Lorand University (ELTE), the Los Alamos National Laboratory, and the Gordon and Betty Moore Foundation.

The Legacy Surveys consist of three individual and complementary projects: the Dark Energy Camera Legacy Survey (DECaLS; Proposal ID \#2014B-0404; PIs D. Schlegel and A. Dey), the Beijing-Arizona Sky Survey (BASS; NOAO Prop. ID \#2015A-0801; PIs Zhou Xu and Xiaohui Fan), and the Mayall z-band Legacy Survey (MzLS; Prop. ID \#2016A-0453; PI A. Dey). DECaLS, BASS and MzLS together include data obtained, respectively, at the Blanco telescope, Cerro Tololo Inter-American Observatory, NSF’s NOIRLab; the Bok telescope, Steward Observatory, University of Arizona; and the Mayall telescope, Kitt Peak National Observatory, NOIRLab. Pipeline processing and analyses of the data were supported by NOIRLab and the Lawrence Berkeley National Laboratory (LBNL). The Legacy Surveys project is honored to be permitted to conduct astronomical research on Iolkam Du’ag (Kitt Peak), a mountain with particular significance to the Tohono O’odham Nation.

NOIRLab is operated by the Association of Universities for Research in Astronomy (AURA) under a cooperative agreement with the U.S. LBNL is managed by the Regents of the University of California under contract to the U.S. Department of Energy.

This project used data obtained with the Dark Energy Camera (DECam), which was constructed by the Dark Energy Survey (DES) collaboration. Funding for the DES Projects has been provided by the U.S. Department of Energy, the U.S. NSF, the Ministry of Science and Education of Spain, the Science and Technology Facilities Council of the United Kingdom, the Higher Education Funding Council for England, the National Center for Supercomputing Applications at the University of Illinois at Urbana-Champaign, the Kavli Institute of Cosmological Physics at the University of Chicago, Center for Cosmology and Astro-Particle Physics at the Ohio State University, the Mitchell Institute for Fundamental Physics and Astronomy at Texas A\&M University, Financiadora de Estudos e Projetos, Fundacao Carlos Chagas Filho de Amparo, Financiadora de Estudos e Projetos, Fundacao Carlos Chagas Filho de Amparo a Pesquisa do Estado do Rio de Janeiro, Conselho Nacional de Desenvolvimento Cientifico e Tecnologico and the Ministerio da Ciencia, Tecnologia e Inovacao, the Deutsche Forschungsgemeinschaft and the Collaborating Institutions in the Dark Energy Survey. The Collaborating Institutions are Argonne National Laboratory, the University of California at Santa Cruz, the University of Cambridge, Centro de Investigaciones Energeticas, Medioambientales y Tecnologicas-Madrid, the University of Chicago, University College London, the DES-Brazil Consortium, the University of Edinburgh, the Eidgenossische Technische Hochschule (ETH) Zurich, Fermi National Accelerator Laboratory, the University of Illinois at Urbana-Champaign, the Institut de Ciencies de l’Espai (IEEC/CSIC), the Institut de Fisica d’Altes Energies, Lawrence Berkeley National Laboratory, the Ludwig Maximilians Universitat Munchen and the associated Excellence Cluster Universe, the University of Michigan, NSF’s NOIRLab, the University of Nottingham, the Ohio State University, the University of Pennsylvania, the University of Portsmouth, SLAC National Accelerator Laboratory, Stanford University, the University of Sussex, and Texas A\&M University.

BASS is a key project of the Telescope Access Program (TAP), which has been funded by the National Astronomical Observatories of China, the Chinese Academy of Sciences (the Strategic Priority Research Program “The Emergence of Cosmological Structures” Grant \# XDB09000000), and the Special Fund for Astronomy from the Ministry of Finance. The BASS is also supported by the External Cooperation Program of Chinese Academy of Sciences (Grant \# 114A11KYSB20160057), and Chinese National Natural Science Foundation (grant \#12120101003, \#11433005).

The Legacy Surveys imaging of the DESI footprint is supported by the Director, Office of Science, Office of High Energy Physics of the U.S. Department of Energy under Contract No. DE-AC02-05CH1123, by the National Energy Research Scientific Computing Center, a DOE Office of Science User Facility under the same contract; and by the U.S. NSF, Division of Astronomical Sciences under Contract No. AST-0950945 to NOAO.

This publication makes use of data products from the {\it Near-Earth Object Wide-field Infrared Survey Explorer (NEOWISE)}, which is a joint project of the Jet Propulsion Laboratory/California Institute of Technology and the University of California, Los Angeles. {\it NEOWISE} is funded by NASA.

This work is based on data from eROSITA, the soft X-ray instrument aboard SRG, a joint Russian-German science mission supported by the Russian Space Agency (Roskosmos), in the interests of the Russian Academy of Sciences represented by its Space Research Institute (IKI), and the Deutsches Zentrum für Luft- und Raumfahrt (DLR). The SRG spacecraft was built by Lavochkin Association (NPOL) and its subcontractors, and is operated by NPOL with support from the Max Planck Institute for Extraterrestrial Physics (MPE). The development and construction of the eROSITA X-ray instrument was led by MPE, with contributions from the Dr. Karl Remeis Observatory Bamberg \& ECAP (FAU Erlangen-Nuernberg), the University of Hamburg Observatory, the Leibniz Institute for Astrophysics Potsdam (AIP), and the Institute for Astronomy and Astrophysics of the University of Tübingen, with the support of DLR and the Max Planck Society. The Argelander Institute for Astronomy of the University of Bonn and the Ludwig Maximilians Universität Munich also participated in the science preparation for eROSITA.

\facilities{Blanco (DECaLS), Bok (BASS), DCT (LMI), GALEX, Gemini:North, Gemini:South, GMRT, IRAM:NOEMA, Keck:I (LRIS), Keck:II (KCWI, DEIMOS), Liverpool:2m, Magellan:Baade (IMACS), Mayall (MzLS), MMT (Binospec), NOT (ALFOSC), PO:1.2m, PO:1.5m, Sloan, SOAR (GHTS), Swift, VLA, NEOWISE}

\bibliography{mybib}

@ARTICLE{1998bwradio,
       author = {{Kulkarni}, S.~R. and {Frail}, D.~A. and {Wieringa}, M.~H. and {Ekers}, R.~D. and {Sadler}, E.~M. and {Wark}, R.~M. and {Higdon}, J.~L. and {Phinney}, E.~S. and {Bloom}, J.~S.},
        title = "{Radio emission from the unusual supernova 1998bw and its association with the {\ensuremath{\gamma}}-ray burst of 25 April 1998}",
      journal = {\nat},
         year = 1998,
        month = oct,
       volume = {395},
       number = {6703},
        pages = {663-669},
          doi = {10.1038/27139},
       adsurl = {https://ui.adsabs.harvard.edu/abs/1998Natur.395..663K}
}

@ARTICLE{atlas,
       author = {{Tonry}, J.~L. and {Denneau}, L. and {Heinze}, A.~N. and {Stalder}, B. and
         {Smith}, K.~W. and {Smartt}, S.~J. and {Stubbs}, C.~W. and {Weiland
        }, H.~J. and {Rest}, A.},
        title = "{ATLAS: A High-cadence All-sky Survey System}",
      journal = {\pasp},
         year = 2018,
        month = jun,
       volume = {130},
       number = {988},
        pages = {064505},
          doi = {10.1088/1538-3873/aabadf},
archivePrefix = {arXiv},
       eprint = {1802.00879},
 primaryClass = {astro-ph.IM},
       adsurl = {https://ui.adsabs.harvard.edu/abs/2018PASP..130f4505T}
}

@ARTICLE{Kraft1991,
       author = {{Kraft}, Ralph P. and {Burrows}, David N. and {Nousek}, John A.},
        title = "{Determination of Confidence Limits for Experiments with Low Numbers of Counts}",
      journal = {\apj},
         year = 1991,
        month = jun,
       volume = {374},
        pages = {344},
          doi = {10.1086/170124},
       adsurl = {https://ui.adsabs.harvard.edu/abs/1991ApJ...374..344K}
}

@ARTICLE{abmag,
       author = {{Oke}, J.~B. and {Gunn}, J.~E.},
        title = "{Secondary standard stars for absolute spectrophotometry.}",
      journal = {\apj},
         year = 1983,
        month = mar,
       volume = {266},
        pages = {713-717},
          doi = {10.1086/160817},
       adsurl = {https://ui.adsabs.harvard.edu/abs/1983ApJ...266..713O}
}

@ARTICLE{Andreoni2021,
       author = {{Andreoni}, Igor and {Coughlin}, Michael W. and {Kool}, Erik C. and {Kasliwal}, Mansi M. and {Kumar}, Harsh and {Bhalerao}, Varun and {Carracedo}, Ana Sagu{\'e}s and {Ho}, Anna Y.~Q. and {Pang}, Peter T.~H. and {Saraogi}, Divita and {Sharma}, Kritti and {Shenoy}, Vedant and {Burns}, Eric and {Ahumada}, Tom{\'a}s and {Anand}, Shreya and {Singer}, Leo P. and {Perley}, Daniel A. and {De}, Kishalay and {Fremling}, U.~C. and {Bellm}, Eric C. and {Bulla}, Mattia and {Crellin-Quick}, Arien and {Dietrich}, Tim and {Drake}, Andrew and {Duev}, Dmitry A. and {Goobar}, Ariel and {Graham}, Matthew J. and {Kaplan}, David L. and {Kulkarni}, S.~R. and {Laher}, Russ R. and {Mahabal}, Ashish A. and {Shupe}, David L. and {Sollerman}, Jesper and {Walters}, Richard and {Yao}, Yuhan},
        title = "{Fast-transient Searches in Real Time with ZTFReST: Identification of Three Optically Discovered Gamma-Ray Burst Afterglows and New Constraints on the Kilonova Rate}",
      journal = {\apj},
         year = 2021,
        month = sep,
       volume = {918},
       number = {2},
          eid = {63},
        pages = {63},
          doi = {10.3847/1538-4357/ac0bc7},
archivePrefix = {arXiv},
       eprint = {2104.06352},
 primaryClass = {astro-ph.HE},
       adsurl = {https://ui.adsabs.harvard.edu/abs/2021ApJ...918...63A}
}

@ARTICLE{astro2022abfc_disc,
       author = {{Fremling}, C.},
        title = "{ZTF Transient Discovery Report for 2022-11-21}",
      journal = {Transient Name Server Discovery Report},
         year = 2022,
        month = nov,
       volume = {2022-3387},
        pages = {1},
       adsurl = {https://ui.adsabs.harvard.edu/abs/2022TNSTR3387....1F}
}

@ARTICLE{astro2022abfc_vla,
       author = {{Ho}, A.~Y.~Q. and {Liu}, C. and {Chen}, P. and {Perley}, D. and {Wang}, K. and {Altunin}, I.},
        title = "{Gemini, Swift, and VLA Observations of AT2022abfc, a Radio-loud Fast Optical Transient Coincident with a z=0.212 Galaxy}",
      journal = {Transient Name Server AstroNote},
         year = 2022,
        month = dec,
       volume = {275},
        pages = {1},
       adsurl = {https://ui.adsabs.harvard.edu/abs/2022TNSAN.275....1H}
}

@ARTICLE{astro2023fhn_chandra,
       author = {{Chrimes}, A. and {Jonker}, P. and {Levan}, A. and {Coppejans}, D. and {Malesani}, D.~B.},
        title = "{AT2023fhn Chandra X-ray detection}",
      journal = {Transient Name Server AstroNote},
         year = 2023,
        month = may,
       volume = {105},
        pages = {1},
       adsurl = {https://ui.adsabs.harvard.edu/abs/2023TNSAN.105....1C}
}

@ARTICLE{astro2023fhn_disc,
       author = {{Fremling}, C.},
        title = "{ZTF Transient Discovery Report for 2023-04-12}",
      journal = {Transient Name Server Discovery Report},
         year = 2023,
        month = apr,
       volume = {2023-775},
        pages = {1},
       adsurl = {https://ui.adsabs.harvard.edu/abs/2023TNSTR.775....1F}
}

@ARTICLE{astro2023fhn_p200,
       author = {{Ho}, A.~Y.~Q. and {Liu}, C. and {Andreoni}, I. and {Coughlin}, M. and {Qin}, Y. and {Perley}, D.},
        title = "{ZTF, Gemini, and P200 Observations of the Candidate Luminous Fast Blue Optical Transient AT2023fhn}",
      journal = {Transient Name Server AstroNote},
         year = 2023,
        month = apr,
       volume = {93},
        pages = {1},
       adsurl = {https://ui.adsabs.harvard.edu/abs/2023TNSAN..93....1H}
}

@ARTICLE{astro2023fhn_vla,
       author = {{Ho}, A.~Y.~Q.},
        title = "{VLA X-band detection of AT2023fhn}",
      journal = {Transient Name Server AstroNote},
         year = 2023,
        month = jun,
       volume = {174},
        pages = {1},
       adsurl = {https://ui.adsabs.harvard.edu/abs/2023TNSAN.174....1H}
}

@ARTICLE{astro2023hkw_disc,
       author = {{Fremling}, C.},
        title = "{ZTF Transient Discovery Report for 2023-05-01}",
      journal = {Transient Name Server Discovery Report},
        year = 2023,
        month = may,
       volume = {2023-981},
        pages = {1},
       adsurl = {https://ui.adsabs.harvard.edu/abs/2023TNSTR.981....1F}
}

@ARTICLE{astro2023hkw_keck,
       author = {{Li}, M.~L. and {Ho}, A.~Y.~Q. and {Filippenko}, A.~V. and {Brink}, T.~G. and {Yang}, Y. and {Zheng}, W. and {Risin}, S. and {Perley}, D. and {Hinds}, K. and {Andreoni}, I. and {Coughlin}, M.},
        title = "{ZTF, LT, and Keck/DEIMOS Observations of the Candidate Fast Blue Optical Transient (FBOT) AT2023hkw}",
      journal = {Transient Name Server AstroNote},
         year = 2023,
        month = may,
       volume = {114},
        pages = {1},
       adsurl = {https://ui.adsabs.harvard.edu/abs/2023TNSAN.114....1L}
}

@ARTICLE{astro2023hkw_vla,
       author = {{Ho}, A.~Y.~Q.},
        title = "{VLA X-band detection of AT2023hkw}",
      journal = {Transient Name Server AstroNote},
         year = 2023,
        month = jun,
       volume = {173},
        pages = {1},
       adsurl = {https://ui.adsabs.harvard.edu/abs/2023TNSAN.173....1H}
}

@ARTICLE{astro2023vth_disc,
       author = {{Fremling}, C.},
        title = "{ZTF Transient Discovery Report for 2023-10-24}",
      journal = {Transient Name Server Discovery Report},
         year = 2023,
        month = oct,
       volume = {2023-2714},
        pages = {1},
       adsurl = {https://ui.adsabs.harvard.edu/abs/2023TNSTR2714....1F}
}

@ARTICLE{astro2023vth_noema,
       author = {{Ho}, A.~Y.~Q. and {Bremer}, M.},
        title = "{NOEMA 100 GHz detection of AT2023vth}",
      journal = {Transient Name Server AstroNote},
         year = 2023,
        month = nov,
       volume = {318},
        pages = {1},
       adsurl = {https://ui.adsabs.harvard.edu/abs/2023TNSAN.318....1H}
}

@ARTICLE{astro2023vth_vla,
       author = {{Ho}, A.~Y.~Q.},
        title = "{VLA X-band detection of AT2023vth}",
      journal = {Transient Name Server AstroNote},
         year = 2023,
        month = nov,
       volume = {317},
        pages = {1},
       adsurl = {https://ui.adsabs.harvard.edu/abs/2023TNSAN.317....1H}
}

@ARTICLE{astro2023vth_ztf,
       author = {{Sevilla}, J. and {Li}, M.~L. and {Ho}, A.~Y.~Q.},
        title = "{ZTF Observations of the Candidate Fast Blue Optical Transient AT2023vth}",
      journal = {Transient Name Server AstroNote},
         year = 2023,
        month = nov,
       volume = {297},
        pages = {1},
       adsurl = {https://ui.adsabs.harvard.edu/abs/2023TNSAN.297....1S}
}

@ARTICLE{astro2024qfm_vla,
       author = {{Ho}, A.~Y.~Q.},
        title = "{VLA Observation of AT2024qfm}",
      journal = {Transient Name Server AstroNote},
         year = 2024,
        month = aug,
       volume = {211},
        pages = {1},
       adsurl = {https://ui.adsabs.harvard.edu/abs/2024TNSAN.211....1H}
}

@ARTICLE{TubinArenas2024,
       author = {{Tub{\'\i}n-Arenas}, Dus{\'a}n and {Krumpe}, Mirko and {Lamer}, Georg and {Haase}, Jonas and {Sanders}, Jeremy and {Brunner}, Hermann and {Homan}, David and {Schwope}, Axel and {Georgakakis}, Antonis and {Poppenhaeger}, Katja and {Traulsen}, Iris and {K{\"o}nig}, Ole and {Merloni}, Andrea and {Gueguen}, Alain and {Strong}, Andrew and {Liu}, Zhu},
        title = "{The eROSITA upper limits. Description and access to the data}",
      journal = {\aap},
         year = 2024,
        month = feb,
       volume = {682},
          eid = {A35},
        pages = {A35},
          doi = {10.1051/0004-6361/202346773},
archivePrefix = {arXiv},
       eprint = {2401.17305},
 primaryClass = {astro-ph.HE},
       adsurl = {https://ui.adsabs.harvard.edu/abs/2024A&A...682A..35T}
}

@ARTICLE{Merloni2024,
       author = {{Merloni}, A. and {Lamer}, G. and {Liu}, T. and {Ramos-Ceja}, M.~E. and {Brunner}, H. and {Bulbul}, E. and {Dennerl}, K. and {Doroshenko}, V. and {Freyberg}, M.~J. and {Friedrich}, S. and {Gatuzz}, E. and {Georgakakis}, A. and {Haberl}, F. and {Igo}, Z. and {Kreykenbohm}, I. and {Liu}, A. and {Maitra}, C. and {Malyali}, A. and {Mayer}, M.~G.~F. and {Nandra}, K. and {Predehl}, P. and {Robrade}, J. and {Salvato}, M. and {Sanders}, J.~S. and {Stewart}, I. and {Tub{\'\i}n-Arenas}, D. and {Weber}, P. and {Wilms}, J. and {Arcodia}, R. and {Artis}, E. and {Aschersleben}, J. and {Avakyan}, A. and {Aydar}, C. and {Bahar}, Y.~E. and {Balzer}, F. and {Becker}, W. and {Berger}, K. and {Boller}, T. and {Bornemann}, W. and {Br{\"u}ggen}, M. and {Brusa}, M. and {Buchner}, J. and {Burwitz}, V. and {Camilloni}, F. and {Clerc}, N. and {Comparat}, J. and {Coutinho}, D. and {Czesla}, S. and {Dannhauer}, S.~M. and {Dauner}, L. and {Dauser}, T. and {Dietl}, J. and {Dolag}, K. and {Dwelly}, T. and {Egg}, K. and {Ehl}, E. and {Freund}, S. and {Friedrich}, P. and {Gaida}, R. and {Garrel}, C. and {Ghirardini}, V. and {Gokus}, A. and {Gr{\"u}nwald}, G. and {Grandis}, S. and {Grotova}, I. and {Gruen}, D. and {Gueguen}, A. and {H{\"a}mmerich}, S. and {Hamaus}, N. and {Hasinger}, G. and {Haubner}, K. and {Homan}, D. and {Ider Chitham}, J. and {Joseph}, W.~M. and {Joyce}, A. and {K{\"o}nig}, O. and {Kaltenbrunner}, D.~M. and {Khokhriakova}, A. and {Kink}, W. and {Kirsch}, C. and {Kluge}, M. and {Knies}, J. and {Krippendorf}, S. and {Krumpe}, M. and {Kurpas}, J. and {Li}, P. and {Liu}, Z. and {Locatelli}, N. and {Lorenz}, M. and {M{\"u}ller}, S. and {Magaudda}, E. and {Mannes}, C. and {McCall}, H. and {Meidinger}, N. and {Michailidis}, M. and {Migkas}, K. and {Mu{\~n}oz-Giraldo}, D. and {Musiimenta}, B. and {Nguyen-Dang}, N.~T. and {Ni}, Q. and {Olechowska}, A. and {Ota}, N. and {Pacaud}, F. and {Pasini}, T. and {Perinati}, E. and {Pires}, A.~M. and {Pommranz}, C. and {Ponti}, G. and {Poppenhaeger}, K. and {P{\"u}hlhofer}, G. and {Rau}, A. and {Reh}, M. and {Reiprich}, T.~H. and {Roster}, W. and {Saeedi}, S. and {Santangelo}, A. and {Sasaki}, M. and {Schmitt}, J. and {Schneider}, P.~C. and {Schrabback}, T. and {Schuster}, N. and {Schwope}, A. and {Seppi}, R. and {Serim}, M.~M. and {Shreeram}, S. and {Sokolova-Lapa}, E. and {Starck}, H. and {Stelzer}, B. and {Stierhof}, J. and {Suleimanov}, V. and {Tenzer}, C. and {Traulsen}, I. and {Tr{\"u}mper}, J. and {Tsuge}, K. and {Urrutia}, T. and {Veronica}, A. and {Waddell}, S.~G.~H. and {Willer}, R. and {Wolf}, J. and {Yeung}, M.~C.~H. and {Zainab}, A. and {Zangrandi}, F. and {Zhang}, X. and {Zhang}, Y. and {Zheng}, X.},
        title = "{The SRG/eROSITA all-sky survey. First X-ray catalogues and data release of the western Galactic hemisphere}",
      journal = {\aap},
         year = 2024,
        month = feb,
       volume = {682},
          eid = {A34},
        pages = {A34},
          doi = {10.1051/0004-6361/202347165},
archivePrefix = {arXiv},
       eprint = {2401.17274},
 primaryClass = {astro-ph.HE},
       adsurl = {https://ui.adsabs.harvard.edu/abs/2024A&A...682A..34M}
}

@ARTICLE{eROSITA,
       author = {{Predehl}, P. and {Andritschke}, R. and {Arefiev}, V. and {Babyshkin}, V. and {Batanov}, O. and {Becker}, W. and {B{\"o}hringer}, H. and {Bogomolov}, A. and {Boller}, T. and {Borm}, K. and {Bornemann}, W. and {Br{\"a}uninger}, H. and {Br{\"u}ggen}, M. and {Brunner}, H. and {Brusa}, M. and {Bulbul}, E. and {Buntov}, M. and {Burwitz}, V. and {Burkert}, W. and {Clerc}, N. and {Churazov}, E. and {Coutinho}, D. and {Dauser}, T. and {Dennerl}, K. and {Doroshenko}, V. and {Eder}, J. and {Emberger}, V. and {Eraerds}, T. and {Finoguenov}, A. and {Freyberg}, M. and {Friedrich}, P. and {Friedrich}, S. and {F{\"u}rmetz}, M. and {Georgakakis}, A. and {Gilfanov}, M. and {Granato}, S. and {Grossberger}, C. and {Gueguen}, A. and {Gureev}, P. and {Haberl}, F. and {H{\"a}lker}, O. and {Hartner}, G. and {Hasinger}, G. and {Huber}, H. and {Ji}, L. and {Kienlin}, A. v. and {Kink}, W. and {Korotkov}, F. and {Kreykenbohm}, I. and {Lamer}, G. and {Lomakin}, I. and {Lapshov}, I. and {Liu}, T. and {Maitra}, C. and {Meidinger}, N. and {Menz}, B. and {Merloni}, A. and {Mernik}, T. and {Mican}, B. and {Mohr}, J. and {M{\"u}ller}, S. and {Nandra}, K. and {Nazarov}, V. and {Pacaud}, F. and {Pavlinsky}, M. and {Perinati}, E. and {Pfeffermann}, E. and {Pietschner}, D. and {Ramos-Ceja}, M.~E. and {Rau}, A. and {Reiffers}, J. and {Reiprich}, T.~H. and {Robrade}, J. and {Salvato}, M. and {Sanders}, J. and {Santangelo}, A. and {Sasaki}, M. and {Scheuerle}, H. and {Schmid}, C. and {Schmitt}, J. and {Schwope}, A. and {Shirshakov}, A. and {Steinmetz}, M. and {Stewart}, I. and {Str{\"u}der}, L. and {Sunyaev}, R. and {Tenzer}, C. and {Tiedemann}, L. and {Tr{\"u}mper}, J. and {Voron}, V. and {Weber}, P. and {Wilms}, J. and {Yaroshenko}, V.},
        title = "{The eROSITA X-ray telescope on SRG}",
      journal = {\aap},
         year = 2021,
        month = mar,
       volume = {647},
          eid = {A1},
        pages = {A1},
          doi = {10.1051/0004-6361/202039313},
archivePrefix = {arXiv},
       eprint = {2010.03477},
 primaryClass = {astro-ph.HE},
       adsurl = {https://ui.adsabs.harvard.edu/abs/2021A&A...647A...1P}
}

@ARTICLE{astro2024aehp_gemini,
       author = {{Sevilla}, J. and {Ho}, A.~Y.~Q. and {Srinivasaragavan}, G.},
        title = "{ZTF and Gemini Observations of the Fast-fading Luminous Optical Transient AT2024aehp}",
      journal = {Transient Name Server AstroNote},
         year = 2024,
        month = dec,
       volume = {386},
        pages = {1},
       adsurl = {https://ui.adsabs.harvard.edu/abs/2024TNSAN.386....1S}
}

@ARTICLE{Margutti2024_AT2024qfm,
       author = {{Margutti}, R. and {Nayana}, A.~J. and {Sears}, H.},
        title = "{Swift-XRT detection at the location of the FBOT candidate AT2024qfm}",
      journal = {The Astronomer's Telegram},
         year = 2024,
        month = aug,
       volume = {16748},
        pages = {1},
       adsurl = {https://ui.adsabs.harvard.edu/abs/2024ATel16748....1M}
}

@ARTICLE{astro2024aehp_disc,
       author = {{Munoz-Arancibia}, A. and {Pignata}, G. and {Bauer}, F.~E. and {Forster}, F. and {Mourao}, A. and {Hernandez-Garcia}, L. and {Galbany}, L. and {Silva-Farfan}, J. and {Dastidar}, R. and {Brandt}, J.~P. and {Alvarez}, A. and {Cabrera-Vives}, G. and {Carrasco-Davis}, R. and {Estevez}, P.~A. and {Gamboa}, B. and {Huijse}, P. and {Larranaga}, H. and {Mansilla}, C. and {Medina}, K. and {Moreno}, D. and {Munoz}, E. and {Reyes}, I. and {Sanchez-Saez}, P. and {Rodriguez-Mancini}, D. and {Catelan}, M. and {Eyheramendy}, S. and {Graham}, M.~J.},
        title = "{ALeRCE/ZTF Transient Discovery Report for 2024-12-19}",
      journal = {Transient Name Server Discovery Report},
         year = 2024,
        month = dec,
       volume = {2024-4959},
        pages = {1},
       adsurl = {https://ui.adsabs.harvard.edu/abs/2024TNSTR4959....1M}
}

@ARTICLE{astro2024qfm_disc,
       author = {{Forster}, F. and {Bauer}, F.~E. and {Pignata}, G. and {Munoz-Arancibia}, A. and {Mourao}, A. and {Hernandez-Garcia}, L. and {Galbany}, L. and {Silva-Farfan}, J. and {Dastidar}, R. and {Brandt}, J.~P. and {Alvarez}, A. and {Cabrera-Vives}, G. and {Carrasco-Davis}, R. and {Estevez}, P.~A. and {Gamboa}, B. and {Huijse}, P. and {Larranaga}, H. and {Mansilla}, C. and {Medina}, K. and {Moreno}, D. and {Munoz}, E. and {Pizarro}, E. and {Reyes}, I. and {Sanchez-Saez}, P. and {Rodriguez-Mancini}, D. and {Catelan}, M. and {Eyheramendy}, S. and {Graham}, M.~J.},
        title = "{ALeRCE/ZTF Transient Discovery Report for 2024-07-28}",
      journal = {Transient Name Server Discovery Report},
         year = 2024,
        month = jul,
       volume = {2024-2624},
        pages = {1},
       adsurl = {https://ui.adsabs.harvard.edu/abs/2024TNSTR2624....1F}
}

@ARTICLE{berger2003,
       author = {{Berger}, E. and {Kulkarni}, S.~R. and {Pooley}, G. and {Frail}, D.~A. and {McIntyre}, V. and {Wark}, R.~M. and {Sari}, R. and {Soderberg}, A.~M. and {Fox}, D.~W. and {Yost}, S. and {Price}, P.~A.},
        title = "{A common origin for cosmic explosions inferred from calorimetry of GRB030329}",
      journal = {\nat},
         year = 2003,
        month = nov,
       volume = {426},
       number = {6963},
        pages = {154-157},
          doi = {10.1038/nature01998},
archivePrefix = {arXiv},
       eprint = {astro-ph/0308187},
 primaryClass = {astro-ph},
       adsurl = {https://ui.adsabs.harvard.edu/abs/2003Natur.426..154B}
}

@ARTICLE{bhandari2022,
       author = {{Bhandari}, Shivani and {Heintz}, Kasper E. and {Aggarwal}, Kshitij and {Marnoch}, Lachlan and {Day}, Cherie K. and {Sydnor}, Jessica and {Burke-Spolaor}, Sarah and {Law}, Casey J. and {Xavier Prochaska}, J. and {Tejos}, Nicolas and {Bannister}, Keith W. and {Butler}, Bryan J. and {Deller}, Adam T. and {Ekers}, R.~D. and {Flynn}, Chris and {Fong}, Wen-fai and {James}, Clancy W. and {Lazio}, T. Joseph W. and {Luo}, Rui and {Mahony}, Elizabeth K. and {Ryder}, Stuart D. and {Sadler}, Elaine M. and {Shannon}, Ryan M. and {Han}, JinLin and {Lee}, Kejia and {Zhang}, Bing},
        title = "{Characterizing the Fast Radio Burst Host Galaxy Population and its Connection to Transients in the Local and Extragalactic Universe}",
      journal = {\aj},
         year = 2022,
        month = feb,
       volume = {163},
       number = {2},
          eid = {69},
        pages = {69},
          doi = {10.3847/1538-3881/ac3aec},
archivePrefix = {arXiv},
       eprint = {2108.01282},
 primaryClass = {astro-ph.HE},
       adsurl = {https://ui.adsabs.harvard.edu/abs/2022AJ....163...69B}
}

@ARTICLE{blagorodnova2017,
       author = {{Blagorodnova}, N. and {Gezari}, S. and {Hung}, T. and {Kulkarni}, S.~R. and {Cenko}, S.~B. and {Pasham}, D.~R. and {Yan}, L. and {Arcavi}, I. and {Ben-Ami}, S. and {Bue}, B.~D. and {Cantwell}, T. and {Cao}, Y. and {Castro-Tirado}, A.~J. and {Fender}, R. and {Fremling}, C. and {Gal-Yam}, A. and {Ho}, A.~Y.~Q. and {Horesh}, A. and {Hosseinzadeh}, G. and {Kasliwal}, M.~M. and {Kong}, A.~K.~H. and {Laher}, R.~R. and {Leloudas}, G. and {Lunnan}, R. and {Masci}, F.~J. and {Mooley}, K. and {Neill}, J.~D. and {Nugent}, P. and {Powell}, M. and {Valeev}, A.~F. and {Vreeswijk}, P.~M. and {Walters}, R. and {Wozniak}, P.},
        title = "{iPTF16fnl: A Faint and Fast Tidal Disruption Event in an E+A Galaxy}",
      journal = {\apj},
         year = 2017,
        month = jul,
       volume = {844},
       number = {1},
          eid = {46},
        pages = {46},
          doi = {10.3847/1538-4357/aa7579},
archivePrefix = {arXiv},
       eprint = {1703.00965},
 primaryClass = {astro-ph.HE},
       adsurl = {https://ui.adsabs.harvard.edu/abs/2017ApJ...844...46B}
}

@ARTICLE{blanchard2016,
       author = {{Blanchard}, Peter K. and {Berger}, Edo and {Fong}, Wen-fai},
        title = "{The Offset and Host Light Distributions of Long Gamma-Ray Bursts: A New View From HST Observations of Swift Bursts}",
      journal = {\apj},
         year = 2016,
        month = feb,
       volume = {817},
       number = {2},
          eid = {144},
        pages = {144},
          doi = {10.3847/0004-637X/817/2/144},
archivePrefix = {arXiv},
       eprint = {1509.07866},
 primaryClass = {astro-ph.HE},
       adsurl = {https://ui.adsabs.harvard.edu/abs/2016ApJ...817..144B}
}

@ARTICLE{blanchard2017,
       author = {{Blanchard}, P.~K. and {Nicholl}, M. and {Berger}, E. and {Guillochon}, J. and {Margutti}, R. and {Chornock}, R. and {Alexander}, K.~D. and {Leja}, J. and {Drout}, M.~R.},
        title = "{PS16dtm: A Tidal Disruption Event in a Narrow-line Seyfert 1 Galaxy}",
      journal = {\apj},
         year = 2017,
        month = jul,
       volume = {843},
       number = {2},
          eid = {106},
        pages = {106},
          doi = {10.3847/1538-4357/aa77f7},
archivePrefix = {arXiv},
       eprint = {1703.07816},
 primaryClass = {astro-ph.HE},
       adsurl = {https://ui.adsabs.harvard.edu/abs/2017ApJ...843..106B}
}

@PHDTHESIS{briggs1995,
       author = {{Briggs}, Daniel Shenon},
        title = "{High fidelity deconvolution of moderately resolved sources}",
       school = {New Mexico Institute of Mining and Technology},
         year = 1995,
        month = jan,
       adsurl = {https://ui.adsabs.harvard.edu/abs/1995PhDT.......238B}
}

@ARTICLE{bright2022,
       author = {{Bright}, Joe S. and {Margutti}, Raffaella and {Matthews}, David and {Brethauer}, Daniel and {Coppejans}, Deanne and {Wieringa}, Mark H. and {Metzger}, Brian D. and {DeMarchi}, Lindsay and {Laskar}, Tanmoy and {Romero}, Charles and {Alexander}, Kate D. and {Horesh}, Assaf and {Migliori}, Giulia and {Chornock}, Ryan and {Berger}, E. and {Bietenholz}, Michael and {Devlin}, Mark J. and {Dicker}, Simon R. and {Jacobson-Gal{\'a}n}, W.~V. and {Mason}, Brian S. and {Milisavljevic}, Dan and {Motta}, Sara E. and {Mroczkowski}, Tony and {Ramirez-Ruiz}, Enrico and {Rhodes}, Lauren and {Sarazin}, Craig L. and {Sfaradi}, Itai and {Sievers}, Jonathan},
        title = "{Radio and X-Ray Observations of the Luminous Fast Blue Optical Transient AT 2020xnd}",
      journal = {\apj},
         year = 2022,
        month = feb,
       volume = {926},
       number = {2},
          eid = {112},
        pages = {112},
          doi = {10.3847/1538-4357/ac4506},
archivePrefix = {arXiv},
       eprint = {2110.05514},
 primaryClass = {astro-ph.HE},
       adsurl = {https://ui.adsabs.harvard.edu/abs/2022ApJ...926..112B}
}

@SOFTWARE{carta,
       author = {{Comrie}, Angus and {Wang}, Kuo-Song and {Hsu}, Shou-Chieh and {Moraghan}, Anthony and {Harris}, Pamela and {Pang}, Qi and {Pi{\'n}ska}, Adrianna and {Chiang}, Cheng-Chin and {Chang}, Tien-Hao and {Hwang}, Yu-Hsuan and {Jan}, Hengtai and {Lin}, Ming-Yi and {Simmonds}, Rob},
        title = "{CARTA: The Cube Analysis and Rendering Tool for Astronomy}",
         year = 2021,
        month = jun,
          eid = {10.5281/zenodo.3377984},
          doi = {10.5281/zenodo.3377984},
      version = {2.0.0},
    publisher = {Zenodo},
       adsurl = {https://ui.adsabs.harvard.edu/abs/2021zndo...3377984C}
}

@article{casa,
doi = {10.1088/1538-3873/ac9642},
url = {https://doi.org/10.1088/1538-3873/ac9642},
year = {2022},
month = {nov},
publisher = {The Astronomical Society of the Pacific},
volume = {134},
number = {1041},
pages = {114501},
author = {The CASA Team and Bean, Ben and Bhatnagar, Sanjay and Castro, Sandra and Meyer, Jennifer Donovan and Emonts, Bjorn and Garcia, Enrique and Garwood, Robert and Golap, Kumar and Villalba, Justo Gonzalez and Harris, Pamela and Hayashi, Yohei and Hoskins, Josh and Hsieh, Mingyu and Jagannathan, Preshanth and Kawasaki, Wataru and Keimpema, Aard and Kettenis, Mark and Lopez, Jorge and Marvil, Joshua and Masters, Joseph and McNichols, Andrew and Mehringer, David and Miel, Renaud and Moellenbrock, George and Montesino, Federico and Nakazato, Takeshi and Ott, Juergen and Petry, Dirk and Pokorny, Martin and Raba, Ryan and Rau, Urvashi and Schiebel, Darrell and Schweighart, Neal and Sekhar, Srikrishna and Shimada, Kazuhiko and Small, Des and Steeb, Jan-Willem and Sugimoto, Kanako and Suoranta, Ville and Tsutsumi, Takahiro and van Bemmel, Ilse M. and Verkouter, Marjolein and Wells, Akeem and Xiong, Wei and Szomoru, Arpad and Griffith, Morgan and Glendenning, Brian and Kern, Jeff},
title = {CASA, the Common Astronomy Software Applications for Radio Astronomy},
journal = {Publications of the Astronomical Society of the Pacific}
}

@ARTICLE{calzetti2000,
       author = {{Calzetti}, Daniela and {Armus}, Lee and {Bohlin}, Ralph C. and {Kinney}, Anne L. and {Koornneef}, Jan and {Storchi-Bergmann}, Thaisa},
        title = "{The Dust Content and Opacity of Actively Star-forming Galaxies}",
      journal = {\apj},
         year = 2000,
        month = apr,
       volume = {533},
       number = {2},
        pages = {682-695},
          doi = {10.1086/308692},
archivePrefix = {arXiv},
       eprint = {astro-ph/9911459},
 primaryClass = {astro-ph},
       adsurl = {https://ui.adsabs.harvard.edu/abs/2000ApJ...533..682C}
}

@ARTICLE{campana2006b,
       author = {{Campana}, S. and {Mangano}, V. and {Blustin}, A.~J. and {Brown}, P. and {Burrows}, D.~N. and {Chincarini}, G. and {Cummings}, J.~R. and {Cusumano}, G. and {Della Valle}, M. and {Malesani}, D. and {M{\'e}sz{\'a}ros}, P. and {Nousek}, J.~A. and {Page}, M. and {Sakamoto}, T. and {Waxman}, E. and {Zhang}, B. and {Dai}, Z.~G. and {Gehrels}, N. and {Immler}, S. and {Marshall}, F.~E. and {Mason}, K.~O. and {Moretti}, A. and {O'Brien}, P.~T. and {Osborne}, J.~P. and {Page}, K.~L. and {Romano}, P. and {Roming}, P.~W.~A. and {Tagliaferri}, G. and {Cominsky}, L.~R. and {Giommi}, P. and {Godet}, O. and {Kennea}, J.~A. and {Krimm}, H. and {Angelini}, L. and {Barthelmy}, S.~D. and {Boyd}, P.~T. and {Palmer}, D.~M. and {Wells}, A.~A. and {White}, N.~E.},
        title = "{The association of GRB 060218 with a supernova and the evolution of the shock wave}",
      journal = {\nat},
         year = 2006,
        month = aug,
       volume = {442},
       number = {7106},
        pages = {1008-1010},
          doi = {10.1038/nature04892},
archivePrefix = {arXiv},
       eprint = {astro-ph/0603279},
 primaryClass = {astro-ph},
       adsurl = {https://ui.adsabs.harvard.edu/abs/2006Natur.442.1008C}
}

@ARTICLE{cendes2024,
       author = {{Cendes}, Y. and {Berger}, E. and {Alexander}, K.~D. and {Chornock}, R. and {Margutti}, R. and {Metzger}, B. and {Wieringa}, M.~H. and {Bietenholz}, M.~F. and {Hajela}, A. and {Laskar}, T. and {Stroh}, M.~C. and {Terreran}, G.},
        title = "{Ubiquitous Late Radio Emission from Tidal Disruption Events}",
      journal = {\apj},
         year = 2024,
        month = aug,
       volume = {971},
       number = {2},
          eid = {185},
        pages = {185},
          doi = {10.3847/1538-4357/ad5541},
archivePrefix = {arXiv},
       eprint = {2308.13595},
 primaryClass = {astro-ph.HE},
       adsurl = {https://ui.adsabs.harvard.edu/abs/2024ApJ...971..185C}
}

@ARTICLE{chabrier2003,
       author = {{Chabrier}, Gilles},
        title = "{Galactic Stellar and Substellar Initial Mass Function}",
      journal = {\pasp},
         year = 2003,
        month = jul,
       volume = {115},
       number = {809},
        pages = {763-795},
          doi = {10.1086/376392},
archivePrefix = {arXiv},
       eprint = {astro-ph/0304382},
 primaryClass = {astro-ph},
       adsurl = {https://ui.adsabs.harvard.edu/abs/2003PASP..115..763C}
}

@INPROCEEDINGS{chandra,
       author = {{Weisskopf}, Martin C. and {Tananbaum}, Harvey D. and {Van Speybroeck}, Leon P. and {O'Dell}, Stephen L.},
        title = "{Chandra X-ray Observatory (CXO): overview}",
    booktitle = {X-Ray Optics, Instruments, and Missions III},
         year = 2000,
       editor = {{Truemper}, Joachim E. and {Aschenbach}, Bernd},
       series = {Society of Photo-Optical Instrumentation Engineers (SPIE) Conference Series},
       volume = {4012},
        month = jul,
        pages = {2-16},
          doi = {10.1117/12.391545},
archivePrefix = {arXiv},
       eprint = {astro-ph/0004127},
 primaryClass = {astro-ph},
       adsurl = {https://ui.adsabs.harvard.edu/abs/2000SPIE.4012....2W}
}

@ARTICLE{chevalier,
       author = {{Chevalier}, Roger A.},
        title = "{Synchrotron Self-Absorption in Radio Supernovae}",
      journal = {\apj},
         year = 1998,
        month = may,
       volume = {499},
       number = {2},
        pages = {810-819},
          doi = {10.1086/305676},
       adsurl = {https://ui.adsabs.harvard.edu/abs/1998ApJ...499..810C}
}

@ARTICLE{chevalier2006,
       author = {{Chevalier}, Roger A. and {Fransson}, Claes},
        title = "{Circumstellar Emission from Type Ib and Ic Supernovae}",
      journal = {\apj},
         year = 2006,
        month = nov,
       volume = {651},
       number = {1},
        pages = {381-391},
          doi = {10.1086/507606},
archivePrefix = {arXiv},
       eprint = {astro-ph/0607196},
 primaryClass = {astro-ph},
       adsurl = {https://ui.adsabs.harvard.edu/abs/2006ApJ...651..381C}
}

@ARTICLE{chrimes2024,
       author = {{Chrimes}, A.~A. and {Coppejans}, D.~L. and {Jonker}, P.~G. and {Levan}, A.~J. and {Groot}, P.~J. and {Mummery}, A. and {Stanway}, E.~R.},
        title = "{Multi-wavelength observations of the luminous fast blue optical transient AT 2023fhn: Up to {\ensuremath{\sim}}200 days post-explosion}",
      journal = {\aap},
         year = 2024,
        month = nov,
       volume = {691},
          eid = {A329},
        pages = {A329},
          doi = {10.1051/0004-6361/202451172},
archivePrefix = {arXiv},
       eprint = {2406.13821},
 primaryClass = {astro-ph.HE},
       adsurl = {https://ui.adsabs.harvard.edu/abs/2024A&A...691A.329C}
}

@ARTICLE{Chrimes2025,
       author = {{Chrimes}, A.~A. and {Jonker}, P.~G. and {Levan}, A.~J. and {Mummery}, A.},
        title = "{Luminous Fast Blue Optical Transients as very massive star core-collapse events}",
      journal = {arXiv e-prints},
         year = 2025,
        month = oct,
          eid = {arXiv:2510.03402},
        pages = {arXiv:2510.03402},
          doi = {10.48550/arXiv.2510.03402},
archivePrefix = {arXiv},
       eprint = {2510.03402},
 primaryClass = {astro-ph.HE},
       adsurl = {https://ui.adsabs.harvard.edu/abs/2025arXiv251003402C}
}

@ARTICLE{chrimesfhn,
       author = {{Chrimes}, A.~A. and {Jonker}, P.~G. and {Levan}, A.~J. and {Coppejans}, D.~L. and {Gaspari}, N. and {Gompertz}, B.~P. and {Groot}, P.~J. and {Malesani}, D.~B. and {Mummery}, A. and {Stanway}, E.~R. and {Wiersema}, K.},
        title = "{AT2023fhn (the Finch): a luminous fast blue optical transient at a large offset from its host galaxy}",
      journal = {\mnras},
         year = 2024,
        month = jan,
       volume = {527},
       number = {1},
        pages = {L47-L53},
          doi = {10.1093/mnrasl/slad145},
archivePrefix = {arXiv},
       eprint = {2307.01771},
 primaryClass = {astro-ph.HE},
       adsurl = {https://ui.adsabs.harvard.edu/abs/2024MNRAS.527L..47C}
}

@ARTICLE{coppejans2020,
       author = {{Coppejans}, D.~L. and {Margutti}, R. and {Terreran}, G. and {Nayana}, A.~J. and {Coughlin}, E.~R. and {Laskar}, T. and {Alexander}, K.~D. and {Bietenholz}, M. and {Caprioli}, D. and {Chandra}, P. and {Drout}, M.~R. and {Frederiks}, D. and {Frohmaier}, C. and {Hurley}, K.~H. and {Kochanek}, C.~S. and {MacLeod}, M. and {Meisner}, A. and {Nugent}, P.~E. and {Ridnaia}, A. and {Sand}, D.~J. and {Svinkin}, D. and {Ward}, C. and {Yang}, S. and {Baldeschi}, A. and {Chilingarian}, I.~V. and {Dong}, Y. and {Esquivia}, C. and {Fong}, W. and {Guidorzi}, C. and {Lundqvist}, P. and {Milisavljevic}, D. and {Paterson}, K. and {Reichart}, D.~E. and {Shappee}, B. and {Stroh}, M.~C. and {Valenti}, S. and {Zauderer}, B.~A. and {Zhang}, B.},
        title = "{A Mildly Relativistic Outflow from the Energetic, Fast-rising Blue Optical Transient CSS161010 in a Dwarf Galaxy}",
      journal = {\apjl},
         year = 2020,
        month = may,
       volume = {895},
       number = {1},
          eid = {L23},
        pages = {L23},
          doi = {10.3847/2041-8213/ab8cc7},
archivePrefix = {arXiv},
       eprint = {2003.10503},
 primaryClass = {astro-ph.HE},
       adsurl = {https://ui.adsabs.harvard.edu/abs/2020ApJ...895L..23C}
}

@ARTICLE{corsi2014,
       author = {{Corsi}, A. and {Ofek}, E.~O. and {Gal-Yam}, A. and {Frail}, D.~A. and {Kulkarni}, S.~R. and {Fox}, D.~B. and {Kasliwal}, M.~M. and {Sullivan}, M. and {Horesh}, A. and {Carpenter}, J. and {Maguire}, K. and {Arcavi}, I. and {Cenko}, S.~B. and {Cao}, Y. and {Mooley}, K. and {Pan}, Y. -C. and {Sesar}, B. and {Sternberg}, A. and {Xu}, D. and {Bersier}, D. and {James}, P. and {Bloom}, J.~S. and {Nugent}, P.~E.},
        title = "{A Multi-wavelength Investigation of the Radio-loud Supernova PTF11qcj and its Circumstellar Environment}",
      journal = {\apj},
         year = 2014,
        month = feb,
       volume = {782},
       number = {1},
          eid = {42},
        pages = {42},
          doi = {10.1088/0004-637X/782/1/42},
archivePrefix = {arXiv},
       eprint = {1307.2366},
 primaryClass = {astro-ph.HE},
       adsurl = {https://ui.adsabs.harvard.edu/abs/2014ApJ...782...42C}
}

@ARTICLE{coughlin2023,
       author = {{Coughlin}, Michael W. and {Bloom}, Joshua S. and {Nir}, Guy and {Antier}, Sarah and {du Laz}, Theophile Jegou and {van der Walt}, St{\'e}fan and {Crellin-Quick}, Arien and {Culino}, Thomas and {Duev}, Dmitry A. and {Goldstein}, Daniel A. and {Healy}, Brian F. and {Karambelkar}, Viraj and {Lilleboe}, Jada and {Shin}, Kyung Min and {Singer}, Leo P. and {Ahumada}, Tom{\'a}s and {Anand}, Shreya and {Bellm}, Eric C. and {Dekany}, Richard and {Graham}, Matthew J. and {Kasliwal}, Mansi M. and {Kostadinova}, Ivona and {Kiendrebeogo}, R. Weizmann and {Kulkarni}, Shrinivas R. and {Jenkins}, Sydney and {LeBaron}, Natalie and {Mahabal}, Ashish A. and {Neill}, James D. and {Parazin}, B. and {Peloton}, Julien and {Perley}, Daniel A. and {Riddle}, Reed and {Rusholme}, Ben and {van Santen}, Jakob and {Sollerman}, Jesper and {Stein}, Robert and {Turpin}, D. and {Wold}, Avery and {Amat}, Carla and {Bonnefon}, Adrien and {Bonnefoy}, Adrien and {Flament}, Manon and {Kerkow}, Frank and {Kishore}, Sulekha and {Jani}, Shloke and {Mahanty}, Stephen K. and {Liu}, C{\'e}line and {Llinares}, Laura and {Makarison}, Jolyane and {Olli{\'e}ric}, Alix and {Perez}, In{\`e}s and {Pont}, Lydie and {Sharma}, Vyom},
        title = "{A Data Science Platform to Enable Time-domain Astronomy}",
      journal = {\apjs},
         year = 2023,
        month = aug,
       volume = {267},
       number = {2},
          eid = {31},
        pages = {31},
          doi = {10.3847/1538-4365/acdee1},
archivePrefix = {arXiv},
       eprint = {2305.00108},
 primaryClass = {astro-ph.IM},
       adsurl = {https://ui.adsabs.harvard.edu/abs/2023ApJS..267...31C}
}

@ARTICLE{de2020,
       author = {{De}, Kishalay and {Kasliwal}, Mansi M. and {Tzanidakis}, Anastasios and {Fremling}, U. Christoffer and {Adams}, Scott and {Aloisi}, Robert and {Andreoni}, Igor and {Bagdasaryan}, Ashot and {Bellm}, Eric C. and {Bildsten}, Lars and {Cannella}, Christopher and {Cook}, David O. and {Delacroix}, Alexandre and {Drake}, Andrew and {Duev}, Dmitry and {Dugas}, Alison and {Frederick}, Sara and {Gal-Yam}, Avishay and {Goldstein}, Daniel and {Golkhou}, V. Zach and {Graham}, Matthew J. and {Hale}, David and {Hankins}, Matthew and {Helou}, George and {Ho}, Anna Y.~Q. and {Irani}, Ido and {Jencson}, Jacob E. and {Kaplan}, David L. and {Kaye}, Stephen and {Kulkarni}, S.~R. and {Kupfer}, Thomas and {Laher}, Russ R. and {Leadbeater}, Robin and {Lunnan}, Ragnhild and {Masci}, Frank J. and {Miller}, Adam A. and {Neill}, James D. and {Ofek}, Eran O. and {Perley}, Daniel A. and {Polin}, Abigail and {Prince}, Thomas A. and {Quataert}, Eliot and {Reiley}, Dan and {Riddle}, Reed L. and {Rusholme}, Ben and {Sharma}, Yashvi and {Shupe}, David L. and {Sollerman}, Jesper and {Tartaglia}, Leonardo and {Walters}, Richard and {Yan}, Lin and {Yao}, Yuhan},
        title = "{The Zwicky Transient Facility Census of the Local Universe. I. Systematic Search for Calcium-rich Gap Transients Reveals Three Related Spectroscopic Subclasses}",
      journal = {\apj},
         year = 2020,
        month = dec,
       volume = {905},
       number = {1},
          eid = {58},
        pages = {58},
          doi = {10.3847/1538-4357/abb45c},
archivePrefix = {arXiv},
       eprint = {2004.09029},
 primaryClass = {astro-ph.HE},
       adsurl = {https://ui.adsabs.harvard.edu/abs/2020ApJ...905...58D}
}

@INPROCEEDINGS{deimos,
       author = {{Faber}, Sandra M. and {Phillips}, Andrew C. and {Kibrick}, Robert I. and {Alcott}, Barry and {Allen}, Steven L. and {Burrous}, Jim and {Cantrall}, T. and {Clarke}, De and {Coil}, Alison L. and {Cowley}, David J. and {Davis}, Marc and {Deich}, William T.~S. and {Dietsch}, Ken and {Gilmore}, David K. and {Harper}, Carol A. and {Hilyard}, David F. and {Lewis}, Jeffrey P. and {McVeigh}, Molly and {Newman}, Jeffrey and {Osborne}, Jack and {Schiavon}, Ricardo and {Stover}, Richard J. and {Tucker}, Dean and {Wallace}, Vernon and {Wei}, Mingzhi and {Wirth}, Gregory and {Wright}, Christopher A.},
        title = "{The DEIMOS spectrograph for the Keck II Telescope: integration and testing}",
    booktitle = {Instrument Design and Performance for Optical/Infrared Ground-based Telescopes},
         year = 2003,
       editor = {{Iye}, Masanori and {Moorwood}, Alan F.~M.},
       series = {Society of Photo-Optical Instrumentation Engineers (SPIE) Conference Series},
       volume = {4841},
        month = mar,
        pages = {1657-1669},
          doi = {10.1117/12.460346},
       adsurl = {https://ui.adsabs.harvard.edu/abs/2003SPIE.4841.1657F}
}

@ARTICLE{dragons,
       author = {{Labrie}, K. and {Simpson}, C. and {Cardenes}, R. and {Turner}, J. and {Soraisam}, M. and {Quint}, B. and {Oberdorf}, O. and {Placco}, V.~M. and {Berke}, D. and {Smirnova}, O. and {Conseil}, S. and {Vacca}, W.~D. and {Thomas-Osip}, J.},
        title = "{DRAGONS-A Quick Overview}",
      journal = {Research Notes of the American Astronomical Society},
         year = 2023,
        month = oct,
       volume = {7},
       number = {10},
          eid = {214},
        pages = {214},
          doi = {10.3847/2515-5172/ad0044},
archivePrefix = {arXiv},
       eprint = {2310.03048},
 primaryClass = {astro-ph.IM},
       adsurl = {https://ui.adsabs.harvard.edu/abs/2023RNAAS...7..214L}
}

@ARTICLE{drout2014,
       author = {{Drout}, M.~R. and {Chornock}, R. and {Soderberg}, A.~M. and {Sanders}, N.~E. and {McKinnon}, R. and {Rest}, A. and {Foley}, R.~J. and {Milisavljevic}, D. and {Margutti}, R. and {Berger}, E. and {Calkins}, M. and {Fong}, W. and {Gezari}, S. and {Huber}, M.~E. and {Kankare}, E. and {Kirshner}, R.~P. and {Leibler}, C. and {Lunnan}, R. and {Mattila}, S. and {Marion}, G.~H. and {Narayan}, G. and {Riess}, A.~G. and {Roth}, K.~C. and {Scolnic}, D. and {Smartt}, S.~J. and {Tonry}, J.~L. and {Burgett}, W.~S. and {Chambers}, K.~C. and {Hodapp}, K.~W. and {Jedicke}, R. and {Kaiser}, N. and {Magnier}, E.~A. and {Metcalfe}, N. and {Morgan}, J.~S. and {Price}, P.~A. and {Waters}, C.},
        title = "{Rapidly Evolving and Luminous Transients from Pan-STARRS1}",
      journal = {\apj},
         year = 2014,
        month = oct,
       volume = {794},
       number = {1},
          eid = {23},
        pages = {23},
          doi = {10.1088/0004-637X/794/1/23},
archivePrefix = {arXiv},
       eprint = {1405.3668},
 primaryClass = {astro-ph.HE},
       adsurl = {https://ui.adsabs.harvard.edu/abs/2014ApJ...794...23D}
}

@ARTICLE{duev2019,
       author = {{Duev}, Dmitry A. and {Mahabal}, Ashish and {Masci}, Frank J. and {Graham}, Matthew J. and {Rusholme}, Ben and {Walters}, Richard and {Karmarkar}, Ishani and {Frederick}, Sara and {Kasliwal}, Mansi M. and {Rebbapragada}, Umaa and {Ward}, Charlotte},
        title = "{Real-bogus classification for the Zwicky Transient Facility using deep learning}",
      journal = {\mnras},
         year = 2019,
        month = nov,
       volume = {489},
       number = {3},
        pages = {3582-3590},
          doi = {10.1093/mnras/stz2357},
archivePrefix = {arXiv},
       eprint = {1907.11259},
 primaryClass = {astro-ph.IM},
       adsurl = {https://ui.adsabs.harvard.edu/abs/2019MNRAS.489.3582D}
}

@Manual{ecume,
    title = {Ecume: Equality of 2 (or k) Continuous Univariate and Multivariate
Distributions},
    author = {Hector {Roux de Bezieux}},
    year = {2024},
    note = {R package version 0.9.2},
    url = {https://CRAN.R-project.org/package=Ecume},
    doi = {10.32614/CRAN.package.Ecume},
  }

@ARTICLE{evans1989,
       author = {{Evans}, Charles R. and {Kochanek}, Christopher S.},
        title = "{The Tidal Disruption of a Star by a Massive Black Hole}",
      journal = {\apjl},
         year = 1989,
        month = nov,
       volume = {346},
        pages = {L13},
          doi = {10.1086/185567},
       adsurl = {https://ui.adsabs.harvard.edu/abs/1989ApJ...346L..13E}
}

@ARTICLE{Perley2026,
       author = {{Perley}, Daniel A. and {Ho}, Anna Y.~Q. and {McGrath}, Zo{\"e} and {Camilo}, Michael and {Sevilla}, Cassie and {Chen}, Ping and {Schroeder}, Genevieve and {Govreen-Segal}, Taya and {Bochenek}, Aleksandra and {Qin}, Yu-Jing and {Gillanders}, James H. and {Amend}, Benjamin and {Anderson}, Joseph P. and {Andreoni}, Igor and {Aryan}, Amar and {Bellm}, Eric C. and {Bloom}, Joshua S. and {de Boer}, Thomas and {Carney}, Jonathan and {Caiazzo}, Ilaria and {Chambers}, Ken C. and {Charalampopoulos}, Panos and {Chen}, Ting-Wan and {Chen}, Tracy X. and {Coughlin}, Eric R. and {Coughlin}, Michael and {Dennefeld}, Michel and {Dimitriadis}, Georgios and {Fremling}, Christoffer and {Frostig}, Danielle and {Gal-Yam}, Avishay and {Galbany}, Llu{\'\i}s and {Gangopadhyay}, Anjashay and {Ghendrih}, Melzie and {Graham}, Matthew J. and {Gromadzki}, Mariusz and {Groom}, Steven L. and {Guti{\'e}rrez}, Claudia P. and {Hinds}, K.-Ryan and {Huber}, Mark E. and {Inserra}, Cosimo and {Kaiser}, Benjamin C. and {Kasliwal}, Mansi M. and {Koivisto}, Niilo E. and {Lin}, Chien-Cheng and {Liu}, Chang and {Lowe}, Thomas B. and {Magnier}, Eugene and {Mahabal}, Ashish A. and {Milligan}, Andrew and {Minguez}, Paloma and {Mo}, Geoffrey and {M{\"u}ller-Bravo}, Tom{\'a}s E. and {Nicholl}, Matt and {Pessi}, Priscila J. and {Pignata}, Giuliano and {Purdum}, Josiah and {Rehemtulla}, Nabeel and {Rich}, R. Michael and {Sahu}, Anwesha and {Singh}, Avinash and {Smartt}, Stephen J. and {Sollerman}, Jesper and {Srinivasaragavan}, Gokul and {Srivastav}, Shubham and {Stein}, Robert D. and {Schulze}, Steve and {Tweddle}, Jack W. and {Wainscoat}, Richard and {Wise}, Jacob L. and {Yan}, Lin and {Young}, David R.},
        title = "{AT2024wpp: An Extremely Luminous Fast Ultraviolet Transient Powered by Accretion onto a Black Hole}",
      journal = {arXiv e-prints},
         year = 2026,
        month = jan,
          eid = {arXiv:2601.03337},
        pages = {arXiv:2601.03337},
archivePrefix = {arXiv},
       eprint = {2601.03337},
 primaryClass = {astro-ph.HE},
       adsurl = {https://ui.adsabs.harvard.edu/abs/2026arXiv260103337P}
}

@ARTICLE{evla,
       author = {{Perley}, R.~A. and {Chandler}, C.~J. and {Butler}, B.~J. and {Wrobel}, J.~M.},
        title = "{The Expanded Very Large Array: A New Telescope for New Science}",
      journal = {\apjl},
         year = 2011,
        month = sep,
       volume = {739},
       number = {1},
          eid = {L1},
        pages = {L1},
          doi = {10.1088/2041-8205/739/1/L1},
archivePrefix = {arXiv},
       eprint = {1106.0532},
 primaryClass = {astro-ph.IM},
       adsurl = {https://ui.adsabs.harvard.edu/abs/2011ApJ...739L...1P}
}

@ARTICLE{ferguson2025,
       author = {{Ferguson}, Ross and {Margalit}, Ben},
        title = "{Numerical Modeling of Relativistic Effects in Synchrotron-Emitting Shocks}",
      journal = {arXiv e-prints},
         year = 2025,
        month = sep,
          eid = {arXiv:2509.16313},
        pages = {arXiv:2509.16313},
          doi = {10.48550/arXiv.2509.16313},
archivePrefix = {arXiv},
       eprint = {2509.16313},
 primaryClass = {astro-ph.HE},
       adsurl = {https://ui.adsabs.harvard.edu/abs/2025arXiv250916313F}
}

@ARTICLE{fong2022,
       author = {{Fong}, Wen-fai and {Nugent}, Anya E. and {Dong}, Yuxin and {Berger}, Edo and {Paterson}, Kerry and {Chornock}, Ryan and {Levan}, Andrew and {Blanchard}, Peter and {Alexander}, Kate D. and {Andrews}, Jennifer and {Cobb}, Bethany E. and {Cucchiara}, Antonino and {Fox}, Derek and {Fryer}, Chris L. and {Gordon}, Alexa C. and {Kilpatrick}, Charles D. and {Lunnan}, Ragnhild and {Margutti}, Raffaella and {Miller}, Adam and {Milne}, Peter and {Nicholl}, Matt and {Perley}, Daniel and {Rastinejad}, Jillian and {Escorial}, Alicia Rouco and {Schroeder}, Genevieve and {Smith}, Nathan and {Tanvir}, Nial and {Terreran}, Giacomo},
        title = "{Short GRB Host Galaxies. I. Photometric and Spectroscopic Catalogs, Host Associations, and Galactocentric Offsets}",
      journal = {\apj},
         year = 2022,
        month = nov,
       volume = {940},
       number = {1},
          eid = {56},
        pages = {56},
          doi = {10.3847/1538-4357/ac91d0},
archivePrefix = {arXiv},
       eprint = {2206.01763},
 primaryClass = {astro-ph.GA},
       adsurl = {https://ui.adsabs.harvard.edu/abs/2022ApJ...940...56F}
}

@ARTICLE{french2020,
       author = {{French}, K. Decker and {Wevers}, Thomas and {Law-Smith}, Jamie and {Graur}, Or and {Zabludoff}, Ann I.},
        title = "{The Host Galaxies of Tidal Disruption Events}",
      journal = {\ssr},
         year = 2020,
        month = mar,
       volume = {216},
       number = {3},
          eid = {32},
        pages = {32},
          doi = {10.1007/s11214-020-00657-y},
archivePrefix = {arXiv},
       eprint = {2003.02863},
 primaryClass = {astro-ph.HE},
       adsurl = {https://ui.adsabs.harvard.edu/abs/2020SSRv..216...32F}
}

@ARTICLE{Fulton2024_AT2024qfm,
       author = {{Fulton}, M. and {Chen}, T.~W. and {Smartt}, S.~J. and {Srivastav}, S. and {Gillanders}, J. and {Stevance}, H. and {Rhodes}, L. and {Schmidt}, B. and {Angus}, C. and {Smith}, K.~W. and {Young}, D.~R. and {Nicholl}, M. and {Moore}, T. and {McCollum}, M. and {Weston}, J. and {Sheng}, X. and {Aamer}, A. and {Ramsden}, P. and {Williams}, R. and {Francis}, G.},
        title = "{AT2024qfm : a fast, blue, extragalactic transient in ZTF alert stream}",
      journal = {Transient Name Server AstroNote},
         year = 2024,
        month = jul,
       volume = {206},
        pages = {1},
       adsurl = {https://ui.adsabs.harvard.edu/abs/2024TNSAN.206....1F}
}

@ARTICLE{Bietenholz2021,
       author = {{Bietenholz}, M.~F. and {Bartel}, N. and {Argo}, M. and {Dua}, R. and {Ryder}, S. and {Soderberg}, A.},
        title = "{The Radio Luminosity-risetime Function of Core-collapse Supernovae}",
      journal = {\apj},
         year = 2021,
        month = feb,
       volume = {908},
       number = {1},
          eid = {75},
        pages = {75},
          doi = {10.3847/1538-4357/abccd9},
archivePrefix = {arXiv},
       eprint = {2011.11737},
 primaryClass = {astro-ph.HE},
       adsurl = {https://ui.adsabs.harvard.edu/abs/2021ApJ...908...75B}
}

@ARTICLE{galex,
       author = {{Martin}, D. Christopher and {Fanson}, James and {Schiminovich}, David and {Morrissey}, Patrick and {Friedman}, Peter G. and {Barlow}, Tom A. and {Conrow}, Tim and {Grange}, Robert and {Jelinsky}, Patrick N. and {Milliard}, Bruno and {Siegmund}, Oswald H.~W. and {Bianchi}, Luciana and {Byun}, Yong-Ik and {Donas}, Jose and {Forster}, Karl and {Heckman}, Timothy M. and {Lee}, Young-Wook and {Madore}, Barry F. and {Malina}, Roger F. and {Neff}, Susan G. and {Rich}, R. Michael and {Small}, Todd and {Surber}, Frank and {Szalay}, Alex S. and {Welsh}, Barry and {Wyder}, Ted K.},
        title = "{The Galaxy Evolution Explorer: A Space Ultraviolet Survey Mission}",
      journal = {\apjl},
         year = 2005,
        month = jan,
       volume = {619},
       number = {1},
        pages = {L1-L6},
          doi = {10.1086/426387},
archivePrefix = {arXiv},
       eprint = {astro-ph/0411302},
 primaryClass = {astro-ph},
       adsurl = {https://ui.adsabs.harvard.edu/abs/2005ApJ...619L...1M}
}

@ARTICLE{gallazzi2005,
       author = {{Gallazzi}, Anna and {Charlot}, St{\'e}phane and {Brinchmann}, Jarle and {White}, Simon D.~M. and {Tremonti}, Christy A.},
        title = "{The ages and metallicities of galaxies in the local universe}",
      journal = {\mnras},
         year = 2005,
        month = sep,
       volume = {362},
       number = {1},
        pages = {41-58},
          doi = {10.1111/j.1365-2966.2005.09321.x},
archivePrefix = {arXiv},
       eprint = {astro-ph/0506539},
 primaryClass = {astro-ph},
       adsurl = {https://ui.adsabs.harvard.edu/abs/2005MNRAS.362...41G}
}

@ARTICLE{Gehrels2004,
       author = {{Gehrels}, N. and {Chincarini}, G. and {Giommi}, P. and {Mason}, K.~O. and {Nousek}, J.~A. and {Wells}, A.~A. and {White}, N.~E. and {Barthelmy}, S.~D. and {Burrows}, D.~N. and {Cominsky}, L.~R. and {Hurley}, K.~C. and {Marshall}, F.~E. and {M{\'e}sz{\'a}ros}, P. and {Roming}, P.~W.~A. and {Angelini}, L. and {Barbier}, L.~M. and {Belloni}, T. and {Campana}, S. and {Caraveo}, P.~A. and {Chester}, M.~M. and {Citterio}, O. and {Cline}, T.~L. and {Cropper}, M.~S. and {Cummings}, J.~R. and {Dean}, A.~J. and {Feigelson}, E.~D. and {Fenimore}, E.~E. and {Frail}, D.~A. and {Fruchter}, A.~S. and {Garmire}, G.~P. and {Gendreau}, K. and {Ghisellini}, G. and {Greiner}, J. and {Hill}, J.~E. and {Hunsberger}, S.~D. and {Krimm}, H.~A. and {Kulkarni}, S.~R. and {Kumar}, P. and {Lebrun}, F. and {Lloyd-Ronning}, N.~M. and {Markwardt}, C.~B. and {Mattson}, B.~J. and {Mushotzky}, R.~F. and {Norris}, J.~P. and {Osborne}, J. and {Paczynski}, B. and {Palmer}, D.~M. and {Park}, H. -S. and {Parsons}, A.~M. and {Paul}, J. and {Rees}, M.~J. and {Reynolds}, C.~S. and {Rhoads}, J.~E. and {Sasseen}, T.~P. and {Schaefer}, B.~E. and {Short}, A.~T. and {Smale}, A.~P. and {Smith}, I.~A. and {Stella}, L. and {Tagliaferri}, G. and {Takahashi}, T. and {Tashiro}, M. and {Townsley}, L.~K. and {Tueller}, J. and {Turner}, M.~J.~L. and {Vietri}, M. and {Voges}, W. and {Ward}, M.~J. and {Willingale}, R. and {Zerbi}, F.~M. and {Zhang}, W.~W.},
        title = "{The Swift Gamma-Ray Burst Mission}",
      journal = {\apj},
         year = 2004,
        month = aug,
       volume = {611},
       number = {2},
        pages = {1005-1020},
          doi = {10.1086/422091},
archivePrefix = {arXiv},
       eprint = {astro-ph/0405233},
 primaryClass = {astro-ph},
       adsurl = {https://ui.adsabs.harvard.edu/abs/2004ApJ...611.1005G}
}

@ARTICLE{gezari2021,
       author = {{Gezari}, Suvi},
        title = "{Tidal Disruption Events}",
      journal = {\araa},
         year = 2021,
        month = sep,
       volume = {59},
        pages = {21-58},
          doi = {10.1146/annurev-astro-111720-030029},
archivePrefix = {arXiv},
       eprint = {2104.14580},
 primaryClass = {astro-ph.HE},
       adsurl = {https://ui.adsabs.harvard.edu/abs/2021ARA&A..59...21G}
}

@INPROCEEDINGS{ghts,
       author = {{Clemens}, J. Christopher and {Crain}, J. Adam and {Anderson}, Robert},
        title = "{The Goodman spectrograph}",
    booktitle = {Ground-based Instrumentation for Astronomy},
         year = 2004,
       editor = {{Moorwood}, Alan F.~M. and {Iye}, Masanori},
       series = {Society of Photo-Optical Instrumentation Engineers (SPIE) Conference Series},
       volume = {5492},
        month = sep,
        pages = {331-340},
          doi = {10.1117/12.550069},
       adsurl = {https://ui.adsabs.harvard.edu/abs/2004SPIE.5492..331C}
}

@software{gildas,
       author = {{Gildas Team}},
        title = "{GILDAS: Grenoble Image and Line Data Analysis Software}",
 howpublished = {Astrophysics Source Code Library, record ascl:1305.010},
         year = 2013,
        month = may,
          eid = {ascl:1305.010},
       adsurl = {https://ui.adsabs.harvard.edu/abs/2013ascl.soft05010G}
}

@ARTICLE{Gillanders2024_AT2024qfm,
       author = {{Gillanders}, J.~H. and {Huber}, M. and {Chambers}, K. and {Smartt}, S.~J. and {Srivastav}, S. and {Fulton}, M. and {Chen}, T.~W. and {Schmidt}, B. and {Angus}, C.~R. and {Moore}, T. and {Nicholl}, M. and {Smith}, K.~W. and {Young}, D.~R. and {Stevance}, H.},
        title = "{Follow-up observations and redshift confirmation of AT2024qfm - a luminous FBOT}",
      journal = {Transient Name Server AstroNote},
         year = 2024,
        month = aug,
       volume = {210},
        pages = {1},
       adsurl = {https://ui.adsabs.harvard.edu/abs/2024TNSAN.210....1G}
}

@ARTICLE{gmos-n,
       author = {{Hook}, I.~M. and {J{\o}rgensen}, Inger and {Allington-Smith}, J.~R. and {Davies}, R.~L. and {Metcalfe}, N. and {Murowinski}, R.~G. and {Crampton}, D.},
        title = "{The Gemini-North Multi-Object Spectrograph: Performance in Imaging, Long-Slit, and Multi-Object Spectroscopic Modes}",
      journal = {\pasp},
         year = 2004,
        month = may,
       volume = {116},
       number = {819},
        pages = {425-440},
          doi = {10.1086/383624},
       adsurl = {https://ui.adsabs.harvard.edu/abs/2004PASP..116..425H}
}

@INPROCEEDINGS{gmos-s,
       author = {{Gimeno}, German and {Roth}, Katherine and {Chiboucas}, Kristin and {Hibon}, Pascale and {Boucher}, Luc and {White}, John and {Rippa}, Matthew and {Labrie}, Kathleen and {Turner}, James and {Hanna}, Kevin and {Lazo}, Manuel and {P{\'e}rez}, Gabriel and {Rogers}, Rolando and {Rojas}, Roberto and {Placco}, Vinicius and {Murowinski}, Richard},
        title = "{On-sky commissioning of Hamamatsu CCDs in GMOS-S}",
    booktitle = {Ground-based and Airborne Instrumentation for Astronomy VI},
         year = 2016,
       editor = {{Evans}, Christopher J. and {Simard}, Luc and {Takami}, Hideki},
       series = {Society of Photo-Optical Instrumentation Engineers (SPIE) Conference Series},
       volume = {9908},
        month = aug,
          eid = {99082S},
        pages = {99082S},
          doi = {10.1117/12.2233883},
       adsurl = {https://ui.adsabs.harvard.edu/abs/2016SPIE.9908E..2SG}
}

@ARTICLE{gottlieb2022,
       author = {{Gottlieb}, Ore and {Tchekhovskoy}, Alexander and {Margutti}, Raffaella},
        title = "{Shocked jets in CCSNe can power the zoo of fast blue optical transients}",
      journal = {\mnras},
         year = 2022,
        month = jul,
       volume = {513},
       number = {3},
        pages = {3810-3817},
          doi = {10.1093/mnras/stac910},
archivePrefix = {arXiv},
       eprint = {2201.04636},
 primaryClass = {astro-ph.HE},
       adsurl = {https://ui.adsabs.harvard.edu/abs/2022MNRAS.513.3810G}
}

@ARTICLE{gphoton,
       author = {{Million}, Chase and {Fleming}, Scott W. and {Shiao}, Bernie and {Seibert}, Mark and {Loyd}, Parke and {Tucker}, Michael and {Smith}, Myron and {Thompson}, Randy and {White}, Richard L.},
        title = "{gPhoton: The GALEX Photon Data Archive}",
      journal = {\apj},
         year = 2016,
        month = dec,
       volume = {833},
       number = {2},
          eid = {292},
        pages = {292},
          doi = {10.3847/1538-4357/833/2/292},
archivePrefix = {arXiv},
       eprint = {1609.09492},
 primaryClass = {astro-ph.IM},
       adsurl = {https://ui.adsabs.harvard.edu/abs/2016ApJ...833..292M}
}

@ARTICLE{gmrt,
       author = {{Swarup}, G. and {Ananthakrishnan}, S. and {Kapahi}, V.~K. and {Rao}, A.~P. and {Subrahmanya}, C.~R. and {Kulkarni}, V.~K.},
        title = "{The Giant Metre-Wave Radio Telescope}",
      journal = {Current Science},
         year = 1991,
        month = jan,
       volume = {60},
        pages = {95},
       adsurl = {https://ui.adsabs.harvard.edu/abs/1991CSci...60...95S}
}

@ARTICLE{Gutierrez2024,
       author = {{Guti{\'e}rrez}, Claudia P. and {Mattila}, Seppo and {Lundqvist}, Peter and {Dessart}, Luc and {Gonz{\'a}lez-Gait{\'a}n}, Santiago and {Jonker}, Peter G. and {Dong}, Subo and {Coppejans}, Deanne and {Chen}, Ping and {Charalampopoulos}, Panos and {Elias-Rosa}, Nancy and {Reynolds}, Thomas M. and {Kochanek}, Christopher and {Fraser}, Morgan and {Pastorello}, Andrea and {Gromadzki}, Mariusz and {Neustadt}, Jack and {Benetti}, Stefano and {Kankare}, Erkki and {Kangas}, Tuomas and {Kotak}, Rubina and {Stritzinger}, Maximilian D. and {Wevers}, Thomas and {Zhang}, Bing and {Bersier}, David and {Bose}, Subhash and {Buckley}, David A.~H. and {Dastidar}, Raya and {Gangopadhyay}, Anjasha and {Hamanowicz}, Aleksandra and {Kollmeier}, Juna A. and {Mao}, Jirong and {Misra}, Kuntal and {Potter}, Stephen. B. and {Prieto}, Jose L. and {Romero-Colmenero}, Encarni and {Singh}, Mridweeka and {Somero}, Auni and {Terreran}, Giacomo and {Vaisanen}, Petri and {Wyrzykowski}, {\L}ukasz},
        title = "{CSS 161010: A Luminous Fast Blue Optical Transient with Broad Blueshifted Hydrogen Lines}",
      journal = {\apj},
         year = 2024,
        month = dec,
       volume = {977},
       number = {2},
          eid = {162},
        pages = {162},
          doi = {10.3847/1538-4357/ad89a5},
archivePrefix = {arXiv},
       eprint = {2408.04698},
 primaryClass = {astro-ph.HE},
       adsurl = {https://ui.adsabs.harvard.edu/abs/2024ApJ...977..162G}
}

@ARTICLE{ho2019,
       author = {{Ho}, Anna Y.~Q. and {Phinney}, E. Sterl and {Ravi}, Vikram and {Kulkarni}, S.~R. and {Petitpas}, Glen and {Emonts}, Bjorn and {Bhalerao}, V. and {Blundell}, Ray and {Cenko}, S. Bradley and {Dobie}, Dougal and {Howie}, Ryan and {Kamraj}, Nikita and {Kasliwal}, Mansi M. and {Murphy}, Tara and {Perley}, Daniel A. and {Sridharan}, T.~K. and {Yoon}, Ilsang},
        title = "{AT2018cow: A Luminous Millimeter Transient}",
      journal = {\apj},
         year = 2019,
        month = jan,
       volume = {871},
       number = {1},
          eid = {73},
        pages = {73},
          doi = {10.3847/1538-4357/aaf473},
archivePrefix = {arXiv},
       eprint = {1810.10880},
 primaryClass = {astro-ph.HE},
       adsurl = {https://ui.adsabs.harvard.edu/abs/2019ApJ...871...73H}
}

@ARTICLE{hinkle2021,
       author = {{Hinkle}, Jason T. and {Holoien}, T.~W.-S. and {Auchettl}, K. and {Shappee}, B.~J. and {Neustadt}, J.~M.~M. and {Payne}, A.~V. and {Brown}, J.~S. and {Kochanek}, C.~S. and {Stanek}, K.~Z. and {Graham}, M.~J. and {Tucker}, M.~A. and {Do}, A. and {Anderson}, J.~P. and {Bose}, S. and {Chen}, P. and {Coulter}, D.~A. and {Dimitriadis}, G. and {Dong}, Subo and {Foley}, R.~J. and {Huber}, M.~E. and {Hung}, T. and {Kilpatrick}, C.~D. and {Pignata}, G. and {Piro}, A.~L. and {Rojas-Bravo}, C. and {Siebert}, M.~R. and {Stalder}, B. and {Thompson}, Todd A. and {Tonry}, J.~L. and {Vallely}, P.~J. and {Wisniewski}, J.~P.},
        title = "{Discovery and follow-up of ASASSN-19dj: an X-ray and UV luminous TDE in an extreme post-starburst galaxy}",
      journal = {\mnras},
         year = 2021,
        month = jan,
       volume = {500},
       number = {2},
        pages = {1673-1696},
          doi = {10.1093/mnras/staa3170},
archivePrefix = {arXiv},
       eprint = {2006.06690},
 primaryClass = {astro-ph.HE},
       adsurl = {https://ui.adsabs.harvard.edu/abs/2021MNRAS.500.1673H}
}

@ARTICLE{ho2020blt,
       author = {{Ho}, Anna Y.~Q. and {Perley}, Daniel A. and {Beniamini}, Paz and {Cenko}, S. Bradley and {Kulkarni}, S.~R. and {Andreoni}, Igor and {Singer}, Leo P. and {De}, Kishalay and {Kasliwal}, Mansi M. and {Fremling}, Christoffer and {Bellm}, Eric C. and {Dekany}, Richard and {Delacroix}, Alexandre and {Duev}, Dmitry A. and {Goldstein}, Daniel A. and {Golkhou}, V. Zach and {Goobar}, Ariel and {Graham}, Matthew J. and {Hale}, David and {Kupfer}, Thomas and {Laher}, Russ R. and {Masci}, Frank J. and {Miller}, Adam A. and {Neill}, James D. and {Riddle}, Reed and {Rusholme}, Ben and {Shupe}, David L. and {Smith}, Roger and {Sollerman}, Jesper and {van Roestel}, Jan},
        title = "{ZTF20aajnksq (AT 2020blt): A Fast Optical Transient at z {\ensuremath{\approx}} 2.9 with No Detected Gamma-Ray Burst Counterpart}",
      journal = {\apj},
         year = 2020,
        month = dec,
       volume = {905},
       number = {2},
          eid = {98},
        pages = {98},
          doi = {10.3847/1538-4357/abc34d},
archivePrefix = {arXiv},
       eprint = {2006.10761},
 primaryClass = {astro-ph.HE},
       adsurl = {https://ui.adsabs.harvard.edu/abs/2020ApJ...905...98H}
}

@ARTICLE{ho2022,
       author = {{Ho}, Anna Y.~Q. and {Perley}, Daniel A. and {Yao}, Yuhan and {Svinkin}, Dmitry and {de Ugarte Postigo}, A. and {Perley}, R.~A. and {Kann}, D. Alexander and {Burns}, Eric and {Andreoni}, Igor and {Bellm}, Eric C. and {Bissaldi}, Elisabetta and {Bloom}, Joshua S. and {Brink}, Thomas G. and {Dekany}, Richard and {Drake}, Andrew J. and {Ag{\"u}{\'\i} Fern{\'a}ndez}, Jos{\'e} Feliciano and {Filippenko}, Alexei V. and {Frederiks}, Dmitry and {Graham}, Matthew J. and {Hristov}, Boyan A. and {Kasliwal}, Mansi M. and {Kulkarni}, S.~R. and {Kumar}, Harsh and {Laher}, Russ R. and {Lysenko}, Alexandra L. and {Mailyan}, Bagrat and {Malacaria}, Christian and {Miller}, A.~A. and {Poolakkil}, S. and {Riddle}, Reed and {Ridnaia}, Anna and {Rusholme}, Ben and {Savchenko}, Volodymyr and {Sollerman}, Jesper and {Th{\"o}ne}, Christina and {Tsvetkova}, Anastasia and {Ulanov}, Mikhail and {von Kienlin}, Andreas},
        title = "{Cosmological Fast Optical Transients with the Zwicky Transient Facility: A Search for Dirty Fireballs}",
      journal = {\apj},
         year = 2022,
        month = oct,
       volume = {938},
       number = {1},
          eid = {85},
        pages = {85},
          doi = {10.3847/1538-4357/ac8bd0},
archivePrefix = {arXiv},
       eprint = {2201.12366},
 primaryClass = {astro-ph.HE},
       adsurl = {https://ui.adsabs.harvard.edu/abs/2022ApJ...938...85H}
}

@ARTICLE{ho2022xnd,
       author = {{Ho}, Anna Y.~Q. and {Margalit}, Ben and {Bremer}, Michael and {Perley}, Daniel A. and {Yao}, Yuhan and {Dobie}, Dougal and {Kaplan}, David L. and {O'Brien}, Andrew and {Petitpas}, Glen and {Zic}, Andrew},
        title = "{Luminous Millimeter, Radio, and X-Ray Emission from ZTF 20acigmel (AT 2020xnd)}",
      journal = {\apj},
         year = 2022,
        month = jun,
       volume = {932},
       number = {2},
          eid = {116},
        pages = {116},
          doi = {10.3847/1538-4357/ac4e97},
archivePrefix = {arXiv},
       eprint = {2110.05490},
 primaryClass = {astro-ph.HE},
       adsurl = {https://ui.adsabs.harvard.edu/abs/2022ApJ...932..116H}
}

@ARTICLE{ho2023,
       author = {{Ho}, Anna Y.~Q. and {Perley}, Daniel A. and {Gal-Yam}, Avishay and {Lunnan}, Ragnhild and {Sollerman}, Jesper and {Schulze}, Steve and {Das}, Kaustav K. and {Dobie}, Dougal and {Yao}, Yuhan and {Fremling}, Christoffer and {Adams}, Scott and {Anand}, Shreya and {Andreoni}, Igor and {Bellm}, Eric C. and {Bruch}, Rachel J. and {Burdge}, Kevin B. and {Castro-Tirado}, Alberto J. and {Dahiwale}, Aishwarya and {De}, Kishalay and {Dekany}, Richard and {Drake}, Andrew J. and {Duev}, Dmitry A. and {Graham}, Matthew J. and {Helou}, George and {Kaplan}, David L. and {Karambelkar}, Viraj and {Kasliwal}, Mansi M. and {Kool}, Erik C. and {Kulkarni}, S.~R. and {Mahabal}, Ashish A. and {Medford}, Michael S. and {Miller}, A.~A. and {Nordin}, Jakob and {Ofek}, Eran and {Petitpas}, Glen and {Riddle}, Reed and {Sharma}, Yashvi and {Smith}, Roger and {Stewart}, Adam J. and {Taggart}, Kirsty and {Tartaglia}, Leonardo and {Tzanidakis}, Anastasios and {Winters}, Jan Martin},
        title = "{A Search for Extragalactic Fast Blue Optical Transients in ZTF and the Rate of AT2018cow-like Transients}",
      journal = {\apj},
         year = 2023,
        month = jun,
       volume = {949},
       number = {2},
          eid = {120},
        pages = {120},
          doi = {10.3847/1538-4357/acc533},
archivePrefix = {arXiv},
       eprint = {2105.08811},
 primaryClass = {astro-ph.HE},
       adsurl = {https://ui.adsabs.harvard.edu/abs/2023ApJ...949..120H}
}

@ARTICLE{ho2020koala,
       author = {{Ho}, Anna Y.~Q. and {Perley}, Daniel A. and {Kulkarni}, S.~R. and {Dong}, Dillon Z.~J. and {De}, Kishalay and {Chandra}, Poonam and {Andreoni}, Igor and {Bellm}, Eric C. and {Burdge}, Kevin B. and {Coughlin}, Michael and {Dekany}, Richard and {Feeney}, Michael and {Frederiks}, Dmitry D. and {Fremling}, Christoffer and {Golkhou}, V. Zach and {Graham}, Matthew J. and {Hale}, David and {Helou}, George and {Horesh}, Assaf and {Kasliwal}, Mansi M. and {Laher}, Russ R. and {Masci}, Frank J. and {Miller}, A.~A. and {Porter}, Michael and {Ridnaia}, Anna and {Rusholme}, Ben and {Shupe}, David L. and {Soumagnac}, Maayane T. and {Svinkin}, Dmitry S.},
        title = "{The Koala: A Fast Blue Optical Transient with Luminous Radio Emission from a Starburst Dwarf Galaxy at z = 0.27}",
      journal = {\apj},
         year = 2020,
        month = may,
       volume = {895},
       number = {1},
          eid = {49},
        pages = {49},
          doi = {10.3847/1538-4357/ab8bcf},
archivePrefix = {arXiv},
       eprint = {2003.01222},
 primaryClass = {astro-ph.HE},
       adsurl = {https://ui.adsabs.harvard.edu/abs/2020ApJ...895...49H}
}

@ARTICLE{ho2023tsd,
       author = {{Ho}, Anna Y.~Q. and {Perley}, Daniel A. and {Chen}, Ping and {Schulze}, Steve and {Dhillon}, Vik and {Kumar}, Harsh and {Suresh}, Aswin and {Swain}, Vishwajeet and {Bremer}, Michael and {Smartt}, Stephen J. and {Anderson}, Joseph P. and {Anupama}, G.~C. and {Awiphan}, Supachai and {Barway}, Sudhanshu and {Bellm}, Eric C. and {Ben-Ami}, Sagi and {Bhalerao}, Varun and {de Boer}, Thomas and {Brink}, Thomas G. and {Burruss}, Rick and {Chandra}, Poonam and {Chen}, Ting-Wan and {Chen}, Wen-Ping and {Cooke}, Jeff and {Coughlin}, Michael W. and {Das}, Kaustav K. and {Drake}, Andrew J. and {Filippenko}, Alexei V. and {Freeburn}, James and {Fremling}, Christoffer and {Fulton}, Michael D. and {Gal-Yam}, Avishay and {Galbany}, Llu{\'\i}s and {Gao}, Hua and {Graham}, Matthew J. and {Gromadzki}, Mariusz and {Guti{\'e}rrez}, Claudia P. and {Hinds}, K. -Ryan and {Inserra}, Cosimo and {A J}, Nayana and {Karambelkar}, Viraj and {Kasliwal}, Mansi M. and {Kulkarni}, Shri and {M{\"u}ller-Bravo}, Tom{\'a}s E. and {Magnier}, Eugene A. and {Mahabal}, Ashish A. and {Moore}, Thomas and {Ngeow}, Chow-Choong and {Nicholl}, Matt and {Ofek}, Eran O. and {Omand}, Conor M.~B. and {Onori}, Francesca and {Pan}, Yen-Chen and {Pessi}, Priscila J. and {Petitpas}, Glen and {Polishook}, David and {Poshyachinda}, Saran and {Pursiainen}, Miika and {Riddle}, Reed and {Rodriguez}, Antonio C. and {Rusholme}, Ben and {Segre}, Enrico and {Sharma}, Yashvi and {Smith}, Ken W. and {Sollerman}, Jesper and {Srivastav}, Shubham and {Strotjohann}, Nora Linn and {Suhr}, Mark and {Svinkin}, Dmitry and {Wang}, Yanan and {Wiseman}, Philip and {Wold}, Avery and {Yang}, Sheng and {Yang}, Yi and {Yao}, Yuhan and {Young}, David R. and {Zheng}, WeiKang},
        title = "{Minutes-duration optical flares with supernova luminosities}",
      journal = {\nat},
         year = 2023,
        month = nov,
       volume = {623},
       number = {7989},
        pages = {927-931},
          doi = {10.1038/s41586-023-06673-6},
archivePrefix = {arXiv},
       eprint = {2311.10195},
 primaryClass = {astro-ph.HE},
       adsurl = {https://ui.adsabs.harvard.edu/abs/2023Natur.623..927H}
}

@ARTICLE{horesh2021,
       author = {{Horesh}, A. and {Cenko}, S.~B. and {Arcavi}, I.},
        title = "{Delayed radio flares from a tidal disruption event}",
      journal = {Nature Astronomy},
         year = 2021,
        month = may,
       volume = {5},
        pages = {491-497},
          doi = {10.1038/s41550-021-01300-8},
archivePrefix = {arXiv},
       eprint = {2102.11290},
 primaryClass = {astro-ph.HE},
       adsurl = {https://ui.adsabs.harvard.edu/abs/2021NatAs...5..491H}
}

@MISC{imacs,
       author = {{Dressler}, Alan},
        title = "{Enhanced Performance for the IMACS F/4 Camera Upgrade}",
 howpublished = {NSF Award Number 1038894. Directorate for Mathematical and Physical Sciences, Division Of Astronomical Sciences. 2010.},
         year = 2010,
        month = aug,
        pages = {38894},
       adsurl = {https://ui.adsabs.harvard.edu/abs/2010nsf....1038894D}
}

@ARTICLE{inkenhaag2025,
       author = {{Inkenhaag}, Anne and {Levan}, Andrew J. and {Mummery}, Andrew and {Jonker}, Peter G.},
        title = "{AT 2018cow at {\ensuremath{\sim}}5 years: additional evidence for a tidal disruption origin}",
      journal = {\mnras},
         year = 2025,
        month = nov,
       volume = {544},
       number = {1},
        pages = {L108-L112},
          doi = {10.1093/mnrasl/slaf107},
archivePrefix = {arXiv},
       eprint = {2510.08505},
 primaryClass = {astro-ph.HE},
       adsurl = {https://ui.adsabs.harvard.edu/abs/2025MNRAS.544L.108I}
}

@ARTICLE{inserra2019,
       author = {{Inserra}, C.},
        title = "{Observational properties of extreme supernovae}",
      journal = {Nature Astronomy},
         year = 2019,
        month = aug,
       volume = {3},
        pages = {697-705},
          doi = {10.1038/s41550-019-0854-4},
archivePrefix = {arXiv},
       eprint = {1908.02314},
 primaryClass = {astro-ph.HE},
       adsurl = {https://ui.adsabs.harvard.edu/abs/2019NatAs...3..697I}
}

@INPROCEEDINGS{iraf,
       author = {{Tody}, Doug},
        title = "{The IRAF Data Reduction and Analysis System}",
    booktitle = {Instrumentation in astronomy VI},
         year = 1986,
       editor = {{Crawford}, David L.},
       series = {Society of Photo-Optical Instrumentation Engineers (SPIE) Conference Series},
       volume = {627},
        month = jan,
        pages = {733},
          doi = {10.1117/12.968154},
       adsurl = {https://ui.adsabs.harvard.edu/abs/1986SPIE..627..733T}
}

@ARTICLE{kansky2019,
       author = {{Kansky}, Jan and {Chilingarian}, Igor and {Fabricant}, Daniel and {Matthews}, Anne and {Moran}, Sean and {Paegert}, Martin and {Duane Gibson}, J. and {Porter}, Dallan and {Roll}, John},
        title = "{Binospec Software System}",
      journal = {\pasp},
         year = 2019,
        month = jul,
       volume = {131},
       number = {1001},
        pages = {075005},
          doi = {10.1088/1538-3873/ab1ceb},
archivePrefix = {arXiv},
       eprint = {1905.03321},
 primaryClass = {astro-ph.IM},
       adsurl = {https://ui.adsabs.harvard.edu/abs/2019PASP..131g5005K}
}

@ARTICLE{kcwi,
       author = {{Morrissey}, Patrick and {Matuszewski}, Matuesz and {Martin}, D. Christopher and {Neill}, James D. and {Epps}, Harland and {Fucik}, Jason and {Weber}, Bob and {Darvish}, Behnam and {Adkins}, Sean and {Allen}, Steve and {Bartos}, Randy and {Belicki}, Justin and {Cabak}, Jerry and {Callahan}, Shawn and {Cowley}, Dave and {Crabill}, Marty and {Deich}, Willian and {Delecroix}, Alex and {Doppman}, Greg and {Hilyard}, David and {James}, Ean and {Kaye}, Steve and {Kokorowski}, Michael and {Kwok}, Shui and {Lanclos}, Kyle and {Milner}, Steve and {Moore}, Anna and {O'Sullivan}, Donal and {Parihar}, Prachi and {Park}, Sam and {Phillips}, Andrew and {Rizzi}, Luca and {Rockosi}, Constance and {Rodriguez}, Hector and {Salaun}, Yves and {Seaman}, Kirk and {Sheikh}, David and {Weiss}, Jason and {Zarzaca}, Ray},
        title = "{The Keck Cosmic Web Imager Integral Field Spectrograph}",
      journal = {\apj},
         year = 2018,
        month = sep,
       volume = {864},
       number = {1},
          eid = {93},
        pages = {93},
          doi = {10.3847/1538-4357/aad597},
archivePrefix = {arXiv},
       eprint = {1807.10356},
 primaryClass = {astro-ph.IM},
       adsurl = {https://ui.adsabs.harvard.edu/abs/2018ApJ...864...93M}
}

@ARTICLE{kelly2012,
       author = {{Kelly}, Patrick L. and {Kirshner}, Robert P.},
        title = "{Core-collapse Supernovae and Host Galaxy Stellar Populations}",
      journal = {\apj},
         year = 2012,
        month = nov,
       volume = {759},
       number = {2},
          eid = {107},
        pages = {107},
          doi = {10.1088/0004-637X/759/2/107},
archivePrefix = {arXiv},
       eprint = {1110.1377},
 primaryClass = {astro-ph.CO},
       adsurl = {https://ui.adsabs.harvard.edu/abs/2012ApJ...759..107K}
}

@ARTICLE{Klencki2025,
       author = {{Klencki}, Jakub and {Metzger}, Brian D.},
        title = "{Luminous Fast Blue Optical Transients as ``Failed'' Gravitational Wave Sources: Helium Core$-$Black Hole Mergers Following Delayed Dynamical Instability}",
      journal = {arXiv e-prints},
         year = 2025,
        month = oct,
          eid = {arXiv:2510.09745},
        pages = {arXiv:2510.09745},
          doi = {10.48550/arXiv.2510.09745},
archivePrefix = {arXiv},
       eprint = {2510.09745},
 primaryClass = {astro-ph.HE},
       adsurl = {https://ui.adsabs.harvard.edu/abs/2025arXiv251009745K}
}

@ARTICLE{Kremer2023,
       author = {{Kremer}, Kyle and {Mockler}, Brenna and {Piro}, Anthony L. and {Lombardi}, James C.},
        title = "{Wind-reprocessed transients from stellar-mass black hole Tidal Disruption Events}",
      journal = {\mnras},
         year = 2023,
        month = oct,
       volume = {524},
       number = {4},
        pages = {6358-6373},
          doi = {10.1093/mnras/stad2239},
archivePrefix = {arXiv},
       eprint = {2305.08905},
 primaryClass = {astro-ph.HE},
       adsurl = {https://ui.adsabs.harvard.edu/abs/2023MNRAS.524.6358K}
}

@ARTICLE{Kremer2021,
       author = {{Kremer}, Kyle and {Lu}, Wenbin and {Piro}, Anthony L. and {Chatterjee}, Sourav and {Rasio}, Frederic A. and {Ye}, Claire S.},
        title = "{Fast Optical Transients from Stellar-mass Black Hole Tidal Disruption Events in Young Star Clusters}",
      journal = {\apj},
         year = 2021,
        month = apr,
       volume = {911},
       number = {2},
          eid = {104},
        pages = {104},
          doi = {10.3847/1538-4357/abeb14},
archivePrefix = {arXiv},
       eprint = {2012.02796},
 primaryClass = {astro-ph.HE},
       adsurl = {https://ui.adsabs.harvard.edu/abs/2021ApJ...911..104K}
}

@ARTICLE{kuin2019,
       author = {{Kuin}, N. Paul M. and {Wu}, Kinwah and {Oates}, Samantha and {Lien}, Amy and {Emery}, Sam and {Kennea}, Jamie A. and {de Pasquale}, Massimiliano and {Han}, Qin and {Brown}, Peter J. and {Tohuvavohu}, Aaron and {Breeveld}, Alice and {Burrows}, David N. and {Cenko}, S. Bradley and {Campana}, Sergio and {Levan}, Andrew and {Markwardt}, Craig and {Osborne}, Julian P. and {Page}, Mat J. and {Page}, Kim L. and {Sbarufatti}, Boris and {Siegel}, Michael and {Troja}, Eleonora},
        title = "{Swift spectra of AT2018cow: a white dwarf tidal disruption event?}",
      journal = {\mnras},
         year = 2019,
        month = aug,
       volume = {487},
       number = {2},
        pages = {2505-2521},
          doi = {10.1093/mnras/stz053},
archivePrefix = {arXiv},
       eprint = {1808.08492},
 primaryClass = {astro-ph.HE},
       adsurl = {https://ui.adsabs.harvard.edu/abs/2019MNRAS.487.2505K}
}

@ARTICLE{Kulkarni2021,
       author = {{Kulkarni}, S.~R. and {Harrison}, Fiona A. and {Grefenstette}, Brian W. and {Earnshaw}, Hannah P. and {Andreoni}, Igor and {Berg}, Danielle A. and {Bloom}, Joshua S. and {Cenko}, S. Bradley and {Chornock}, Ryan and {Christiansen}, Jessie L. and {Coughlin}, Michael W. and {Wuollet Criswell}, Alexander and {Darvish}, Behnam and {Das}, Kaustav K. and {De}, Kishalay and {Dessart}, Luc and {Dixon}, Don and {Dorsman}, Bas and {El-Badry}, Kareem and {Evans}, Christopher and {Ford}, K.~E. Saavik and {Fremling}, Christoffer and {Gansicke}, Boris T. and {Gezari}, Suvi and {Goetberg}, Y. and {Green}, Gregory M. and {Graham}, Matthew J. and {Heida}, Marianne and {Ho}, Anna Y.~Q. and {Jaodand}, Amruta D. and {Johns-Krull}, Christopher M. and {Kasliwal}, Mansi M. and {Lazzarini}, Margaret and {Lu}, Wenbin and {Margutti}, Raffaella and {Martin}, D. Christopher and {Masters}, Daniel Charles and {McKernan}, Barry and {Naze}, Yael and {Nissanke}, Samaya M. and {Parazin}, B. and {Perley}, Daniel A. and {Phinney}, E. Sterl and {Piro}, Anthony L. and {Raaijmakers}, G. and {Rauw}, Gregor and {Rodriguez}, Antonio C. and {Sana}, Hugues and {Senchyna}, Peter and {Singer}, Leo P. and {Spake}, Jessica J. and {Stassun}, Keivan G. and {Stern}, Daniel and {Teplitz}, Harry I. and {Weisz}, Daniel R. and {Yao}, Yuhan},
        title = "{Science with the Ultraviolet Explorer (UVEX)}",
      journal = {arXiv e-prints},
         year = 2021,
        month = nov,
          eid = {arXiv:2111.15608},
        pages = {arXiv:2111.15608},
          doi = {10.48550/arXiv.2111.15608},
archivePrefix = {arXiv},
       eprint = {2111.15608},
 primaryClass = {astro-ph.GA},
       adsurl = {https://ui.adsabs.harvard.edu/abs/2021arXiv211115608K}
}

@ARTICLE{laskar2016,
       author = {{Laskar}, T. and {Alexander}, K.~D. and {Berger}, E.},
        title = "{GRB 161219B: Rapid ALMA Observations \& Detection.}",
      journal = {GRB Coordinates Network},
         year = 2016,
        month = jan,
       volume = {20328},
        pages = {1},
       adsurl = {https://ui.adsabs.harvard.edu/abs/2016GCN.20328....1L}
}

@ARTICLE{laskar2018,
       author = {{Laskar}, T. and {Berger}, E. and {Chornock}, R. and {Fong}, W. and {Margutti}, R. and {Mundell}, C.~G. and {Schady}, P.},
        title = "{GRB 181201A: VLA detection.}",
      journal = {GRB Coordinates Network},
         year = 2018,
        month = jan,
       volume = {23519},
        pages = {1},
       adsurl = {https://ui.adsabs.harvard.edu/abs/2018GCN.23519....1L}
}

@ARTICLE{lawsmith2017,
       author = {{Law-Smith}, Jamie and {Ramirez-Ruiz}, Enrico and {Ellison}, Sara L. and {Foley}, Ryan J.},
        title = "{Tidal Disruption Event Host Galaxies in the Context of the Local Galaxy Population}",
      journal = {\apj},
         year = 2017,
        month = nov,
       volume = {850},
       number = {1},
          eid = {22},
        pages = {22},
          doi = {10.3847/1538-4357/aa94c7},
archivePrefix = {arXiv},
       eprint = {1707.01559},
 primaryClass = {astro-ph.HE},
       adsurl = {https://ui.adsabs.harvard.edu/abs/2017ApJ...850...22L}
}

@INPROCEEDINGS{ldt,
       author = {{Bida}, Thomas A. and {Dunham}, Edward W. and {Massey}, Philip and {Roe}, Henry G.},
        title = "{First-generation instrumentation for the Discovery Channel Telescope}",
    booktitle = {Ground-based and Airborne Instrumentation for Astronomy V},
         year = 2014,
       editor = {{Ramsay}, Suzanne K. and {McLean}, Ian S. and {Takami}, Hideki},
       series = {Society of Photo-Optical Instrumentation Engineers (SPIE) Conference Series},
       volume = {9147},
        month = jul,
          eid = {91472N},
        pages = {91472N},
          doi = {10.1117/12.2056872},
       adsurl = {https://ui.adsabs.harvard.edu/abs/2014SPIE.9147E..2NB}
}

@ARTICLE{lebaron2025,
       author = {{LeBaron}, Natalie and {Margutti}, Raffaella and {Chornock}, Ryan and {Nayana}, A.~J. and {Aspegren}, Olivia and {Lu}, Wenbin and {Metzger}, Brian and {Kasen}, Daniel and {Brink}, Thomas and {Campana}, Sergio and {D'Avanzo}, Paolo and {Faber}, Jakob and {Ferro}, Matteo and {Filippenko}, Alex and {Foley}, Ryan and {Guo}, Xinze and {Hammerstein}, Erica and {Jha}, Saurabh and {Kilpatrick}, Charles and {Migliori}, Giulia and {Milisavljevic}, Dan and {Patra}, Kishore and {Sears}, Huei and {Swift}, Jonathan and {Tinyanont}, Samaporn and {Ravi}, Vikram and {Yao}, Yuhan and {Alexander}, Kate and {Arunachalam}, Prasiddha and {Berger}, Edo and {Bright}, Joe and {Cynamon}, Chuck and {Davis}, Kyle and {Garretson}, Braden and {Guhathakurta}, Puragra and {Jacobson-Galan}, Wynn and {Jones}, David and {Kaur}, Ravjit and {Kimura}, Stefan and {Laskar}, Tanmoy and {Nu{\~n}ez}, Morgan and {Schwab}, Michaela and {Soraisam}, Monika and {Suzuki}, Nao and {Taggart}, Kirsty and {Wiston}, Eli and {Yang}, Yi and {Zheng}, WeiKang},
        title = "{The Most Luminous Known Fast Blue Optical Transient AT 2024wpp: Unprecedented Evolution and Properties in the Ultraviolet to the Near-Infrared}",
      journal = {arXiv e-prints},
         year = 2025,
        month = aug,
          eid = {arXiv:2509.00951},
        pages = {arXiv:2509.00951},
          doi = {10.48550/arXiv.2509.00951},
archivePrefix = {arXiv},
       eprint = {2509.00951},
 primaryClass = {astro-ph.HE},
       adsurl = {https://ui.adsabs.harvard.edu/abs/2025arXiv250900951L}
}

@ARTICLE{legacy,
       author = {{Dey}, Arjun and {Schlegel}, David J. and {Lang}, Dustin and {Blum}, Robert and {Burleigh}, Kaylan and {Fan}, Xiaohui and {Findlay}, Joseph R. and {Finkbeiner}, Doug and {Herrera}, David and {Juneau}, St{\'e}phanie and {Landriau}, Martin and {Levi}, Michael and {McGreer}, Ian and {Meisner}, Aaron},
        title = "{Overview of the DESI Legacy Imaging Surveys}",
      journal = {\aj},
         year = 2019,
        month = may,
       volume = {157},
       number = {5},
          eid = {168},
        pages = {168},
          doi = {10.3847/1538-3881/ab089d},
archivePrefix = {arXiv},
       eprint = {1804.08657},
 primaryClass = {astro-ph.IM},
       adsurl = {https://ui.adsabs.harvard.edu/abs/2019AJ....157..168D}
}

@ARTICLE{lpipe,
       author = {{Perley}, Daniel A.},
        title = "{Fully Automated Reduction of Longslit Spectroscopy with the Low Resolution Imaging Spectrometer at the Keck Observatory}",
      journal = {\pasp},
         year = 2019,
        month = aug,
       volume = {131},
       number = {1002},
        pages = {084503},
          doi = {10.1088/1538-3873/ab215d},
archivePrefix = {arXiv},
       eprint = {1903.07629},
 primaryClass = {astro-ph.IM},
       adsurl = {https://ui.adsabs.harvard.edu/abs/2019PASP..131h4503P}
}

@ARTICLE{lris,
       author = {{Oke}, J.~B. and {Cohen}, J.~G. and {Carr}, M. and {Cromer}, J. and {Dingizian}, A. and {Harris}, F.~H. and {Labrecque}, S. and {Lucinio}, R. and {Schaal}, W. and {Epps}, H. and {Miller}, J.},
        title = "{The Keck Low-Resolution Imaging Spectrometer}",
      journal = {\pasp},
         year = 1995,
        month = apr,
       volume = {107},
        pages = {375},
          doi = {10.1086/133562},
       adsurl = {https://ui.adsabs.harvard.edu/abs/1995PASP..107..375O}
}

@ARTICLE{lsst,
       author = {{Ivezi{\'c}}, {\v{Z}}eljko and {Kahn}, Steven M. and {Tyson}, J. Anthony and {Abel}, Bob and {Acosta}, Emily and {Allsman}, Robyn and {Alonso}, David and {AlSayyad}, Yusra and {Anderson}, Scott F. and {Andrew}, John and {Angel}, James Roger P. and {Angeli}, George Z. and {Ansari}, Reza and {Antilogus}, Pierre and {Araujo}, Constanza and {Armstrong}, Robert and {Arndt}, Kirk T. and {Astier}, Pierre and {Aubourg}, {\'E}ric and {Auza}, Nicole and {Axelrod}, Tim S. and {Bard}, Deborah J. and {Barr}, Jeff D. and {Barrau}, Aurelian and {Bartlett}, James G. and {Bauer}, Amanda E. and {Bauman}, Brian J. and {Baumont}, Sylvain and {Bechtol}, Ellen and {Bechtol}, Keith and {Becker}, Andrew C. and {Becla}, Jacek and {Beldica}, Cristina and {Bellavia}, Steve and {Bianco}, Federica B. and {Biswas}, Rahul and {Blanc}, Guillaume and {Blazek}, Jonathan and {Blandford}, Roger D. and {Bloom}, Josh S. and {Bogart}, Joanne and {Bond}, Tim W. and {Booth}, Michael T. and {Borgland}, Anders W. and {Borne}, Kirk and {Bosch}, James F. and {Boutigny}, Dominique and {Brackett}, Craig A. and {Bradshaw}, Andrew and {Brandt}, William Nielsen and {Brown}, Michael E. and {Bullock}, James S. and {Burchat}, Patricia and {Burke}, David L. and {Cagnoli}, Gianpietro and {Calabrese}, Daniel and {Callahan}, Shawn and {Callen}, Alice L. and {Carlin}, Jeffrey L. and {Carlson}, Erin L. and {Chandrasekharan}, Srinivasan and {Charles-Emerson}, Glenaver and {Chesley}, Steve and {Cheu}, Elliott C. and {Chiang}, Hsin-Fang and {Chiang}, James and {Chirino}, Carol and {Chow}, Derek and {Ciardi}, David R. and {Claver}, Charles F. and {Cohen-Tanugi}, Johann and {Cockrum}, Joseph J. and {Coles}, Rebecca and {Connolly}, Andrew J. and {Cook}, Kem H. and {Cooray}, Asantha and {Covey}, Kevin R. and {Cribbs}, Chris and {Cui}, Wei and {Cutri}, Roc and {Daly}, Philip N. and {Daniel}, Scott F. and {Daruich}, Felipe and {Daubard}, Guillaume and {Daues}, Greg and {Dawson}, William and {Delgado}, Francisco and {Dellapenna}, Alfred and {de Peyster}, Robert and {de Val-Borro}, Miguel and {Digel}, Seth W. and {Doherty}, Peter and {Dubois}, Richard and {Dubois-Felsmann}, Gregory P. and {Durech}, Josef and {Economou}, Frossie and {Eifler}, Tim and {Eracleous}, Michael and {Emmons}, Benjamin L. and {Fausti Neto}, Angelo and {Ferguson}, Henry and {Figueroa}, Enrique and {Fisher-Levine}, Merlin and {Focke}, Warren and {Foss}, Michael D. and {Frank}, James and {Freemon}, Michael D. and {Gangler}, Emmanuel and {Gawiser}, Eric and {Geary}, John C. and {Gee}, Perry and {Geha}, Marla and {Gessner}, Charles J.~B. and {Gibson}, Robert R. and {Gilmore}, D. Kirk and {Glanzman}, Thomas and {Glick}, William and {Goldina}, Tatiana and {Goldstein}, Daniel A. and {Goodenow}, Iain and {Graham}, Melissa L. and {Gressler}, William J. and {Gris}, Philippe and {Guy}, Leanne P. and {Guyonnet}, Augustin and {Haller}, Gunther and {Harris}, Ron and {Hascall}, Patrick A. and {Haupt}, Justine and {Hernandez}, Fabio and {Herrmann}, Sven and {Hileman}, Edward and {Hoblitt}, Joshua and {Hodgson}, John A. and {Hogan}, Craig and {Howard}, James D. and {Huang}, Dajun and {Huffer}, Michael E. and {Ingraham}, Patrick and {Innes}, Walter R. and {Jacoby}, Suzanne H. and {Jain}, Bhuvnesh and {Jammes}, Fabrice and {Jee}, M. James and {Jenness}, Tim and {Jernigan}, Garrett and {Jevremovi{\'c}}, Darko and {Johns}, Kenneth and {Johnson}, Anthony S. and {Johnson}, Margaret W.~G. and {Jones}, R. Lynne and {Juramy-Gilles}, Claire and {Juri{\'c}}, Mario and {Kalirai}, Jason S. and {Kallivayalil}, Nitya J. and {Kalmbach}, Bryce and {Kantor}, Jeffrey P. and {Karst}, Pierre and {Kasliwal}, Mansi M. and {Kelly}, Heather and {Kessler}, Richard and {Kinnison}, Veronica and {Kirkby}, David and {Knox}, Lloyd and {Kotov}, Ivan V. and {Krabbendam}, Victor L. and {Krughoff}, K. Simon and {Kub{\'a}nek}, Petr and {Kuczewski}, John and {Kulkarni}, Shri and {Ku}, John and {Kurita}, Nadine R. and {Lage}, Craig S. and {Lambert}, Ron and {Lange}, Travis and {Langton}, J. Brian and {Le Guillou}, Laurent and {Levine}, Deborah and {Liang}, Ming and {Lim}, Kian-Tat and {Lintott}, Chris J. and {Long}, Kevin E. and {Lopez}, Margaux and {Lotz}, Paul J. and {Lupton}, Robert H. and {Lust}, Nate B. and {MacArthur}, Lauren A. and {Mahabal}, Ashish and {Mandelbaum}, Rachel and {Markiewicz}, Thomas W. and {Marsh}, Darren S. and {Marshall}, Philip J. and {Marshall}, Stuart and {May}, Morgan and {McKercher}, Robert and {McQueen}, Michelle and {Meyers}, Joshua and {Migliore}, Myriam and {Miller}, Michelle and {Mills}, David J.},
        title = "{LSST: From Science Drivers to Reference Design and Anticipated Data Products}",
      journal = {\apj},
         year = 2019,
        month = mar,
       volume = {873},
       number = {2},
          eid = {111},
        pages = {111},
          doi = {10.3847/1538-4357/ab042c},
archivePrefix = {arXiv},
       eprint = {0805.2366},
 primaryClass = {astro-ph},
       adsurl = {https://ui.adsabs.harvard.edu/abs/2019ApJ...873..111I}
}

@INPROCEEDINGS{lt,
       author = {{Steele}, Iain A. and {Smith}, Robert J. and {Rees}, Paul C. and {Baker}, Ian P. and {Bates}, S.~D. and {Bode}, Michael F. and {Bowman}, Mark K. and {Carter}, Dave and {Etherton}, Jason and {Ford}, Martyn J. and {Fraser}, Stephen N. and {Gomboc}, A. and {Lett}, Robert D.~J. and {Mansfield}, Anthony G. and {Marchant}, Jonathon M. and {Medrano-Cerda}, Gustavo A. and {Mottram}, Christopher J. and {Raback}, D. and {Scott}, A.~B. and {Tomlinson}, M.~D. and {Zamanov}, R.},
        title = "{The Liverpool Telescope: performance and first results}",
    booktitle = {Ground-based Telescopes},
         year = 2004,
       editor = {{Oschmann}, Jacobus M., Jr.},
       series = {Society of Photo-Optical Instrumentation Engineers (SPIE) Conference Series},
       volume = {5489},
        month = oct,
        pages = {679-692},
          doi = {10.1117/12.551456},
       adsurl = {https://ui.adsabs.harvard.edu/abs/2004SPIE.5489..679S}
}

@ARTICLE{lunnan2015,
       author = {{Lunnan}, R. and {Chornock}, R. and {Berger}, E. and {Rest}, A. and {Fong}, W. and {Scolnic}, D. and {Jones}, D.~O. and {Soderberg}, A.~M. and {Challis}, P.~M. and {Drout}, M.~R. and {Foley}, R.~J. and {Huber}, M.~E. and {Kirshner}, R.~P. and {Leibler}, C. and {Marion}, G.~H. and {McCrum}, M. and {Milisavljevic}, D. and {Narayan}, G. and {Sanders}, N.~E. and {Smartt}, S.~J. and {Smith}, K.~W. and {Tonry}, J.~L. and {Burgett}, W.~S. and {Chambers}, K.~C. and {Flewelling}, H. and {Kudritzki}, R.-P. and {Wainscoat}, R.~J. and {Waters}, C.},
        title = "{Zooming In on the Progenitors of Superluminous Supernovae With the HST}",
      journal = {\apj},
         year = 2015,
        month = may,
       volume = {804},
       number = {2},
          eid = {90},
        pages = {90},
          doi = {10.1088/0004-637X/804/2/90},
archivePrefix = {arXiv},
       eprint = {1411.1060},
 primaryClass = {astro-ph.HE},
       adsurl = {https://ui.adsabs.harvard.edu/abs/2015ApJ...804...90L}
}

@ARTICLE{lyman2017,
       author = {{Lyman}, J.~D. and {Levan}, A.~J. and {Tanvir}, N.~R. and {Fynbo}, J.~P.~U. and {McGuire}, J.~T.~W. and {Perley}, D.~A. and {Angus}, C.~R. and {Bloom}, J.~S. and {Conselice}, C.~J. and {Fruchter}, A.~S. and {Hjorth}, J. and {Jakobsson}, P. and {Starling}, R.~L.~C.},
        title = "{The host galaxies and explosion sites of long-duration gamma ray bursts: Hubble Space Telescope near-infrared imaging}",
      journal = {\mnras},
         year = 2017,
        month = may,
       volume = {467},
       number = {2},
        pages = {1795-1817},
          doi = {10.1093/mnras/stx220},
archivePrefix = {arXiv},
       eprint = {1701.05925},
 primaryClass = {astro-ph.GA},
       adsurl = {https://ui.adsabs.harvard.edu/abs/2017MNRAS.467.1795L}
}

@ARTICLE{mahabal2019,
       author = {{Mahabal}, Ashish and {Rebbapragada}, Umaa and {Walters}, Richard and {Masci}, Frank J. and {Blagorodnova}, Nadejda and {van Roestel}, Jan and {Ye}, Quan-Zhi and {Biswas}, Rahul and {Burdge}, Kevin and {Chang}, Chan-Kao and {Duev}, Dmitry A. and {Golkhou}, V. Zach and {Miller}, Adam A. and {Nordin}, Jakob and {Ward}, Charlotte and {Adams}, Scott and {Bellm}, Eric C. and {Branton}, Doug and {Bue}, Brian and {Cannella}, Chris and {Connolly}, Andrew and {Dekany}, Richard and {Feindt}, Ulrich and {Hung}, Tiara and {Fortson}, Lucy and {Frederick}, Sara and {Fremling}, C. and {Gezari}, Suvi and {Graham}, Matthew and {Groom}, Steven and {Kasliwal}, Mansi M. and {Kulkarni}, Shrinivas and {Kupfer}, Thomas and {Lin}, Hsing Wen and {Lintott}, Chris and {Lunnan}, Ragnhild and {Parejko}, John and {Prince}, Thomas A. and {Riddle}, Reed and {Rusholme}, Ben and {Saunders}, Nicholas and {Sedaghat}, Nima and {Shupe}, David L. and {Singer}, Leo P. and {Soumagnac}, Maayane T. and {Szkody}, Paula and {Tachibana}, Yutaro and {Tirumala}, Kushal and {van Velzen}, Sjoert and {Wright}, Darryl},
        title = "{Machine Learning for the Zwicky Transient Facility}",
      journal = {\pasp},
         year = 2019,
        month = mar,
       volume = {131},
       number = {997},
        pages = {038002},
          doi = {10.1088/1538-3873/aaf3fa},
archivePrefix = {arXiv},
       eprint = {1902.01936},
 primaryClass = {astro-ph.IM},
       adsurl = {https://ui.adsabs.harvard.edu/abs/2019PASP..131c8002M}
}

@ARTICLE{mandigo-stoba2022,
       author = {{Mandigo-Stoba}, Milan Sharma and {Fremling}, Christoffer and {Kasliwal}, Mansi},
        title = "{DBSP\_DRP: A Python package for automated spectroscopic data reduction of DBSP data}",
      journal = {The Journal of Open Source Software},
         year = 2022,
        month = feb,
       volume = {7},
       number = {70},
          eid = {3612},
        pages = {3612},
          doi = {10.21105/joss.03612},
archivePrefix = {arXiv},
       eprint = {2107.12339},
 primaryClass = {astro-ph.IM},
       adsurl = {https://ui.adsabs.harvard.edu/abs/2022JOSS....7.3612M}
}

@ARTICLE{margutti,
       author = {{Margutti}, R. and {Metzger}, B.~D. and {Chornock}, R. and {Vurm}, I. and {Roth}, N. and {Grefenstette}, B.~W. and {Savchenko}, V. and {Cartier}, R. and {Steiner}, J.~F. and {Terreran}, G. and {Margalit}, B. and {Migliori}, G. and {Milisavljevic}, D. and {Alexander}, K.~D. and {Bietenholz}, M. and {Blanchard}, P.~K. and {Bozzo}, E. and {Brethauer}, D. and {Chilingarian}, I.~V. and {Coppejans}, D.~L. and {Ducci}, L. and {Ferrigno}, C. and {Fong}, W. and {G{\"o}tz}, D. and {Guidorzi}, C. and {Hajela}, A. and {Hurley}, K. and {Kuulkers}, E. and {Laurent}, P. and {Mereghetti}, S. and {Nicholl}, M. and {Patnaude}, D. and {Ubertini}, P. and {Banovetz}, J. and {Bartel}, N. and {Berger}, E. and {Coughlin}, E.~R. and {Eftekhari}, T. and {Frederiks}, D.~D. and {Kozlova}, A.~V. and {Laskar}, T. and {Svinkin}, D.~S. and {Drout}, M.~R. and {MacFadyen}, A. and {Paterson}, K.},
        title = "{An Embedded X-Ray Source Shines through the Aspherical AT 2018cow: Revealing the Inner Workings of the Most Luminous Fast-evolving Optical Transients}",
      journal = {\apj},
         year = 2019,
        month = feb,
       volume = {872},
       number = {1},
          eid = {18},
        pages = {18},
          doi = {10.3847/1538-4357/aafa01},
archivePrefix = {arXiv},
       eprint = {1810.10720},
 primaryClass = {astro-ph.HE},
       adsurl = {https://ui.adsabs.harvard.edu/abs/2019ApJ...872...18M}
}

@ARTICLE{margalit2021,
       author = {{Margalit}, Ben and {Quataert}, Eliot},
        title = "{Thermal Electrons in Mildly Relativistic Synchrotron Blast Waves}",
      journal = {\apjl},
         year = 2021,
        month = dec,
       volume = {923},
       number = {1},
          eid = {L14},
        pages = {L14},
          doi = {10.3847/2041-8213/ac3d97},
archivePrefix = {arXiv},
       eprint = {2111.00012},
 primaryClass = {astro-ph.HE},
       adsurl = {https://ui.adsabs.harvard.edu/abs/2021ApJ...923L..14M}
}

@ARTICLE{margalit2024,
       author = {{Margalit}, Ben and {Quataert}, Eliot},
        title = "{The Peak Frequency and Luminosity of Synchrotron Emitting Shocks: From Nonrelativistic to Ultrarelativistic Explosions}",
      journal = {\apj},
         year = 2024,
        month = dec,
       volume = {977},
       number = {1},
          eid = {134},
        pages = {134},
          doi = {10.3847/1538-4357/ad8b47},
archivePrefix = {arXiv},
       eprint = {2403.07048},
 primaryClass = {astro-ph.HE},
       adsurl = {https://ui.adsabs.harvard.edu/abs/2024ApJ...977..134M}
}

@ARTICLE{masci2019,
       author = {{Masci}, Frank J. and {Laher}, Russ R. and {Rusholme}, Ben and {Shupe}, David L. and {Groom}, Steven and {Surace}, Jason and {Jackson}},
        title = "{The Zwicky Transient Facility: Data Processing, Products, and Archive}",
      journal = {\pasp},
         year = 2019,
        month = jan,
       volume = {131},
       number = {995},
        pages = {018003},
          doi = {10.1088/1538-3873/aae8ac},
archivePrefix = {arXiv},
       eprint = {1902.01872},
 primaryClass = {astro-ph.IM},
       adsurl = {https://ui.adsabs.harvard.edu/abs/2019PASP..131a8003M}
}

@ARTICLE{matthews2023,
       author = {{Matthews}, D. and {Margutti}, R. and {Metzger}, B.~D. and {Milisavljevic}, D. and {Migliori}, G. and {Laskar}, T. and {Brethauer}, D. and {Berger}, E. and {Chornock}, R. and {Drout}, M. and {Ramirez-Ruiz}, E.},
        title = "{Unprecedented X-Ray Emission from the Fast Blue Optical Transient AT2022tsd}",
      journal = {Research Notes of the American Astronomical Society},
         year = 2023,
        month = jun,
       volume = {7},
       number = {6},
          eid = {126},
        pages = {126},
          doi = {10.3847/2515-5172/acdde1},
archivePrefix = {arXiv},
       eprint = {2306.01114},
 primaryClass = {astro-ph.HE},
       adsurl = {https://ui.adsabs.harvard.edu/abs/2023RNAAS...7..126M}
}

@ARTICLE{metzger2022,
       author = {{Metzger}, Brian D.},
        title = "{Luminous Fast Blue Optical Transients and Type Ibn/Icn SNe from Wolf-Rayet/Black Hole Mergers}",
      journal = {\apj},
         year = 2022,
        month = jun,
       volume = {932},
       number = {2},
          eid = {84},
        pages = {84},
          doi = {10.3847/1538-4357/ac6d59},
archivePrefix = {arXiv},
       eprint = {2203.04331},
 primaryClass = {astro-ph.HE},
       adsurl = {https://ui.adsabs.harvard.edu/abs/2022ApJ...932...84M}
}

@ARTICLE{metzger2023,
       author = {{Metzger}, Brian D. and {Perley}, Daniel A.},
        title = "{Dust Echoes from Luminous Fast Blue Optical Transients}",
      journal = {\apj},
         year = 2023,
        month = feb,
       volume = {944},
       number = {1},
          eid = {74},
        pages = {74},
          doi = {10.3847/1538-4357/acae89},
archivePrefix = {arXiv},
       eprint = {2210.01819},
 primaryClass = {astro-ph.HE},
       adsurl = {https://ui.adsabs.harvard.edu/abs/2023ApJ...944...74M}
}

@ARTICLE{nayana2025,
       author = {{Nayana}, A.~J. and {Margutti}, Raffaella and {Wiston}, Eli and {Laskar}, Tanmoy and {Migliori}, Giulia and {Chornock}, Ryan and {Galvin}, Timothy J. and {LeBaron}, Natalie and {Hajela}, Aprajita and {Christy}, Collin T. and {Sfaradi}, Itai and {Tsuna}, Daichi and {Aspegren}, Olivia and {De Colle}, Fabio and {Metzger}, Brian D. and {Lu}, Wenbin and {Beniamini}, Paz and {Kasen}, Daniel and {Berger}, Edo and {Grefenstette}, Brian W. and {Alexander}, Kate D. and {Anupama}, G.~C. and {Coppejans}, Deanne L. and {Cruz}, Luigi F. and {DeBoer}, David R and {Drout}, Maria R. and {Farah}, Wael and {Huang}, Xiaoshan and {Jacobson-Gal{\'a}n}, W.~V. and {Milisavljevic}, Dan and {Pollak}, Alexander W. and {Roth}, Nathan J. and {Sears}, Huei and {Siemion}, Andrew and {Sheikh}, Sofia Z. and {Steiner}, James F. and {Vurm}, Indrek},
        title = "{The Most Luminous Known Fast Blue Optical Transient AT 2024wpp: Unprecedented Evolution and Properties in the X-rays and Radio}",
      journal = {arXiv e-prints},
         year = 2025,
        month = aug,
          eid = {arXiv:2509.00952},
        pages = {arXiv:2509.00952},
          doi = {10.48550/arXiv.2509.00952},
archivePrefix = {arXiv},
       eprint = {2509.00952},
 primaryClass = {astro-ph.HE},
       adsurl = {https://ui.adsabs.harvard.edu/abs/2025arXiv250900952N}
}

@INPROCEEDINGS{not,
       author = {{Djupvik}, Anlaug Amanda and {Andersen}, Johannes},
        title = "{The Nordic Optical Telescope}",
    booktitle = {Highlights of Spanish Astrophysics V},
         year = 2010,
       editor = {{Diego}, Jose M. and {Goicoechea}, Luis J. and {Gonz{\'a}lez-Serrano}, J. Ignacio and {Gorgas}, Javier},
       series = {Astrophysics and Space Science Proceedings},
       volume = {14},
        month = jan,
        pages = {211},
          doi = {10.1007/978-3-642-11250-8_21},
archivePrefix = {arXiv},
       eprint = {0901.4015},
 primaryClass = {astro-ph.IM},
       adsurl = {https://ui.adsabs.harvard.edu/abs/2010ASSP...14..211D}
}

@ARTICLE{nugent2026,
       author = {{Nugent}, Anya E. and {Villar}, V. Ashley and {Metzger}, Brian D. and {Fryer}, Christopher L. and {Burns}, Eric and {Gordon}, Alexa and {Frostig}, Danielle},
        title = "{The Environments of Luminous Fast Blue Optical Transients: Evidence for a Compact Object and Wolf-Rayet Star Merger Origin}",
      journal = {arXiv e-prints},
         year = 2026,
        month = mar,
          eid = {arXiv:2603.23597},
        pages = {arXiv:2603.23597},
          doi = {10.48550/arXiv.2603.23597},
archivePrefix = {arXiv},
       eprint = {2603.23597},
 primaryClass = {astro-ph.HE},
       adsurl = {https://ui.adsabs.harvard.edu/abs/2026arXiv260323597N}
}

@ARTICLE{ofek2010,
       author = {{Ofek}, E.~O. and {Rabinak}, I. and {Neill}, J.~D. and {Arcavi}, I. and {Cenko}, S.~B. and {Waxman}, E. and {Kulkarni}, S.~R. and {Gal-Yam}, A. and {Nugent}, P.~E. and {Bildsten}, L. and {Bloom}, J.~S. and {Filippenko}, A.~V. and {Forster}, K. and {Howell}, D.~A. and {Jacobsen}, J. and {Kasliwal}, M.~M. and {Law}, N. and {Martin}, C. and {Poznanski}, D. and {Quimby}, R.~M. and {Shen}, K.~J. and {Sullivan}, M. and {Dekany}, R. and {Rahmer}, G. and {Hale}, D. and {Smith}, R. and {Zolkower}, J. and {Velur}, V. and {Walters}, R. and {Henning}, J. and {Bui}, K. and {McKenna}, D.},
        title = "{Supernova PTF 09UJ: A Possible Shock Breakout from a Dense Circumstellar Wind}",
      journal = {\apj},
         year = 2010,
        month = dec,
       volume = {724},
       number = {2},
        pages = {1396-1401},
          doi = {10.1088/0004-637X/724/2/1396},
archivePrefix = {arXiv},
       eprint = {1009.5378},
 primaryClass = {astro-ph.HE},
       adsurl = {https://ui.adsabs.harvard.edu/abs/2010ApJ...724.1396O}
}

@ARTICLE{p60,
       author = {{Cenko}, S. Bradley and {Fox}, Derek B. and {Moon}, Dae-Sik and {Harrison}, Fiona A. and {Kulkarni}, S.~R. and {Henning}, John R. and {Guzman}, C. Dani and {Bonati}, Marco and {Smith}, Roger M. and {Thicksten}, Robert P. and {Doyle}, Michael W. and {Petrie}, Hal L. and {Gal-Yam}, Avishay and {Soderberg}, Alicia M. and {Anagnostou}, Nathaniel L. and {Laity}, Anastasia C.},
        title = "{The Automated Palomar 60 Inch Telescope}",
      journal = {\pasp},
         year = 2006,
        month = oct,
       volume = {118},
       number = {848},
        pages = {1396-1406},
          doi = {10.1086/508366},
archivePrefix = {arXiv},
       eprint = {astro-ph/0608323},
 primaryClass = {astro-ph},
       adsurl = {https://ui.adsabs.harvard.edu/abs/2006PASP..118.1396C}
}

@ARTICLE{p200_dbsp,
       author = {{Oke}, J.~B. and {Gunn}, J.~E.},
        title = "{An Efficient Low Resolution and Moderate Resolution Spectrograph for the Hale Telescope}",
      journal = {\pasp},
         year = 1982,
        month = jun,
       volume = {94},
        pages = {586},
          doi = {10.1086/131027},
       adsurl = {https://ui.adsabs.harvard.edu/abs/1982PASP...94..586O}
}

@ARTICLE{patterson2019,
       author = {{Patterson}, Maria T. and {Bellm}, Eric C. and {Rusholme}, Ben and {Masci}, Frank J. and {Juric}, Mario and {Krughoff}, K. Simon and {Golkhou}, V. Zach and {Graham}, Matthew J. and {Kulkarni}, Shrinivas R. and {Helou}, George and {Zwicky Transient Facility Collaboration}},
        title = "{The Zwicky Transient Facility Alert Distribution System}",
      journal = {\pasp},
         year = 2019,
        month = jan,
       volume = {131},
       number = {995},
        pages = {018001},
          doi = {10.1088/1538-3873/aae904},
archivePrefix = {arXiv},
       eprint = {1902.02227},
 primaryClass = {astro-ph.IM},
       adsurl = {https://ui.adsabs.harvard.edu/abs/2019PASP..131a8001P}
}

@ARTICLE{panstarrs,
       author = {{Chambers}, K.~C. and {Magnier}, E.~A. and {Metcalfe}, N. and {Flewelling}, H.~A. and {Huber}, M.~E. and {Waters}, C.~Z. and {Denneau}, L. and {Draper}, P.~W.\},
        title = "{The Pan-STARRS1 Surveys}",
      journal = {arXiv e-prints},
     keywords = {Astrophysics - Instrumentation and Methods for Astrophysics, Astrophysics - Earth and Planetary Astrophysics, Astrophysics - Astrophysics of Galaxies, Astrophysics - Solar and Stellar Astrophysics},
         year = 2016,
        month = dec,
          eid = {arXiv:1612.05560},
        pages = {arXiv:1612.05560},
          doi = {10.48550/arXiv.1612.05560},
archivePrefix = {arXiv},
       eprint = {1612.05560},
 primaryClass = {astro-ph.IM},
       adsurl = {https://ui.adsabs.harvard.edu/abs/2016arXiv161205560C},
      adsnote = {Provided by the SAO/NASA Astrophysics Data System}
}

@ARTICLE{perley2013_130427a,
       author = {{Perley}, D.~A.},
        title = "{GRB 130427A: CARMA 3mm observations.}",
      journal = {GRB Coordinates Network},
         year = 2013,
        month = jan,
       volume = {14494},
        pages = {1},
       adsurl = {https://ui.adsabs.harvard.edu/abs/2013GCN.14494....1P}
}

@ARTICLE{perley2017_171205a,
       author = {{Perley}, D.~A. and {Schulze}, S. and {de Ugarte Postigo}, A.},
        title = "{GRB 171205A: ALMA observations.}",
      journal = {GRB Coordinates Network},
         year = 2017,
        month = jan,
       volume = {22252},
        pages = {1},
       adsurl = {https://ui.adsabs.harvard.edu/abs/2017GCN.22252....1P}
}

@ARTICLE{perley2019,
       author = {{Perley}, Daniel A. and {Mazzali}, Paolo A. and {Yan}, Lin and {Cenko}, S. Bradley and {Gezari}, Suvi and {Taggart}, Kirsty and {Blagorodnova}, Nadia and {Fremling}, Christoffer and {Mockler}, Brenna and {Singh}, Avinash and {Tominaga}, Nozomu and {Tanaka}, Masaomi and {Watson}, Alan M. and {Ahumada}, Tom{\'a}s and {Anupama}, G.~C. and {Ashall}, Chris and {Becerra}, Rosa L. and {Bersier}, David and {Bhalerao}, Varun and {Bloom}, Joshua S. and {Butler}, Nathaniel R. and {Copperwheat}, Chris and {Coughlin}, Michael W. and {De}, Kishalay and {Drake}, Andrew J. and {Duev}, Dmitry A. and {Frederick}, Sara and {Gonz{\'a}lez}, J. Jes{\'u}s and {Goobar}, Ariel and {Heida}, Marianne and {Ho}, Anna Y.~Q. and {Horst}, John and {Hung}, Tiara and {Itoh}, Ryosuke and {Jencson}, Jacob E. and {Kasliwal}, Mansi M. and {Kawai}, Nobuyuki and {Khanam}, Tanazza and {Kulkarni}, Shrinivas R. and {Kumar}, Brajesh and {Kumar}, Harsh and {Kutyrev}, Alexander S. and {Lee}, William H. and {Maeda}, Keiichi and {Mahabal}, Ashish and {Murata}, Katsuhiro L. and {Neill}, James D. and {Ngeow}, Chow-Choong and {Penprase}, Bryan and {Pian}, Elena and {Quimby}, Robert and {Ramirez-Ruiz}, Enrico and {Richer}, Michael G. and {Rom{\'a}n-Z{\'u}{\~n}iga}, Carlos G. and {Sahu}, D.~K. and {Srivastav}, Shubham and {Socia}, Quentin and {Sollerman}, Jesper and {Tachibana}, Yutaro and {Taddia}, Francesco and {Tinyanont}, Samaporn and {Troja}, Eleonora and {Ward}, Charlotte and {Wee}, Jerrick and {Yu}, Po-Chieh},
        title = "{The fast, luminous ultraviolet transient AT2018cow: extreme supernova, or disruption of a star by an intermediate-mass black hole?}",
      journal = {\mnras},
         year = 2019,
        month = mar,
       volume = {484},
       number = {1},
        pages = {1031-1049},
          doi = {10.1093/mnras/sty3420},
archivePrefix = {arXiv},
       eprint = {1808.00969},
 primaryClass = {astro-ph.HE},
       adsurl = {https://ui.adsabs.harvard.edu/abs/2019MNRAS.484.1031P}
}

@ARTICLE{perley2021,
       author = {{Perley}, Daniel A. and {Ho}, Anna Y.~Q. and {Yao}, Yuhan and {Fremling}, Christoffer and {Anderson}, Joseph P. and {Schulze}, Steve and {Kumar}, Harsh and {Anupama}, G.~C. and {Barway}, Sudhanshu and {Bellm}, Eric C. and {Bhalerao}, Varun and {Chen}, Ting-Wan and {Duev}, Dmitry A. and {Galbany}, Llu{\'\i}s and {Graham}, Matthew J. and {Gromadzki}, Mariusz and {Guti{\'e}rrez}, Claudia P. and {Ihanec}, Nada and {Inserra}, Cosimo and {Kasliwal}, Mansi M. and {Kool}, Erik C. and {Kulkarni}, S.~R. and {Laher}, Russ R. and {Masci}, Frank J. and {Neill}, James D. and {Nicholl}, Matt and {Pursiainen}, Miika and {van Roestel}, Joannes and {Sharma}, Yashvi and {Sollerman}, Jesper and {Walters}, Richard and {Wiseman}, Philip},
        title = "{Real-time discovery of AT2020xnd: a fast, luminous ultraviolet transient with minimal radioactive ejecta}",
      journal = {\mnras},
         year = 2021,
        month = dec,
       volume = {508},
       number = {4},
        pages = {5138-5147},
          doi = {10.1093/mnras/stab2785},
archivePrefix = {arXiv},
       eprint = {2103.01968},
 primaryClass = {astro-ph.HE},
       adsurl = {https://ui.adsabs.harvard.edu/abs/2021MNRAS.508.5138P}
}

@INPROCEEDINGS{perley2023abvizsw,
       author = {{Perley}, Daniel},
        title = "{ZTF19abvizsw: A Gamma-Ray Burst Afterglow Without a Gamma-Ray Burst}",
    booktitle = {American Astronomical Society Meeting Abstracts},
         year = 2023,
       series = {American Astronomical Society Meeting Abstracts},
       volume = {241},
        month = jan,
          eid = {252.05},
        pages = {252.05},
       adsurl = {https://ui.adsabs.harvard.edu/abs/2023AAS...24125205P}
}

@ARTICLE{pian2000,
       author = {{Pian}, E. and {Amati}, L. and {Antonelli}, L.~A. and {Butler}, R.~C. and {Costa}, E. and {Cusumano}, G. and {Danziger}, J. and {Feroci}, M. and {Fiore}, F. and {Frontera}, F. and {Giommi}, P. and {Masetti}, N. and {Muller}, J.~M. and {Nicastro}, L. and {Oosterbroek}, T. and {Orlandini}, M. and {Owens}, A. and {Palazzi}, E. and {Parmar}, A. and {Piro}, L. and {in't Zand}, J.~J.~M. and {Castro-Tirado}, A. and {Coletta}, A. and {Dal Fiume}, D. and {Del Sordo}, S. and {Heise}, J. and {Soffitta}, P. and {Torroni}, V.},
        title = "{BEPPOSAX Observations of GRB 980425: Detection of the Prompt Event and Monitoring of the Error Box}",
      journal = {\apj},
         year = 2000,
        month = jun,
       volume = {536},
       number = {2},
        pages = {778-787},
          doi = {10.1086/308978},
archivePrefix = {arXiv},
       eprint = {astro-ph/9910235},
 primaryClass = {astro-ph},
       adsurl = {https://ui.adsabs.harvard.edu/abs/2000ApJ...536..778P}
}

@ARTICLE{planck2018,
       author = {{Planck Collaboration} and {Aghanim}, N. and {Akrami}, Y. and {Ashdown}, M. and {Aumont}, J. and {Baccigalupi}, C. and {Ballardini}, M. and {Banday}, A.~J. and {Barreiro}, R.~B. and {Bartolo}, N. and {Basak}, S. and {Battye}, R. and {Benabed}, K. and {Bernard}, J. -P. and {Bersanelli}, M. and {Bielewicz}, P. and {Bock}, J.~J. and {Bond}, J.~R. and {Borrill}, J. and {Bouchet}, F.~R. and {Boulanger}, F. and {Bucher}, M. and {Burigana}, C. and {Butler}, R.~C. and {Calabrese}, E. and {Cardoso}, J. -F. and {Carron}, J. and {Challinor}, A. and {Chiang}, H.~C. and {Chluba}, J. and {Colombo}, L.~P.~L. and {Combet}, C. and {Contreras}, D. and {Crill}, B.~P. and {Cuttaia}, F. and {de Bernardis}, P. and {de Zotti}, G. and {Delabrouille}, J. and {Delouis}, J. -M. and {Di Valentino}, E. and {Diego}, J.~M. and {Dor{\'e}}, O. and {Douspis}, M. and {Ducout}, A. and {Dupac}, X. and {Dusini}, S. and {Efstathiou}, G. and {Elsner}, F. and {En{\ss}lin}, T.~A. and {Eriksen}, H.~K. and {Fantaye}, Y. and {Farhang}, M. and {Fergusson}, J. and {Fernandez-Cobos}, R. and {Finelli}, F. and {Forastieri}, F. and {Frailis}, M. and {Fraisse}, A.~A. and {Franceschi}, E. and {Frolov}, A. and {Galeotta}, S. and {Galli}, S. and {Ganga}, K. and {G{\'e}nova-Santos}, R.~T. and {Gerbino}, M. and {Ghosh}, T. and {Gonz{\'a}lez-Nuevo}, J. and {G{\'o}rski}, K.~M. and {Gratton}, S. and {Gruppuso}, A. and {Gudmundsson}, J.~E. and {Hamann}, J. and {Handley}, W. and {Hansen}, F.~K. and {Herranz}, D. and {Hildebrandt}, S.~R. and {Hivon}, E. and {Huang}, Z. and {Jaffe}, A.~H. and {Jones}, W.~C. and {Karakci}, A. and {Keih{\"a}nen}, E. and {Keskitalo}, R. and {Kiiveri}, K. and {Kim}, J. and {Kisner}, T.~S. and {Knox}, L. and {Krachmalnicoff}, N. and {Kunz}, M. and {Kurki-Suonio}, H. and {Lagache}, G. and {Lamarre}, J. -M. and {Lasenby}, A. and {Lattanzi}, M. and {Lawrence}, C.~R. and {Le Jeune}, M. and {Lemos}, P. and {Lesgourgues}, J. and {Levrier}, F. and {Lewis}, A. and {Liguori}, M. and {Lilje}, P.~B. and {Lilley}, M. and {Lindholm}, V. and {L{\'o}pez-Caniego}, M. and {Lubin}, P.~M. and {Ma}, Y. -Z. and {Mac{\'\i}as-P{\'e}rez}, J.~F. and {Maggio}, G. and {Maino}, D. and {Mandolesi}, N. and {Mangilli}, A. and {Marcos-Caballero}, A. and {Maris}, M. and {Martin}, P.~G. and {Martinelli}, M. and {Mart{\'\i}nez-Gonz{\'a}lez}, E. and {Matarrese}, S. and {Mauri}, N. and {McEwen}, J.~D. and {Meinhold}, P.~R. and {Melchiorri}, A. and {Mennella}, A. and {Migliaccio}, M. and {Millea}, M. and {Mitra}, S. and {Miville-Desch{\^e}nes}, M. -A. and {Molinari}, D. and {Montier}, L. and {Morgante}, G. and {Moss}, A. and {Natoli}, P. and {N{\o}rgaard-Nielsen}, H.~U. and {Pagano}, L. and {Paoletti}, D. and {Partridge}, B. and {Patanchon}, G. and {Peiris}, H.~V. and {Perrotta}, F. and {Pettorino}, V. and {Piacentini}, F. and {Polastri}, L. and {Polenta}, G. and {Puget}, J. -L. and {Rachen}, J.~P. and {Reinecke}, M. and {Remazeilles}, M. and {Renzi}, A. and {Rocha}, G. and {Rosset}, C. and {Roudier}, G. and {Rubi{\~n}o-Mart{\'\i}n}, J.~A. and {Ruiz-Granados}, B. and {Salvati}, L. and {Sandri}, M. and {Savelainen}, M. and {Scott}, D. and {Shellard}, E.~P.~S. and {Sirignano}, C. and {Sirri}, G. and {Spencer}, L.~D. and {Sunyaev}, R. and {Suur-Uski}, A. -S. and {Tauber}, J.~A. and {Tavagnacco}, D. and {Tenti}, M. and {Toffolatti}, L. and {Tomasi}, M. and {Trombetti}, T. and {Valenziano}, L. and {Valiviita}, J. and {Van Tent}, B. and {Vibert}, L. and {Vielva}, P. and {Villa}, F. and {Vittorio}, N. and {Wandelt}, B.~D. and {Wehus}, I.~K. and {White}, M. and {White}, S.~D.~M. and {Zacchei}, A. and {Zonca}, A.},
        title = "{Planck 2018 results. VI. Cosmological parameters}",
      journal = {\aap},
         year = 2020,
        month = sep,
       volume = {641},
          eid = {A6},
        pages = {A6},
          doi = {10.1051/0004-6361/201833910},
archivePrefix = {arXiv},
       eprint = {1807.06209},
 primaryClass = {astro-ph.CO},
       adsurl = {https://ui.adsabs.harvard.edu/abs/2020A&A...641A...6P}
}

@ARTICLE{prentice2018,
       author = {{Prentice}, S.~J. and {Maguire}, K. and {Smartt}, S.~J. and {Magee}, M.~R. and {Schady}, P. and {Sim}, S. and {Chen}, T. -W. and {Clark}, P. and {Colin}, C. and {Fulton}, M. and {McBrien}, O. and {O'Neill}, D. and {Smith}, K.~W. and {Ashall}, C. and {Chambers}, K.~C. and {Denneau}, L. and {Flewelling}, H.~A. and {Heinze}, A. and {Holoien}, T.~W. -S. and {Huber}, M.~E. and {Kochanek}, C.~S. and {Mazzali}, P.~A. and {Prieto}, J.~L. and {Rest}, A. and {Shappee}, B.~J. and {Stalder}, B. and {Stanek}, K.~Z. and {Stritzinger}, M.~D. and {Thompson}, T.~A. and {Tonry}, J.~L.},
        title = "{The Cow: Discovery of a Luminous, Hot, and Rapidly Evolving Transient}",
      journal = {\apjl},
         year = 2018,
        month = sep,
       volume = {865},
       number = {1},
          eid = {L3},
        pages = {L3},
          doi = {10.3847/2041-8213/aadd90},
archivePrefix = {arXiv},
       eprint = {1807.05965},
 primaryClass = {astro-ph.HE},
       adsurl = {https://ui.adsabs.harvard.edu/abs/2018ApJ...865L...3P}
}

@ARTICLE{prospector1,
       author = {{Leja}, Joel and {Johnson}, Benjamin D. and {Conroy}, Charlie and {van Dokkum}, Pieter G. and {Byler}, Nell},
        title = "{Deriving Physical Properties from Broadband Photometry with Prospector: Description of the Model and a Demonstration of its Accuracy Using 129 Galaxies in the Local Universe}",
      journal = {\apj},
         year = 2017,
        month = mar,
       volume = {837},
       number = {2},
          eid = {170},
        pages = {170},
          doi = {10.3847/1538-4357/aa5ffe},
archivePrefix = {arXiv},
       eprint = {1609.09073},
 primaryClass = {astro-ph.GA},
       adsurl = {https://ui.adsabs.harvard.edu/abs/2017ApJ...837..170L}
}

@ARTICLE{prospector2,
    author = {{Johnson}, Benjamin D. and {Leja}, Joel and {Conroy}, Charlie and {Speagle}, Joshua S.},
        title = "{Stellar Population Inference with Prospector}",
    journal = {\apjs},
        year = 2021,
        month = jun,
    volume = {254},
    number = {2},
        eid = {22},
        pages = {22},
        doi = {10.3847/1538-4365/abef67},
archivePrefix = {arXiv},
    eprint = {2012.01426},
primaryClass = {astro-ph.GA},
    adsurl = {https://ui.adsabs.harvard.edu/abs/2021ApJS..254...22J}
}

@article{pursiainen2018,
    author = {Pursiainen, M and Childress, M and Smith, M and Prajs, S and Sullivan, M and Davis, T M and Foley, R J and Asorey, J and Calcino, J and Carollo, D and Curtin, C and D’Andrea, C B and Glazebrook, K and Gutierrez, C and Hinton, S R and Hoormann, J K and Inserra, C and Kessler, R and King, A and Kuehn, K and Lewis, G F and Lidman, C and Macaulay, E and Möller, A and Nichol, R C and Sako, M and Sommer, N E and Swann, E and Tucker, B E and Uddin, S A and Wiseman, P and Zhang, B and Abbott, T M C and Abdalla, F B and Allam, S and Annis, J and Avila, S and Brooks, D and Buckley-Geer, E and Burke, D L and Carnero Rosell, A and Carrasco Kind, M and Carretero, J and Castander, F J and Cunha, C E and Davis, C and De Vicente, J and Diehl, H T and Doel, P and Eifler, T F and Flaugher, B and Fosalba, P and Frieman, J and García-Bellido, J and Gruen, D and Gruendl, R A and Gutierrez, G and Hartley, W G and Hollowood, D L and Honscheid, K and James, D J and Jeltema, T and Kuropatkin, N and Li, T S and Lima, M and Maia, M A G and Martini, P and Menanteau, F and Ogando, R L C and Plazas, A A and Roodman, A and Sanchez, E and Scarpine, V and Schindler, R and Smith, R C and Soares-Santos, M and Sobreira, F and Suchyta, E and Swanson, M E C and Tarle, G and Tucker, D L and Walker, A R and (DES Collaboration)},
    title = "{Rapidly evolving transients in the Dark Energy Survey}",
    journal = {Monthly Notices of the Royal Astronomical Society},
    volume = {481},
    number = {1},
    pages = {894-917},
    year = {2018},
    month = {08},
    issn = {0035-8711},
    doi = {10.1093/mnras/sty2309},
    url = {https://doi.org/10.1093/mnras/sty2309},
    eprint = {https://academic.oup.com/mnras/article-pdf/481/1/894/25697676/sty2309.pdf},
}

@ARTICLE{Pursiainen2025,
       author = {{Pursiainen}, M. and {Killestein}, T.~L. and {Kuncarayakti}, H. and {Charalampopoulos}, P. and {Warwick}, B. and {Lyman}, J. and {Kotak}, R. and {Leloudas}, G. and {Coppejans}, D. and {Kravtsov}, T. and {Maeda}, K. and {Nagao}, T. and {Taguchi}, K. and {Ackley}, K. and {Dhillon}, V.~S. and {Galloway}, D.~K. and {Kumar}, A. and {O'Neill}, D. and {Ramsay}, G. and {Steeghs}, D.},
        title = "{Optical evolution of AT 2024wpp: the high-velocity outflows in Cow-like transients are consistent with high spherical symmetry}",
      journal = {\mnras},
         year = 2025,
        month = mar,
       volume = {537},
       number = {4},
        pages = {3298-3309},
          doi = {10.1093/mnras/staf232},
archivePrefix = {arXiv},
       eprint = {2411.03272},
 primaryClass = {astro-ph.HE},
       adsurl = {https://ui.adsabs.harvard.edu/abs/2025MNRAS.537.3298P}
}

@Manual{r,
    title = {R: A Language and Environment for Statistical Computing},
    author = {{R Core Team}},
    organization = {R Foundation for Statistical Computing},
    address = {Vienna, Austria},
    year = {2025},
    url = {https://www.R-project.org/},
  }

@ARTICLE{salas2013,
       author = {{Salas}, P. and {Bauer}, F.~E. and {Stockdale}, C. and {Prieto}, J.~L.},
        title = "{SN 2007bg: the complex circumstellar medium around one of the most radio-luminous broad-lined Type Ic supernovae}",
      journal = {\mnras},
         year = 2013,
        month = jan,
       volume = {428},
       number = {2},
        pages = {1207-1217},
          doi = {10.1093/mnras/sts104},
archivePrefix = {arXiv},
       eprint = {1208.3455},
 primaryClass = {astro-ph.HE},
       adsurl = {https://ui.adsabs.harvard.edu/abs/2013MNRAS.428.1207S}
}

@ARTICLE{sandoval2018,
       author = {{Rivera Sandoval}, L.~E. and {Maccarone}, T.~J. and {Corsi}, A. and {Brown}, P.~J. and {Pooley}, D. and {Wheeler}, J.~C.},
        title = "{X-ray Swift observations of SN 2018cow}",
      journal = {\mnras},
         year = 2018,
        month = oct,
       volume = {480},
       number = {1},
        pages = {L146-L150},
          doi = {10.1093/mnrasl/sly145},
archivePrefix = {arXiv},
       eprint = {1807.06369},
 primaryClass = {astro-ph.HE},
       adsurl = {https://ui.adsabs.harvard.edu/abs/2018MNRAS.480L.146R}
}

@ARTICLE{schlafly,
       author = {{Schlafly}, Edward F. and {Finkbeiner}, Douglas P.},
        title = "{Measuring Reddening with Sloan Digital Sky Survey Stellar Spectra and Recalibrating SFD}",
      journal = {\apj},
         year = 2011,
        month = aug,
       volume = {737},
       number = {2},
          eid = {103},
        pages = {103},
          doi = {10.1088/0004-637X/737/2/103},
archivePrefix = {arXiv},
       eprint = {1012.4804},
 primaryClass = {astro-ph.GA},
       adsurl = {https://ui.adsabs.harvard.edu/abs/2011ApJ...737..103S}
}

@ARTICLE{Schroeder2025_AT2024aehp,
       author = {{Schroeder}, G. and {Chen}, P. and {Sollerman}, J. and {Srinivasaragavan}, G. and {Sevilla}, J. and {Ho}, A.~Y.~Q.},
        title = "{Multiwavelength Observations of AT2024aehp}",
      journal = {Transient Name Server AstroNote},
         year = 2025,
        month = jan,
       volume = {3},
        pages = {1},
       adsurl = {https://ui.adsabs.harvard.edu/abs/2025TNSAN...3....1S}
}

@ARTICLE{Schroeder2025_AT2024aehp_Detection,
       author = {{Schroeder}, G. and {Ho}, A.~Y.~Q.},
        title = "{VLA Continued Observations of AT2024aehp}",
      journal = {Transient Name Server AstroNote},
         year = 2025,
        month = mar,
       volume = {99},
        pages = {1},
       adsurl = {https://ui.adsabs.harvard.edu/abs/2025TNSAN..99....1S}
}

@ARTICLE{schulze2021,
       author = {{Schulze}, Steve and {Yaron}, Ofer and {Sollerman}, Jesper and {Leloudas}, Giorgos and {Gal}, Amit and {Wright}, Angus H. and {Lunnan}, Ragnhild and {Gal-Yam}, Avishay and {Ofek}, Eran O. and {Perley}, Daniel A. and {Filippenko}, Alexei V. and {Kasliwal}, Mansi M. and {Kulkarni}, Shrinivas R. and {Neill}, James D. and {Nugent}, Peter E. and {Quimby}, Robert M. and {Sullivan}, Mark and {Strotjohann}, Nora Linn and {Arcavi}, Iair and {Ben-Ami}, Sagi and {Bianco}, Federica and {Bloom}, Joshua S. and {De}, Kishalay and {Fraser}, Morgan and {Fremling}, Christoffer U. and {Horesh}, Assaf and {Johansson}, Joel and {Kelly}, Patrick L. and {Kne{\v{z}}evi{\'c}}, Nikola and {Kne{\v{z}}evi{\'c}}, Sladjana and {Maguire}, Kate and {Nyholm}, Anders and {Papadogiannakis}, Sem{\'e}li and {Petrushevska}, Tanja and {Rubin}, Adam and {Yan}, Lin and {Yang}, Yi and {Adams}, Scott M. and {Bufano}, Filomena and {Clubb}, Kelsey I. and {Foley}, Ryan J. and {Green}, Yoav and {Harmanen}, Jussi and {Ho}, Anna Y.~Q. and {Hook}, Isobel M. and {Hosseinzadeh}, Griffin and {Howell}, D. Andrew and {Kong}, Albert K.~H. and {Kotak}, Rubina and {Matheson}, Thomas and {McCully}, Curtis and {Milisavljevic}, Dan and {Pan}, Yen-Chen and {Poznanski}, Dovi and {Shivvers}, Isaac and {van Velzen}, Sjoert and {Verbeek}, Kars K.},
        title = "{The Palomar Transient Factory Core-collapse Supernova Host-galaxy Sample. I. Host-galaxy Distribution Functions and Environment Dependence of Core-collapse Supernovae}",
      journal = {\apjs},
         year = 2021,
        month = aug,
       volume = {255},
       number = {2},
          eid = {29},
        pages = {29},
          doi = {10.3847/1538-4365/abff5e},
archivePrefix = {arXiv},
       eprint = {2008.05988},
 primaryClass = {astro-ph.GA},
       adsurl = {https://ui.adsabs.harvard.edu/abs/2021ApJS..255...29S}
}

@ARTICLE{sedm1,
       author = {{Blagorodnova}, Nadejda and {Neill}, James D. and {Walters}, Richard and {Kulkarni}, Shrinivas R. and {Fremling}, Christoffer and {Ben-Ami}, Sagi and {Dekany}, Richard G. and {Fucik}, Jason R. and {Konidaris}, Nick and {Nash}, Reston and {Ngeow}, Chow-Choong and {Ofek}, Eran O. and {O' Sullivan}, Donal and {Quimby}, Robert and {Ritter}, Andreas and {Vyhmeister}, Karl E.},
        title = "{The SED Machine: A Robotic Spectrograph for Fast Transient Classification}",
      journal = {\pasp},
         year = 2018,
        month = mar,
       volume = {130},
       number = {985},
        pages = {035003},
          doi = {10.1088/1538-3873/aaa53f},
archivePrefix = {arXiv},
       eprint = {1710.02917},
 primaryClass = {astro-ph.IM},
       adsurl = {https://ui.adsabs.harvard.edu/abs/2018PASP..130c5003B}
}

@ARTICLE{soderberg2006,
       author = {{Soderberg}, A.~M. and {Chevalier}, R.~A. and {Kulkarni}, S.~R. and {Frail}, D.~A.},
        title = "{The Radio and X-Ray Luminous SN 2003bg and the Circumstellar Density Variations around Radio Supernovae}",
      journal = {\apj},
         year = 2006,
        month = nov,
       volume = {651},
       number = {2},
        pages = {1005-1018},
          doi = {10.1086/507571},
archivePrefix = {arXiv},
       eprint = {astro-ph/0512413},
 primaryClass = {astro-ph},
       adsurl = {https://ui.adsabs.harvard.edu/abs/2006ApJ...651.1005S}
}
\bibliographystyle{aasjournal}


\appendix

\section{Identification and Classification}\label{sec:discovery}

\added{In this section, we describe how each LFBOT was identified in the ZTF alert stream, and describe the initial follow-up observations that led to their classification as an LFBOT.}  ZTF observations of the six LFBOTs in this paper occurred during Phase II (ZTF-II), which was divided into public (50\%), partnership (30\%), and Caltech (20\%) observing time.  ZTF uses three filters: {\em ztf-g, ztf-r}, and {\em ztf-i}.  For comparison with other photometry, we denote these as {\em g, r}, and {\em i}.
ZTF data products are processed by the IPAC ZTF pipeline \citep{masci2019}, which uses image subtraction following the methods of \citet{zackay2016}.  Detections more significant that 5-$\sigma$ are distributed as alerts in the Avro format \citep{patterson2019}.  These alerts are filtered using machine-learning real-bogus metrics \citep{mahabal2019, duev2019}, host-galaxy classifiers \citep{tachibana2018}, and inspection of light-curve properties.  For ZTF-II, the collaboration used the Fritz marshal \citep{vanderwalt2019, duev2019} to coordinate detection, analysis, and follow-up observations of transients.  The Fritz marshal uses the open-source software package \texttt{SkyPortal} \citep{coughlin2023}.

\subsection{AT2022abfc/ZTF22abvrxkk}
AT2022abfc was first detected by ZTF on November 21, 2022 (MJD $=59904.344$; \citealt{astro2022abfc_disc}).  ZTF non-detections prior to the first observation and ZTF detections over the next five days confirmed a rise faster than 1\,mag\,day$^{-1}$ and a \added{decay} rate of 0.2\,mag\,day$^{-1}$.  The transient's fast evolution led to follow-up multiwavelength observations.  A Gemini South spectrum taken December 1, 2022 showed narrow H$\alpha$ emission and \ion{Ca}{2} H\&K absorption lines at $z=0.212$, implying a peak absolute magnitude of $-20.7$ \citep{astro2022abfc_vla}.  A radio follow-up campaign with the VLA 
began with a detection at 10 GHz on December 6 \citep{astro2022abfc_vla}.  Subsequent radio observations showed a brightening of the radio emission.  The transient's fast evolution, high luminosity, and radio emission led to its classification as an LFBOT.

\subsection{AT2023fhn/ZTF23aaeozpp/\added{ATLAS23hnk}}
The discovery narrative for AT2023fhn has already been published by \citet{chrimesfhn}; we reiterate it here for convenience.  ZTF first detected AT2023fhn on April 10, 2023 (MJD $=60044.204$; \citealt{astro2023fhn_disc}). \added{The source was also detected by ATLAS \citep{atlas}, one day later.} Its light curve showed a fast \added{decay} of 0.2 mag day$^{-1}$ and blue optical colors ($g-r \approx -0.5$ mag).  A possible host galaxy's photometric redshift implied an absolute optical magnitude of $M_g <-20$, motivating follow-up observations.  On April 20, a spectrum obtained by the DBSP on the Palomar 5\,m 
telescope yielded H$\alpha$, \ion{N}{2}, and \ion{S}{2} emission lines at $z=0.24$, confirming an absolute magnitude of $M_g=-21.5$ \citep{astro2023fhn_p200}.  The transient was subsequently detected at X-ray and radio wavelengths, respectively by \chandra\ on April 25 \citep{astro2023fhn_chandra} and the VLA (10 GHz) on June 15 \citep{astro2023fhn_vla}.  The blue colors, fast fading, high luminosity, and detected emission throughout the electromagnetic spectrum confirmed it as an LFBOT.  The \chandra\ X-ray detections and a subset of this transient's VLA radio observations are already published by \citet{chrimes2024}.

\subsection{AT2023hkw/ZTF23aaimsja}
ZTF first detected AT2023hkw on May 1, 2023 (MJD $=60065.199$; \citealt{astro2023hkw_disc}).  The transient was flagged for fast evolution from a rise rate greater than 1\,mag\,day$^{-1}$ and a \added{decay} rate of $\sim 0.2$\,mag\,day$^{-1}$.  This resulted in follow-up spectroscopy on May 12 using DEIMOS at the Keck Observatory 
\citep{astro2023hkw_keck}, using 1440\,s of exposure time.  The spectrum showed narrow H$\alpha$ and [\ion{N}{2}] emission lines from the host galaxy at $z=0.339$, implying an absolute magnitude of $\mathrm{M}_g=-21$.  It was also detected by the VLA at 10\,GHz on June 14 \citep{astro2023hkw_vla}.  It was classified as an LFBOT from its luminosity, fast evolution, and radio emission.

\subsection{AT2023vth/ZTF23ableqsp}
ZTF first detected AT2023vth on October 18, 2023 (MJD $=60235.116$; \citealt{astro2023vth_disc}).  It had a fast rise of 1 mag day$^{-1}$, a fast fade of 0.25 mag day$^{-1}$, and blue colors with $g-r \approx -0.4$ mag \citep{astro2023vth_ztf}, making it a promising LFBOT candidate.  Optical spectroscopy taken on November 12 revealed narrow H$\alpha$, [\ion{N}{2}], and [\ion{S}{2}] emission lines from the host galaxy at $z=0.0747$, which implied an absolute magnitude of $M_g\approx -20$ \citep{astro2023fhn_p200}.  The VLA detected radio emission at 10 GHz on November 17 \citep{astro2023vth_vla} and NOEMA found millimeter emission at 100 GHz on November 20 \citep{astro2023vth_noema}.  The presence of multiwavelength emission solidified the transient's LFBOT classification. We also obtained deep late-time VLT observations that will be reported in a separate paper \added{according to a prior arrangement} (Wise et al., in prep.). 

\subsection{AT2024qfm/ZTF24aaxhxhf}

AT2024qfm was first detected by ZTF on July 24, 2024 (MJD = 60515.41; \citealt{astro2024qfm_disc}). The most recent non-detection was 0.98\,d prior at $g>19.27\,$mag. The source was detected again three nights later as part of the ZTF public survey, resulting in the first alert. Over the following two nights the source exhibited significant fading of $0.3$\,mag\,day$^{-1}$. SDSS pre-imaging showed a galaxy at the transient's position with photometric redshift $z=0.190\pm0.054$, implying a high luminosity of $M<-20\,$mag. The rapid fading, host-galaxy counterpart, and high luminosity led the transient to be flagged by the daily scanner on July 27. 

The rapid fading flagged several fast-transient pipelines including Fastfinder \citep{Fulton2024_AT2024qfm} and ZTFReST \citep{Andreoni2021}. The rapid fading was publicly reported by Fastfinder \citep{Fulton2024_AT2024qfm}. The transient itself was reported to the Transient Name Server (TNS\footnote{\url{https://www.wis-tns.org/}}) by the Automatic Learning for the Rapid Classification of Events (ALeRCE) broker \citep{astro2024qfm_disc}.

On August 1, a redshift of $z=0.2270$ was determined by \citet{Gillanders2024_AT2024qfm} and an X-ray detection was reported by \citet{Margutti2024_AT2024qfm}. On August 2, we obtained a long-slit spectrum of AT\,2024qfm with GMOS on Gemini-North under a ToO program\footnote{GN-2024B-Q-128; PI A. Ho} and confirmed the reported redshift. A VLA radio observation was also obtained \added{\citep{astro2024qfm_vla}}
; the radio data will be published as part of a separate paper (Nayana A. J. et al., in prep.). Finally, we obtained 100\,GHz observations using NOEMA, which resulted in a detection. The fast optical light curve, and high luminosities at millimeter and X-ray wavelengths, all confirmed the LFBOT nature of this source. We also obtained deep late-time GTC/HiPERCAM observations that will be reported in a separate paper (Wise et al., in prep.). 

\subsection{AT2024aehp/ZTF24abygbss}

AT2024aehp was first detected by ZTF on December 19, 2024 (MJD = 60663.36) and reported to the TNS by ALeRCE \citep{astro2024aehp_disc}. The most recent non-detection was 3\,d prior.  ZTF photometry over the next two nights revealed a \added{decay} rate of 0.2\,mag\,day$^{-1}$, causing the transient to be flagged by the daily scanner on December 21. The fast fading, \added{blue colors ($g-r \approx -0.3\,$mag)} and possible high peak luminosity from the photometric redshift of the host galaxy (SDSS $z_\mathrm{ph} = 0.17 \pm 0.09$) motivated us to trigger follow-up observations. We acquired spectroscopy of the candidate using GMOS on Gemini South on December 23, 2024 
The spectrum shows narrow H$\alpha$, [\ion{N}{2}], and [\ion{S}{2}] emission lines that correspond to $z=0.170$. We reported the rapid evolution and the high luminosity \added{inferred from the spectroscopy.} \citep{astro2024aehp_gemini}. 

Radio and X-ray observations initially resulted in non-detections, and the optical light curve flattened out in a way uncharacteristic of LFBOTs \citep{Schroeder2025_AT2024aehp}. A subsequent radio epoch at 86 days after peak light resulted in a detection and confirmation of a positive spectral index, suggesting a link with AT2018cow-like LFBOTs \citep{Schroeder2025_AT2024aehp_Detection}. Then, another epoch at 141 days showed brightening in all radio bands by almost an order of magnitude, which is unprecedented compared to the radio emission of other LFBOTs.  This paper focuses on the early observations of AT2024aehp, and the interesting late-time behavior will be examined in separate work.

\section{Data Tables}
\label{sec:appendix-tables}

In this section we provide tables of optical spectroscopy (Table~\ref{tab:spectra}), X-ray observations (Table~\ref{tab:xray}), radio observations (Table~\ref{tab:radio}), and host-galaxy photometry (Table~\ref{tab:galaxy}). We also give our broken power-law fits to the radio SEDs (Table~\ref{tab:synchro_fit}) and the resulting synchrotron parameters (Table~\ref{tab:synchro}).  Finally, we list the optical photometry for each transient in Table~\ref{tab:photometry}. 

\begin{deluxetable*}{ccccccc}[!h]
\setlength{\tabcolsep}{0.45em}
\tablewidth{0.9\textwidth}
\tabletypesize{\footnotesize}
\tablewidth{0pt}
\tablecaption{Optical Spectroscopy}
\tablecolumns{7}
\tablehead{
\colhead{Object}
&\colhead{UTC Date}
&\colhead{$t_{\mathrm{obs}}$ (days)}& \colhead{Telescope/Inst.} &\colhead{Exposures}&\colhead{PI \& Program}&\colhead{Reduction Software}}
\startdata
AT2022abfc &20221201&10&Gemini-S/GMOS&$6\times450\,$s&GS-2022B-Q-126; PI A. Ho&\texttt{DRAGONS}$^*$\\
AT2023fhn &20230419&7&Gemini-S/GMOS&$6\times450\,$s&GS-2023A-Q-127; PI A. Ho
&\texttt{DRAGONS}$^*$\\
AT2023fhn &20230420&8&P200/DBSP&$3\times600\,$s& PI Y. Qin&\texttt{DBSP\_DRP}$^{**}$\\
AT2023fhn &20230426&14&Keck-I/LRIS& $4\times2700\,$s&PI M. M. Kasliwal&\texttt{lpipe}$^{***}$\\
AT2023hkw &20230512&12&Keck-II/DEIMOS&1440s&PI A. Filippenko&IRAF routines and custom codes $^\dagger$\\
AT2023vth &20231022&2&P200/SEDM&$\sim1800\,$s$^{\dagger\dagger}$&---$^{\dagger\dagger}$&---$^{\dagger\dagger}$\\
AT2023vth &20231112&23&Gemini-N/GMOS&$2\times500\,$s&GN-2023B-Q-130; PI A. Chrimes&\texttt{DRAGONS}$^*$\\
AT2024qfm & 20240802 & 6 & Gemini-N/GMOS & $4\times450\,$s&GN-2024B-Q-128; PI A. Ho&\texttt{DRAGONS}$^*$\\
AT2024qfm & 20240809 & 13 & Keck-I/LRIS & $4\times875\,$s&U256; PI Chornock&\texttt{lpipe}$^{***}$\\
AT2024qfm & 20240831  & 35  & MMT/Binospec & 1200$\,$s & UAO-G200-24B; PI A. A. Miller&Binospec and \texttt{pypeit}$^{\dagger\dagger\dagger}$\\
AT2024qfm & 20240902 & 37 & Keck-I/LRIS & $4\times900\,$s&U233; PI Filippenko&\texttt{lpipe}$^{***}$\\
AT2024aehp & 20241224 & 5 & Gemini-S/GMOS  &$4\times450\,$s&GS-2024B-Q-121 ; PI A. Ho&\texttt{DRAGONS}$^*$\\
AT2024aehp & 20250101 & 13 & Keck-II/KCWI & $600 + 3\times180\,$s & PI Y. Qin & \texttt{kskywizard}$^{\dagger\dagger\dagger\dagger}$\\
AT2024aehp & 20250107 & 19 & MMT/Binospec &$4\times500\,$s&PI A. Gal-Yam&Binospec and \texttt{pypeit}$^{\dagger\dagger\dagger}$\\
\enddata
\vspace{2.5mm}
\tablecomments{$^*$: \texttt{DRAGONS} \citep{dragons} is a software package specialized for reducing Gemini data, and performs stacking of flatfields and comparison lamps, bias subtraction, and standard-star flux calibration. \\
$^{**}$: \texttt{DBSP\_DRP} \citep{mandigo-stoba2022} is a Python package based on \texttt{pypeit} \citep{pypeit} that performs automated reduction of DBSP data, including atmospheric corrections andflux calibration. \\
$^{***}$\texttt{lpipe} \citep{lpipe} is an IDL pipeline that performs for fully automated data reduction for the LRIS instrument.  Its steps include bias correction, flat-fielding, and wavelength and flux calibration.\\
$^{\dagger}$: The IRAF routines \citep{silverman2012, iraf}\footnote{\url{https://iraf-community.github.io}} routines and custom Python and IDL codes \footnote{\href{https://github.com/ishivvers/TheKastShiv}{https://github.com/ishivvers/TheKastShiv}} use techniques for CCD processing and spectrum extraction, comparison-lamp wavelength calibration, and standard star flux calibration.\\
$^{\dagger\dagger}$: This spectrum was taken as part of automated follow-up for the ZTF Bright Transient Survey \citep{ztf_bts1}.\\
$^{\dagger\dagger\dagger}$: A Binospec pipeline \citep{kansky2019} was used to perform bias subtraction and flat fielding, while \texttt{pypeit} \citep{pypeit} was used to perform cosmic-ray removal, wavelength calibration and relative-flux calibration from a standard star.\\
$^{\dagger\dagger\dagger\dagger}$: \texttt{kskywizard}\footnote{\url{https://github.com/zhuyunz/KSkyWizard}} is a pipeline that can perform sky subtraction, flux calibration, and telluric correction.}
\label{tab:spectra}
\end{deluxetable*}

\begin{deluxetable*}{cccccccc}[!h]
\setlength{\tabcolsep}{0.25em}
\tablewidth{0.9\textwidth}
\tabletypesize{\footnotesize}
\tablewidth{0pt}
\tablecaption{\swift\ XRT Observations}
\tablecolumns{8}
\tablehead{
\colhead{Object}
&\colhead{UTC Date}
& \colhead{$t_{\mathrm{obs}}$} & \colhead{Flux}& \colhead{Count Rate} & \colhead{$n_{\rm H}$} &\colhead{$L_{\rm X}$}&\colhead{Photon Index}\vspace{-2mm}\\ &&
\colhead{(days)}&\colhead{($10^{-13}$\,erg\,cm$^{-2}$\,s$^{-1}$)}&\colhead{($10^{-3}$\,s$^{-1}$)}&\colhead{($10^{20}$ cm$^{-2}$)}&\colhead{($10^{43}$\,erg\,s$^{-1}$)}}
\startdata
AT2022abfc &20221208 &17& $<2 $ &  $<6$ &   2.49& $<3 $ & ---  \\
AT2023fhn &20230425 &13&$<1.7$ &  $<5$ &   $2.26 $ & $<4 $ & ---\\
AT2023hkw &20230521 &21&$<1.1$ &  $<3.5$ &   $1.05$ & $<4.5 $& ---\\
AT2023vth &20231110 &21&$<1.9$  &  $<3.4$ &   $22.6$  & $<0.23 $& --- \\
AT2024qfm &20240801 &5&$5.0\pm0.8$  &  $13\pm2$ &   $3.83 $  & $8.2\pm1.3 $ &$2.1^{+0.8}_{-0.5} $\\ 
AT2024qfm &20240802 &6&$3.7\pm0.9$  &  $10\pm2$ &   ---  & $6.1\pm1.5 $ &$1.5^{+1.6}_{-0.9} $\\ 
AT2024qfm &20240803 &7&$<4.9$  &  $<12$& --- & $<7.9 $& --- \\ 
AT2024qfm &20240803 &7.5&$2.8^{+1.8}_{-1.3}$  & $7^{+5}_{-3}$ & --- & $4.5^{+2.9}_{-2.1} $ &$2.9^{+3.3}_{-2.9} $\\ 
AT2024qfm &20240805 &9&$<3.1$  &  $<8$ & --- & $<5.0 $ & ---\\ 
AT2024qfm &20240807 &11&$25^{+24}_{-14}$  & $63^{+60}_{-35}$ & ---  & $40^{+38}_{-23} $ & \,\,\, --- *\\ 
AT2024qfm &20240809 &13&$<23$  &  $<58$ & --- & $<37 $& --- \\ 
AT2024qfm &20240809 &13.5&$2.0^{+1.2}_{-0.9}$  & $5^{+3.2}_{-2.4}$ & --- & $3.3^{+2.0}_{-1.4} $& \,\,\, --- * \\ 
AT2024qfm &20240810 &14&$1.7^{+1.0}_{-0.7}$  & $4^{+1.9}_{-2.7}$ & --- & $2.7^{+1.6}_{-1.2} $ &$-0.3^{+1.5}_{-1.8}$\\ 
AT2024qfm &20240813 &17&$2.6^{+1.5}_{-1.1}$  & $7_{-3}^{+4}$ & --- & $4.2^{+2.5}_{-1.9} $& \,\,\, --- * \\ 
AT2024qfm &20241204 &130&$<0.9$  & $<2$ & --- & $<1.5 $ & ---\\ 
AT2024aehp &20241231 &12& $<2.3$  & $<6.2$ &   $2.83$  & $<2.0 $ & ---\\ 
AT2024aehp &20250103 &15&$<3.0$  & $<7.9$ &  --- & $<2.6 $ & ---\\ 
AT2024aehp &20250105 &17&$<2.4$  & $<6.3$ &  --- & $<2.1 $ & ---\\ 
CSS161010 &20161103 & 24 & $1.8^{+1.4}_{-0.9}$  & $4.5^{+3.4}_{-2.3}$ & $4.7$  & $4.8^{+3.6}_{-2.4} \times 10^{-2} $ & \,\,\, --- *\\
CSS161010 & 20161108 & 29 & $<1.2$  &  $<2.9$ & --- & $<3.4 \times 10^{-2} $ & ---\\
CSS161010 & 20161111 & 32 & $<2.0$  &  $<5.0$& --- & $<5.7 \times 10^{-2} $ & ---\\ 
\enddata
\vspace{2.5mm}   
\tablecomments{AT2022abfc values are from \citet{astro2022abfc_vla},  AT2023fhn measurements obtained from \citet{chrimesfhn}, $n_{\rm H}$ for CSS161010 obtained from \citet{coppejans2020}. To convert flux to count rate, we assume a photon power-law index of $\Gamma=2$ for all measurements.  $t$ is determined from the middle of the observation period.  $n_{\rm H}$ is the same for all observations of a given object. \\ $^*$Too few counts for a spectral index.}
\label{tab:xray}
\end{deluxetable*} 

\clearpage

\startlongtable
\begin{deluxetable*}{lcccccc}
\setlength{\tabcolsep}{0.45em}
\tablewidth{0.9\textwidth}
\tabletypesize{\footnotesize}
\tablewidth{0pt}
\tablecaption{Radio and Millimeter Observations}
\tablecolumns{6}
\tablehead{
\colhead{Name} &
\colhead{UTC Date}
& \colhead{$t_{\mathrm{obs}}$ (days)} & \colhead{Band}
& \colhead{$\nu$ (GHz)} & \colhead{Config.}
& \colhead{$f_\nu$ ($\mu$Jy)}}
\startdata
AT2022abfc & 20221206 &15&X&9 &VLA-C&$<17$ \\
AT2022abfc & 20221206 &15&X&11 &VLA-C&$<19$\\
AT2022abfc & 20221223 &32&Ku&15 &VLA-C& $52\pm3$ \\
AT2022abfc & 20221223 &32&X&10 &VLA-C& $38\pm4$\\
AT2022abfc & 20230110 & 53&1&86  &NOEMA-12A & $<348$\\
AT2022abfc & 20230405&135& Ku & 10  &  VLA-B & $45\pm7$ \\
AT2022abfc & 20230405&135& X & 10  &  VLA-B & $43\pm9$ \\
AT2022abfc & 20230602 & 193 & 3 & 0.45 &GMRT & $<150 $\\
AT2022abfc & 20230622 & 214 & 5 & 1.37 &GMRT & 44$\pm$12*\\
AT2022abfc & 20230623 & 215 & 3 & 0.45 &GMRT&  $<165$\\
AT2022abfc & 20230624 &216& Ku &    10 &  VLA-BnA$->$A &  $<27$\\
AT2022abfc & 20230624 &216& S & 3 &  VLA-BnA$->$A &  $<26$\\
AT2022abfc & 20230624 &216& X & 10 &  VLA-BnA$->$A &  $<26$ \\
AT2022abfc & 20240202&438&X&9 &VLA-C &$<17$\\
AT2022abfc & 20240202&438&X&11 &VLA-C &$<19$ \\
AT2023fhn & 20230422&10&1& 86&NOEMA-C&$<90$\\
AT2023fhn & 20230511 &29& X & 9 &  VLA-B & $<19$ \\
AT2023fhn & 20230511 &29& X & 11 &  VLA-B & $<20$ \\
AT2023fhn & 20230615 &64& X & 9 &  VLA-BnA & $43\pm6$ \\
AT2023fhn & 20230615 &64& X & 11 &  VLA-BnA & $53\pm7$ \\
AT2023fhn & 20230709 &88& Ku & 15 &  VLA-A & $75\pm7$ \\
AT2023fhn & 20230825 &135& Ku & 15 &  VLA-A & $98\pm9$ \\
AT2023fhn & 20230825 &135& X & 10 &  VLA-A & $143\pm7$ \\
AT2023fhn & 20230910 &151& 4 & 0.65 & GMRT&  $<300$\\
AT2023fhn & 20230910 &151& 5 & 1.37 &  GMRT& $<180$ \\
AT2023fhn & 20231026 &197& 5 & 1.37 &  GMRT& $85\pm24$ \\
AT2023fhn & 20231103 &205& 4 & 0.65 &  GMRT& $<231$ \\
AT2023fhn & 20231127 &229& 4 & 0.65 &  GMRT& $<210$ \\
AT2023fhn & 20231127 &229& 5 & 1.37 &  GMRT& $112\pm20$ \\
AT2023fhn & 20240624 &440& X & 10 &  VLA-B & $33\pm5$ \\
AT2023hkw & 20230614 &44& X & 9 &  VLA-BnA &  $79\pm6$\\
AT2023hkw & 20230614 &44& X & 11 &  VLA-BnA &  $103\pm7$\\
AT2023hkw & 20230704& 64 & Ka & 33 & VLA-A & $124\pm11$ \\
AT2023hkw & 20230704& 64 & Ku & 10 & VLA-A & $152\pm7$ \\
AT2023hkw & 20230826& 117 & C & 6 & VLA-A & $74\pm7$\\
AT2023hkw & 20230826& 117 & X & 10 & VLA-A & $85\pm8$\\
\enddata
\vspace{2.5mm}   
\tablecomments{The full set of radio and millimeter observations for each LFBOT are available online in a machine-readable table.  These tables include comments for specific observations.  Listed frequencies are in the observer frame.  Some X-band observations are split into two sub-bands, one with a central frequency at 9 GHz and the other at 11 GHz.  We require SNR $>3$ for a detection.  Otherwise, we list the 3-$\sigma$ upper limits for a non-detection.\\
$^*$This measurement was binned together with a GMRT Band-5 observation on June 2, 2023.}
\label{tab:radio}
\end{deluxetable*}

\begin{deluxetable*}{cccccccccc}[!h]
\label{tab:galaxy}
\setlength{\tabcolsep}{0.35em}
\tablewidth{0.9\textwidth}
\tabletypesize{\footnotesize}
\tablewidth{0pt}
\tablecaption{Host-Galaxy Photometry}
\tablecolumns{10}
\tablehead{
\colhead{Object}
&\multicolumn{2}{c}{\galex}&\multicolumn{4}{c}{Legacy Survey/Pan-STARRS}&\multicolumn{3}{c}{{\it NEOWISE}} \\
    \cmidrule(lr){2-3}       
    \cmidrule(lr){4-7}
    \cmidrule(lr){8-10}
&
\colhead{FUV}
&\colhead{NUV}& \colhead{$g$} & \colhead{$r$}& \colhead{$i$}& \colhead{$z$}& \colhead{$w1$}& \colhead{$w2$}& \colhead{$w3$}}
\startdata
AT2022abfc &---&21.550$\pm$0.310&19.057$\pm$0.002&18.124$\pm$0.001&17.698$\pm$0.001&17.464$\pm$0.002&17.356$\pm$0.007&17.568$\pm$0.015&15.875$\pm$0.073
\\
AT2023fhn &---&20.913$\pm$0.200& 19.452$\pm$0.006& 18.912$\pm$0.006& 18.606$\pm$0.009& 18.566$\pm$0.009& 18.796$\pm$0.024&18.755$\pm$0.051&---\\
AT2023hkw &---&21.898$\pm$0.086&20.706$\pm$0.014&19.669$\pm$0.012&---&18.996$\pm$0.010&18.833$\pm$0.022&19.097$\pm$0.051&---\\
AT2023vth &---&21.601$\pm$0.274&19.430$\pm$0.019&18.774$\pm$0.012&18.370$\pm$0.008&18.179$\pm$0.014&19.809$\pm$0.192&20.170$\pm$0.380&---\\
AT2024qfm&---&---& 20.530$\pm$0.007&19.706$\pm$0.006&---&19.135$\pm$0.007&19.156$\pm$0.028&19.218$\pm$0.067&---\\
AT2024aehp&22.158$\pm$0.181&21.902$\pm$0.115&20.881$\pm$0.009&20.454$\pm$0.009&20.240$\pm$0.013&20.232$\pm$0.020&20.761$\pm$0.117&20.468$\pm$0.206&---
\enddata
\tablecomments{Photometry used in our \texttt{prospector} host-galaxy fits.  Empty columns represent non-detections or detections with SNR $< 5$.  The middle columns represent photometry in the Legacy Survey {\em griz} filters, or the Pan-STARRS {\em griz} filters \added{for AT2023vth as its host was outside the Legacy Survey footprint}.  Not listed is {\it HST} photometry for AT2023fhn from \citet{chrimes2024} that was also used in the fit.}
\vspace{2.5mm}
\end{deluxetable*}

\begin{deluxetable*}{cccccc}[!htb]
\setlength{\tabcolsep}{0.55em}
\tablewidth{0.9\textwidth}
\tabletypesize{\footnotesize}
\tablewidth{0pt}
\tablecaption{Host-Galaxy Properties}
\tablecolumns{6}
\tablehead{
\colhead{Object}
&\colhead{Mass ($\log M/M_{\odot}$)}
&\colhead{Metallicity ($\log Z/Z_{\odot}$)}& \colhead{Age (Gyr)} & \colhead{SFR ($M_{\odot}\,\mathrm{yr}^{-1}$)} & \colhead{Host Extinction ($\tau$)}}
\startdata
AT2022abfc &$10.80^{+0.02}_{-0.04}$&0.05&$1.0^{+0.1}_{-0.2}$&$4.3^{+0.6}_{-0.6}$&$0.83^{+0.04}_{-0.04}$\\
AT2023fhn &$10.07^{+0.06}_{-0.06}$&$-0.27$&$0.7^{+0.5}_{-0.2}$&$7.7^{+1.4}_{-1.1}$&$0.59^{+0.05}_{-0.05}$\\
AT2023hkw &$10.7^{+0.1}_{-0.1}$&0.03&$2.7^{+1.2}_{-0.8}$&$2.8^{+1.3}_{-0.8}$&$0.22^{+0.17}_{-0.14}$\\
AT2023vth &$8.95^{+0.03}_{-0.04}$&$-0.60$&$0.62^{+0.13}_{-0.07}$&$0.24^{+0.04}_{-0.03}$&$0.0035^{+0.0055}_{-0.0026}$\\
AT2024qfm &$10.2^{+0.2}_{-0.1}$&$-0.20$&$4.4^{+5.0}_{-3.1}$&$3.3^{+1.1}_{-1.1}$&$0.60^{+0.14}_{-0.13}$\\
AT2024aehp &$8.9^{+0.1}_{-0.1}$&$-0.60$&$0.8^{+0.3}_{-0.3}$&$1.6^{+0.2}_{-0.2}$&$0.63^{+0.05}_{-0.05}$\\
\enddata
\tablecomments{Results of our \texttt{prospector} host-galaxy fits.  Metallicities were not determined by the fitting process, but by using a preliminary run at $\log Z/Z_{\odot}=-0.2$, and then converting the mass estimate to a metallicity using \citet{gallazzi2005}.  For the host galaxy extinction, we report the \texttt{dust2} parameter. FSPS defines this parameter as the optical depth $\tau$ in an extinction factor of $e^{-\tau}$ at the wavelength of 5500 \AA.}
\vspace{2.5mm}  
\label{tab:galaxy_prop}
\end{deluxetable*}

\begin{deluxetable*}{cccccc}[!h]
\setlength{\tabcolsep}{0.55em}
\tablewidth{0.9\textwidth}
\tabletypesize{\footnotesize}
\tablewidth{0pt}
\tablecaption{Radio SED Broken Power-law Fit Parameters}
\tablecolumns{6}
\tablehead{
\colhead{Object}
&\colhead{Observer Frame Epoch (days)}
&\colhead{Rest Frame Epoch (days)}&\colhead{$\nu_{\rm obs}$ (GHz)}& \colhead{$F_{\rm obs}$ ($\mu \mathrm{Jy}$)} & \colhead{$a_2$} }
\startdata
AT2023fhn & 88--96 & 71--77 & $4.0\pm0.3$ & $232\pm14$ & $-0.69\pm0.08$ \\
AT2023fhn & 137--138 & 110--111 &$5.5\pm0.3$ & $387\pm14$ & $-0.87\pm0.08$ \\
AT2023vth & 86--87 & 80--81 & $8.5\pm0.6$ & $(2.6\pm0.2)\times10^3$ & $-1.06\pm0.06$ \\
AT2023vth & 104--129 & 97--120 & $2.7\pm0.2$ & $(1.50\pm0.13)\times10^3$&$-0.55\pm0.04$\\
AT2024aehp & 140--155 & 120--132 & $12.8\pm1.1$&$(7.8\pm0.9)\times10^2$&$-0.67\pm0.09$\\
\enddata
\tablecomments{Results of a broken power-law fit (Equation~\ref{eq:broken}).  The values for the fitted peak frequency and flux density is measured in the observer's frame.}
\vspace{2.5mm}  
\label{tab:synchro_fit}
\end{deluxetable*}

\begin{deluxetable*}{cccccccccc}[!h]\label{tab:synchro}
\setlength{\tabcolsep}{0.35em}
\tablewidth{0.9\textwidth}
\tabletypesize{\footnotesize}
\tablewidth{0pt}
\tablecaption{Inferred Parameters from Synchrotron Self-Absorption Modeling}
\tablecolumns{10}
\tablehead{
\colhead{Object}&\colhead{$t_{\mathrm{rest}}$ }&\colhead{$\nu_{\rm p}$ (rest)}
& \colhead{$R$} & \colhead{$B$}& \colhead{$v/c$}& \colhead{Energy }& \colhead{$n_e$}&\colhead{$\dot{M}$}&\colhead{$m_p n_e R^3$ }\vspace{-2mm}\\ &\colhead{(days)}&\colhead{ (GHz)}
& \colhead{($10^{16}\,$cm)} & \colhead{(G)}& & \colhead{($10^{48}\,$erg)}& \colhead{ ($10^{3}$\,cm$^{-3}$)}&\colhead{ ($10^{-4}\,M_{\odot}/\mathrm{yr}$)}& \colhead{($10^{-5}\, M_{\odot}$)}}
\startdata
AT2022abfc &26&$>19$&$<1.4$&$>1.49$&$<0.21$&$<1.7$&$>5.2$&$>0.35$&$>1.3$\\
AT2023fhn&73&$5.0\pm0.4$&$13.0\pm1.1$&$0.32\pm0.03$&$0.57\pm0.05$&$55\pm6$&$0.022\pm0.007$&$0.12\pm0.02$&$4.1\pm1.7$\\
AT2023fhn&111&$6.8\pm0.4$&$12.0\pm0.7$&$0.41\pm0.02$&$0.38\pm0.02$&$74\pm5$&$0.10\pm0.03$&$0.47\pm0.06$&$15.2\pm4.5$\\
AT2023hkw&33&$>15$&$<4.2$&$>0.95$&$<0.44$&$<16$&$>0.40$&$>0.22$&$>2.4$\\
AT2023hkw&87&$>8.0$&$<9.2$&$>0.5$&$<0.37$&$<47$&$>0.16$&$>0.43$&$>10.3$\\
AT2023vth&81&$9.1\pm0.7$&$7.1\pm0.6$&$0.58\pm0.04$&$0.32\pm0.03$&$31\pm3$&$0.31\pm0.09$&$0.49\pm0.07$&$9.4\pm3.6$\\
AT2023vth&110&$2.9\pm0.2$&$17.6\pm1.3$&$0.19\pm0.01$&$0.53\pm0.04$&$50\pm6$&$0.010\pm0.003$&$0.10\pm0.01$&$4.7\pm1.6$\\
AT2024qfm&10&$>115$&$<0.3$&$>8.32$&$<0.13$&$<0.58$&$>449$&$>1.48$&$>1.3$\\
AT2024aehp&73&$>18$&$<1.5$&$>1.36$&$<0.08$&$<1.7$&$>30$&$>2.22$&$>9.2$\\ 
AT2024aehp&127&$15.0\pm1.3$&$5.5\pm0.5$&$0.91\pm0.08$&$0.16\pm0.02$&$34\pm5$&$3.1\pm1.1$&$3.0\pm0.5$&$44\pm20$\\ 
\enddata
\vspace{2.5mm}   
\tablecomments{The shock parameters we derive using Equations 2--6 for the various epochs of radio observations of each LFBOT.  When we can only constrain the peak of the radio SED by observing one side of the broken-power law, we calculate limits on these parameters using the observation closest to where the peak would be. The right-most column is an estimate of the total mass swept up by the shock assuming a constant $n_e$.}
\end{deluxetable*}

\clearpage

\startlongtable
\begin{deluxetable*}{cccccc}
\setlength{\tabcolsep}{0.55em}
\tablewidth{0.9\textwidth}
\tabletypesize{\footnotesize}
\tablewidth{0pt}
\tablecaption{Optical Photometry}
\tablecolumns{6}
\tablehead{
\colhead{Object}
&\colhead{UTC Time}&\colhead{$t_{\mathrm{rest}}$ (days)}
&\colhead{Facility}& \colhead{Filter} & \colhead{mag}}
\startdata
AT2022abfc&20221114 08:31&$-6$&P48/ZTF&{\em r}&$>19.68$\\ 
AT2022abfc&20221114 10:40&$-6$&P48/ZTF&{\em g}&$>19.65$\\ 
AT2022abfc&20221115 09:09&$-5$&P48/ZTF&{\em r}&$>20.40$\\ 
AT2022abfc&20221115 10:08&$-5$&P48/ZTF&{\em g}&$>20.19$\\ 
AT2022abfc&20221116 08:19&$-4$&P48/ZTF&{\em r}&$>19.36$\\ 
AT2022abfc&20221117 08:01&$-3$&P48/ZTF&{\em g}&$>19.69$\\ 
AT2022abfc&20221117 08:48&$-3$&P48/ZTF&{\em r}&$>19.53$\\ 
AT2022abfc&20221118 08:04&$-2$&P48/ZTF&{\em g}&$>20.51$\\ 
AT2022abfc&20221118 09:32&$-2$&P48/ZTF&{\em r}&$>20.13$\\ 
AT2022abfc&20221119 09:34&$-2$&P48/ZTF&{\em g}&$>19.71$\\ 
AT2022abfc&20221120 08:01&$-1$&P48/ZTF&{\em r}&$>20.35$\\ 
AT2022abfc&20221121 08:15&0&P48/ZTF&{\em r}&$19.46\pm0.11$\\ 
AT2022abfc&20221121 09:03&0&P48/ZTF&{\em g}&$19.49\pm0.14$\\ 
AT2022abfc&20221125 08:32&3&P48/ZTF&{\em r}&$19.55\pm0.16$\\ 
AT2022abfc&20221125 09:32&3&P48/ZTF&{\em g}&$19.71\pm0.20$\\ 
AT2022abfc&20221127 07:32&5&P48/ZTF&{\em r}&$20.32\pm0.32$\\ 
AT2022abfc&20221127 08:29&5&P48/ZTF&{\em g}&$20.21\pm0.19$\\ 
AT2022abfc&20221201 09:15&8&P48/ZTF&{\em r}&$>20.23$\\ 
AT2022abfc&20221210 07:04&16&P48/ZTF&{\em g}&$>19.60$\\ 
AT2022abfc&20221211 07:03&16&P48/ZTF&{\em g}&$>19.85$\\ 
AT2022abfc&20221217 06:01&21&P48/ZTF&{\em g}&$>20.35$\\ 
AT2022abfc&20221217 07:03&21&P48/ZTF&{\em r}&$>20.19$\\ 
AT2022abfc&20221220 06:11&24&P48/ZTF&{\em r}&$>19.21$\\ 
AT2022abfc&20221220 07:03&24&P48/ZTF&{\em g}&$>19.66$\\
AT2023fhn&20230408 06:00&$-3$&P48/ZTF&{\em i}&$>19.54$\\ 
AT2023fhn&20230408 06:29&$-3$&P48/ZTF&{\em g}&$>18.76$\\ 
AT2023fhn&20230408 07:12&$-3$&P48/ZTF&{\em r}&$>19.17$\\ 
AT2023fhn&20230410 04:53&$-2$&P48/ZTF&{\em r}&$19.68\pm0.09$\\ 
AT2023fhn&20230411 04:35&$-1$&P48/ZTF&{\em i}&$19.65\pm0.1$\\ 
AT2023fhn&20230412 03:51&0&P48/ZTF&{\em r}&$19.13\pm0.05$\\ 
AT2023fhn&20230412 04:35&0&P48/ZTF&{\em g}&$18.69\pm0.03$\\ 
AT2023fhn&20230412 05:22&0&P48/ZTF&{\em g}&$18.63\pm0.03$\\ 
AT2023fhn&20230412 05:27&0&P48/ZTF&{\em g}&$18.72\pm0.03$\\ 
AT2023fhn&20230417 05:05&4&P48/ZTF&{\em g}&$20.09\pm0.09$\\ 
AT2023fhn&20230417 05:49&4&P48/ZTF&{\em r}&$20.72\pm0.15$\\ 
AT2023fhn&20230419 04:40&6&P60/SEDM&{\em g}&$>19.30$\\ 
AT2023fhn&20230419 04:44&6&P60/SEDM&{\em r}&$>19.93$\\ 
AT2023fhn&20230419 05:20&6&P48/ZTF&{\em r}&$20.57\pm0.22$\\ 
\enddata
\vspace{2.5mm} 
\tablecomments{The full set of photometry for all six LFBOTs obtained from Fritz marshal are available online in a machine-readable table.  We omit all observations prior to six days before $t_0$ and all ZTF non-detections more than fifty days after $t_0$.  For non-detections, we provide 3-$\sigma$ upper limits.}
\label{tab:photometry}
\end{deluxetable*}

\end{document}